\documentclass[twoside,11pt]{article}

\usepackage{jmlr2e}

\PassOptionsToPackage{usenames,dvipsnames,dvinames}{xcolor}
\usepackage[usenames,dvipsnames,dvinames]{xcolor}
\usepackage{multirow}
\usepackage{microtype}
\usepackage{enumitem}
\usepackage{amsmath}
\usepackage[mathscr]{eucal}
\usepackage{amssymb}
\usepackage{footnote}
\usepackage{graphicx} % more modern
\usepackage{algorithmic}
\usepackage[vlined,ruled]{algorithm2e}
\usepackage{wrapfig}
\usepackage{float}
\usepackage{caption}
\usepackage{multicol}

\usepackage{bbm}
\usepackage{mathtools}
\DeclarePairedDelimiter{\ceil}{\lceil}{\rceil}
\newcommand{\theproblems}{Problems~\ref{prob:maxmin-genmix} and~\ref{prob:minmax-genmix}}
\newcommand{\minmaxprob}{Problem~\ref{prob:minmax-genmix}}
\newcommand{\maxminprob}{Problem~\ref{prob:maxmin-genmix}}
\newcommand{\lovasz}{Lov\'asz }
\newcommand{\lovaszRound}{Lov\'aszRound}
\DeclareMathOperator*{\argmax}{argmax}
\DeclareMathOperator*{\argmin}{argmin}
\providecommand{\doarxiv}{true}
\usepackage{ifthen}
\newboolean{isarxiv}
\setboolean{isarxiv}{\doarxiv} 
\ifthenelse{\boolean{isarxiv}}{%
\newcommand{\arxiv}[1]{#1}
\newcommand{\notarxiv}[1]{}
}{
\newcommand{\arxiv}[1]{}
\newcommand{\notarxiv}[1]{#1}
}

\def\Naive{Na\"{\i}ve\xspace}
\newcommand{\arxivalt}[2]{\ifthenelse{\boolean{isarxiv}}{#1}{#2}}
\newcommand{\arxivaltr}[2]{\ifthenelse{\boolean{isarxiv}}{#2}{#1}}

\usepackage{subcaption}

\providecommand{\dodraft}{false}
\usepackage{messages}

\arxivalt{}
{}

\newcounter{problemnumber}
\renewcommand{\theproblemnumber}{\arabic{problemnumber}}

\usepackage{natbib}
\usepackage[colorlinks=true, citecolor=Violet,linkcolor=Mahogany,urlcolor=blue]{hyperref}
\ShortHeadings{Submodular Partitioning}{Wei, Iyer, Wang, Bai, Bilmes}
\firstpageno{1}

\begin{document}

\title{Mixed Robust/Average Submodular Partitioning: Fast Algorithms,
  Guarantees, and Applications to Parallel Machine Learning and Multi-Label
  Image Segmentation}

\author{\name Kai Wei \email kaiwei@uw.edu \\
       \addr Department of Electrical Engineering\\
       University of Washington\\
       Seattle, WA 98195, USA
       \AND
	  \name Rishabh Iyer \email rkiyer@uw.edu \\ 	  
	  \addr Department of Electrical Engineering\\
       University of Washington\\
       Seattle, WA 98195, USA
       \AND
       	  \name Shengjie Wang \email wangsj@uw.edu\\ 	  
	  \addr Department of Computer Science Engineering\\
       University of Washington\\
       Seattle, WA 98195, USA
		\AND
      \name Wenruo Bai \email wrbai@uw.edu \\ 	  
	  \addr Department of Electrical Engineering\\
       University of Washington\\
       Seattle, WA 98195, USA
       \AND
       \name Jeff Bilmes \email bilmes@uw.edu\\ 	  
	  \addr Department of Electrical Engineering\\
       University of Washington\\
       Seattle, WA 98195, USA       
       }

\editor{}
\maketitle

\begin{abstract}%   <- trailing '%' for backward compatibility of .sty file
  We study two mixed robust/average-case submodular 
  partitioning problems that we collectively call \emph{Submodular
    Partitioning}. These problems generalize both purely robust instances
  of the problem (namely \emph{max-min submodular fair allocation}
  (SFA) \cite{golovin2005max} and \emph{min-max submodular load
    balancing} (SLB) \cite{svitkina2008submodular}) and also
  generalize  average-case instances (that is the \emph{submodular welfare
    problem} (SWP)~\cite{vondrak2008optimal} and \emph{submodular
    multiway partition} (SMP)~\cite{chekuri2011approximation}). While
  the robust versions have been studied in the theory
  community~\cite{goemans2009approximating, golovin2005max,
    khot2007approximation, svitkina2008submodular,
    vondrak2008optimal}, existing work has focused on tight
  approximation guarantees, and the resultant algorithms are not, in
  general, scalable to very large real-world applications. This is in
  contrast to the average case, where most of the algorithms are
  scalable.  In the present paper, we bridge this gap, by proposing
  several new algorithms (including those based on greedy, majorization-minimization,
  minorization-maximization, and relaxation algorithms) that not only
  scale to large sizes but that also achieve theoretical
  approximation guarantees close to the state-of-the-art, and in some cases
  achieve new tight bounds. We
  also provide new scalable algorithms that apply to additive
  combinations of the robust and average-case extreme objectives.  We show
  that these problems have many applications in machine learning (ML).
  This includes: 1) data partitioning and load balancing for
  distributed machine algorithms on parallel machines; 2) data clustering; and
  3) multi-label image segmentation with (only) Boolean submodular
  functions via pixel partitioning.
  %We show comparable hardness results, thereby providing a complete theoretical picture of these problems. 
  We empirically demonstrate the efficacy of our algorithms on
  real-world problems involving data partitioning for distributed
  optimization of standard machine learning objectives (including both
  convex and deep neural network objectives), and also on purely
  unsupervised (i.e., no supervised or semi-supervised learning, and
  no interactive segmentation) image segmentation.  \looseness-1
\end{abstract}


\section{Introduction}
\label{sec:introduction}
\notarxiv{\vspace{-0.5\baselineskip}}

The problem of set partitioning arises in many machine learning (ML)
and data science applications. Given a set $V$ of items, an
$m$-partition $\pi = (A_1^\pi, A_2^\pi, \dots, A_m^\pi)$ is a size $m$
set of subsets of $V$ (called blocks) that are non-intersecting (i.e.,
$A_i^\pi \cap A_j^\pi = \emptyset$ for all $i\neq j$) and covering
(i.e., $\bigcup_{i=1}^m A_i^\pi = V$). The goal of a partitioning
algorithm is to produce a partitioning that is measurably good in some
way, often based on an aggregation of the judgements of the internal
goodness of the resulting blocks.

In data science and machine learning applications, a partitioning is
almost always the end result of a clustering (although in some cases a
clustering might allow for intersecting subsets) in which $V$ is
partitioned into $m$ clusters (which we, in this paper, refer to as
blocks).  Most clustering problems are based on optimizing an
objective that is either a sum, or an average-case,
utility where the goal is to optimize the sum of individual cluster
costs --- this includes the ubiquitous $k$-means procedure
\cite{lloyd1982least,arthur2007k} where the goal is to construct a
partitioning based on choosing a set of cluster centers that minimizes
the total sum of squared distances between each point and its closest
center. More rarely, clustering algorithms may be based on robust
objective functions \cite{garcia2010review}, where the goal is to
optimize the worst-case internal cluster cost (i.e., quality of the
clustering is based solely on the quality of the worst internal cluster cost within
the clustering). The average case vs.\ worst case cluster cost
assessment distinction are extremes along a continuum, although all
existing algorithms operate only at, rather than in between, these
extremes.

There is another way of categorizing clustering algorithms, and that
based on the goal of each resultant cluster. Most of the time,
clustering algorithms attempt to produce clusters containing items
similar to or near each other (e.g., with $k$-means, a cluster
consists of a centroid and a set of nearby points), and dissimilarity
exists, ideally, only between different clusters.  An alternate
possible goal of clustering is to have each block contains as diverse
a set of items as possible, where similarity exist between rather than
within clusters. For example, partitioning the vertices of a graph
into a set of $k$ independent (or, equivalently, stable) sets would
fall into this later category, assuming the graph has edges only
between similar vertices. This subsumes graph $k$-colorability
problems, one of the most well-known of the NP-complete problems,
although in some special cases it is solvable in polynomial time
\cite{brandstadt1996partitions,feder1999complexity,hell2004partitioning}.
This general idea, where the clusters are as internally diverse as
possible using some given measure of diversity, has been called
anticlustering
\cite{valev1983set,valev1998set,spath1986anticlustering} in the past
and, as can be seen from the above, comprises in general some
difficult computational problems.

A third way of further delineating cluster problems is based on how
each cluster's internal quality is measured. In most clustering
procedures, each cluster $A^\pi_i$ for $i \in \{1,2, \dots, m\}$ is
internally judged based on the same function $f$ regardless of which
cluster is being evaluated.  The overall clustering is then an
aggregation of the vector of $m$ scores $f(A^\pi_1),
f(A^\pi_2), \dots, f(A^\pi_m)$. We call this strategy a homogeneous
clustering evaluation since every cluster is internally evaluated
using the same underlying function. An alternate strategy to evaluate
a clustering uses a different function for each cluster. For example,
in assignment problems, there are $m$ individuals each of whom is
assigned a cluster, and each individual might not judge the same
cluster identically. In such cases, we have $m$ functions $f_1, f_2,
\dots, f_m$ that, respectively, internally evaluate the corresponding
cluster, and the overall clustering evaluation is based on an
aggregation of the vector of scores $f_1(A^\pi_1), f_2(A^\pi_2), \dots,
f_m(A^\pi_m)$. We call this a heterogeneous clustering evaluation.

The above three strategies to determine the form of cluster evaluation
(i.e., average vs.\ robust case, internally similar vs.\ diverse
clusters, and homogeneous vs. heterogeneous internal cluster
evaluation) combined with the various ways of judging the internal
cluster evaluations, and ways to aggregate the scores, leads to a
plethora of possible clustering objectives. Even in the simple cases,
however (such as $k$-means, or independent sets of graphs), the
problems are generally hard and/or difficult to approximate.

This paper studies partitioning problems that span the range within
and across the aforementioned three strategies, and all from the
perspective of submodular function based internal cluster evaluations.
In particular, we study problems of the following form:
\refstepcounter{problemnumber}\label{prob:maxmin-genmix}
\begin{align}
\text{Problem \theproblemnumber: }\qquad
\max_{\pi \in \Pi} \Bigl[(1 - \lambda) \min_{i} f_i(A_i^{\pi}) + \frac{\lambda}{m} \sum_{j = 1}^m f_j(A^{\pi}_j)\Bigr],
\intertext{ and } 
% \abovedisplayshortskip=.2ex\belowdisplayshortskip=.2ex\abovedisplayskip=.2ex\belowdisplayskip=.2ex
% 
% \begin{align}
\refstepcounter{problemnumber}\label{prob:minmax-genmix}
\text{Problem \theproblemnumber: }\qquad
\min_{\pi \in \Pi} \Bigl[(1 - \lambda) \max_{i} f_i(A^\pi_i) + \frac{\lambda}{m} \sum_{j = 1}^m f_j(A^\pi_j)\Bigr],
\end{align}
where $0 \leq \lambda \leq 1$, the set of sets $\pi = (A^\pi_1,
A^\pi_2, \cdots, A^\pi_m)$ is an ordered partition of a finite set $V$
(i.e, $\cup_i A^\pi_i = V$ and $\forall i\neq j, A^\pi_i \cap A^\pi_j
= \emptyset$), and $\Pi$ refers to the set of all ordered partitions
of $V$ into $m$ blocks. In contrast to the notion of the partition
often used in the computer science and mathematical communities, we
clarify that an ordered partition $\pi$ is fully characterized by both
its constituent blocks $\{A^\pi_i\}_{i=1}^m$ as well as the ordering
of the blocks --- this allows us to cover assignment
problems, where the $i^\text{th}$ block is assigned to the
$i^\text{th}$, potentially distinct, individual.
The parameter $\lambda$ controls the objective:
$\lambda=1$ corresponds to the average case, $\lambda=0$ to the robust case, and
$0 < \lambda < 1$ is a mixed case.  We are unaware, however, of any
previous work (other than our own) that allows for a mixing between
worst- and average-case objectives in the context of any form of set
partitioning.  For convenience, we also define $\bar \lambda
\triangleq 1-\lambda$ in the below.  In general, \theproblems{} are
hopelessly intractable, even to approximate, but we assume that the
$f_1, f_2, \cdots, f_m$ are all monotone non-decreasing (i.e., $f_i(S)
\leq f_i(T)$ whenever $S \subseteq T$), normalized ($f_i(\emptyset) =
0$), and submodular \cite{fujishige2005submodular} (i.e., $\forall S,
T \subseteq V$, $f_i(S) + f_i(T) \geq f_i(S \cup T) + f_i(S \cap
T)$). These assumptions allow us to develop fast, simple, and scalable
algorithms that have approximation guarantees, as is done in this
paper. These assumptions, moreover, allow us to retain the naturalness
and applicability of Problems~\ref{prob:maxmin-genmix}
and~\ref{prob:minmax-genmix} to a wide variety of practical problems.

Submodularity is a natural property in many real-world ML
applications\arxivalt{~\cite{wei-doc-sum-speech-sel-2013,
    zheng2014submodular, nagano2010minimum,
    wei2014-supervised-large-icassp, krause2008near,
    jegelka2011-nonsubmod-vision,
    das2011submodular_old,krause08robust,wei2015-submodular-data-active}}{~\cite{smola2015graphpartitioning,
    jegelka2011-nonsubmod-vision,krause08robust,wei2015-submodular-data-active}}.
When minimizing, submodularity naturally model notions of interacting
costs and complexity, while when maximizing it readily models notions
of diversity, summarization quality, and information. 
Hence, Problem~\ref{prob:maxmin-genmix} asks for a partition whose blocks
each (to the extent that $\lambda$ is close to 0) and that
collectively (to the extent that $\lambda$ is close to 1) are diverse,
as represented by the submodular functions.  The reason for this is
that submodular functions are natural at representing diversity, and a
subset $A\subseteq V$ is considered diverse, relative to $V$ and
considered amongst other sets also of size $|A|$, if $f(A)$ is as large
as possible.  Problem~\ref{prob:minmax-genmix}, on the other hand, asks
for a partition whose blocks each (to the extent that $\lambda$ is
close to 0) and that collectively (to the extent that $\lambda$ is
close to 1) are internally similar (as is typical in clustering) or
that are redundant, as measured by the set of submodular functions.
Taken together, we call \theproblems{} \emph{Submodular Partitioning}.
As described above, we further categorize these problems depending on
if the $f_i$'s are identical to each other (the \emph{homogeneous}
case) or not (the \emph{heterogeneous} case).\footnote{Similar
  sub-categorizations have been called the ``uniform'' vs.\ the
  ``non-uniform'' case in the
  past~\cite{svitkina2008submodular,goemans2009approximating},
but we utilize the names 
homogeneous and heterogeneous to avoid implying
that we desire the final set of submodular function valuations
be uniform, which they do not need to be in our case.
} The
heterogeneous case clearly generalizes the homogeneous setting, but as
we will see, the additional homogeneous structure can be exploited to
provide more efficient and/or tighter algorithms.\looseness-1

The paper is organized as follows.  We start by motivating
\theproblems{} in the context of machine learning, where there are a
number of relevant applications, two of which are expounded upon in
Sections~\ref{sec:data-part-parall} and~\ref{sec:multi-label-image},
and some of which we ultimately utilize (in
Section~\ref{sec:experiments}) to evaluate on real-world data.  We
then provide algorithms for the robust ($\lambda = 0$) versions of
\maxminprob{} and \minmaxprob{} in
Section~\ref{sec:algorithms_for_SFA_SLB}. Algorithms for
\theproblems{} with general $\lambda$ are given in
Section~\ref{sec:gener-subm-part}. %Section~\ref{sec:applications_both_problems}
We further explore the applications given
section~\ref{sec:data-part-parall} and~\ref{sec:multi-label-image} and provide an empirical validation in
Section~\ref{sec:experiments}. We conclude in
Section~\ref{sec:conclusions}.

% We first introduce a practical and
% scalable greedy algorithm for solving Problem~\ref{prob:maxmin} with
% $\lambda=0$ under the homogeneous setting. We then empirically
% demonstrate the efficacy of the proposed algorithm for obtaining data
% partitions in parallel machine learning schemes, such as ADMM and
% distributed neural network training.
% %apply the proposed algorithms to obtain data partitions for parallel schemes of training statistical models, such as ADMM, and distributed neural network training. 

\subsection{Data Partitioning for Parallel Machine Learning}
\label{sec:data-part-parall}

Many of today's statistical learning methods can take advantage of the
vast and unprecedented amount of training data that now exists and is
readily available, as indeed ``there is no data like more data.''  On
the other hand, big data presents significant computational challenges
to machine learning since, while big data is still getting bigger, it
is expected that we are nearing the end of Moore's
law~\cite{thompson2006moore}, and single threaded computing speed has
unfortunately not significantly improved since about 2003.  In light
of the ever increasing flood of data that is becoming available, it is
hence imperative to develop efficient and scalable methods for large
scale training of statistical models.  One strategy is to develop
smarter and more efficient algorithms, and indeed this fervently is
being pursued in the machine learning community.  Another natural and
complementary strategy, also being widely pursued, is via parallel
and distributed computing.

Since machine learning procedures are performed over sets of data, one
simple way to achieve parallelism is to split the data into chunks
each of which resides on a compute node.  This is the idea behind many
parallel learning approaches such as ADMM~\cite{boyd2011distributed}
and distributed neural network training~\cite{povey2014parallel}, to
name only a few.  Such parallel schemes are often performed where the
data samples are distributed to their compute nodes in an arbitrary or
random fashion.  However, there has apparently been very little work
on how to intelligently split the data to ensure that the resultant
model can be learned in an efficient, or a particularly good, manner.

%As an alternate strategy, if the data is intelligently
%partitioned such that each block of samples can itself lead to a good
%approximate solution, a consensus amongst the distributed results
%could be reached more quickly than when under a poor partitioning.
%Submodular functions can in fact express the value of a subset of
%training data for certain machine learning risk functions,
%e.g.,~\cite{wei2015-submodular-data-active} in the case of
%  classification.  Using these functions within
%Problem~\ref{prob:maxmin}, one can expect a partitioning (by formulating the problem as an instance of Problem~\ref{prob:maxmin}, where
%each block is a good \emph{representative} of the entire set, thereby
%achieving faster convergence in distributed settings.  We demonstrate
%empirically, in Section~\ref{sec:experiments}, that this provides
%better results on several machine learning tasks, including the
%training of deep neural networks.

One way to approach this problem is to consider a class of ``utility''
functions on the training data.  Given a set $V=\{v_1,\dots, v_n\}$ of
training data items, suppose that we have a set function $f: 2^V \to
\mathbb R_+$ that measures the utility of subsets of the data set $V$.
That is, given any $A \subseteq V$, $f(A)$ measures the utility of the
training data subset $A$ for producing a good resulting trained model.
Given a parallel training scheme (e.g., ADMM) with $m$ compute nodes,
we consider an $m$-partition $\pi = ( A_1^\pi, A_2^\pi, \dots, A_m^\pi
)$  of the entire training data $V$, where we send
the $i^\text{th}$ block $A_i^\pi$ of the partition $\pi$ to the
$i^\text{th}$ compute node.  If each compute node $i$ has a block of
the data $A_i^{\pi}$ that is non-representative of the utility of the
whole (i.e., $f(A_i^m) \ll f(V)$), the parallel learning algorithm, at
each iteration, might result in the compute nodes deducing models that
are widely different from each other. Any subsequent
aggregation of the models could then result in a joint model
that is non-representative of the whole, especially in a non-convex
case like deep neural network models. On the other hand, suppose that
an intelligent partition $\pi$ of the training data $V$ is achieved
such that each block $A_i^{\pi}$ is highly representative of the whole
(i.e., $f(A_i^m) \approx f(V),\forall i$). In this case, the models
deduced at the compute nodes will tend to be close to a model 
trained on the entire data, and any aggregation of the resulting
models (say via ADMM) will likely be better.  
An intelligent data partition, in fact,  may have a positive effect in two
ways: 1) it can lead to a better final solution (in the non-convex
case), and 2) faster convergence may be achieved even in the convex
case thanks to the fact that any ``oscillations'' between a
distributed stage (where the compute nodes are operating on local
data) and an aggregation stage (where some form of model average is
computed) could be dampened and hence reduced.

It should be noted that the above process should ideally also be done
in the context of any necessary parallel load balancing, where data is
distributed so that each processing node has roughly the same amount
of compute needing to be performed. Our approach above, by contrast,
might be instead called ``information balancing,'', where we ensure
that each node has a representative amount of information so that
reasonable consensus solutions are constructed.

In this work, based on the above intuition, we mathematically describe
this goal as obtaining an $m$-partition of the data set $V$ such that
the worst-case or the average-case utility among all blocks in the
partition is maximized.  More precisely, this can be formulated
Problem~\ref{prob:maxmin-genmix} above, where the set of sets $\pi =
(A^\pi_1, A^\pi_2, \cdots, A^\pi_m)$ forms a partition of the training
data $V$. The parameter $\lambda$ controls the nature of the
objective.  For example, $\lambda=0$ is the robust case, which is
particularly important in mission critical applications, such as
parallel and distributed computing, where one single poor partition
block can significantly slow down an entire parallel machine (as all
compute nodes might need to spin block while waiting for a slow node to
complete its round of computation). $\lambda=1$ is the average case.
Taking a weighted combination of the both robust and average case
objectives (the mixed case, $0 <\lambda<1$) allows one to balance
between optimizing worst-case and overall performance.  

A critical aspect of Problem~\ref{prob:maxmin-genmix} is that
submodular functions are an ideal class of functions for modeling
information over training data sets. For
example,~\cite{wei2015-submodular-data-active} show that the utility
functions of data subsets for training certain machine learning
classifiers can be derived as submodular functions.  If $f$ is
selected as the class of submodular functions that appropriately model
the notion of utility for a given machine learning setting (which
could be different depending, say, on what form of training one is
doing), solving the homogeneous instance of
Problem~\ref{prob:maxmin-genmix} with $f$ as the objective, then,
addresses a core goal in modern machine learning on big data sets,
namely how to intelligently partition and distribute training data to
multiple compute nodes.

It should be noted that a random partition might have a high
probability of performing well, and might have exponentially small
probability of producing a partition that performs very poorly.  This
could be used as an argument for utilizing just a random data
partitioning procedure.  On the other hand, there are also quite
likely a small number of partitions that perform exceedingly well and
a random partition also has a very small probability of achieving on
one of these high quality partitions. Our long-term quest is to
develop methods that increase the likelihood that we can discover one
of these rare but high performing partitions. Moreover, there are
other applications (i.e., locality maximization via bipartite graphs,
where the goal is to organize the data so as to improve data locality,
as in the work of
\cite{smola2015graphpartitioning,alistarhStreamingMinMax_nips2015}
where a random partition is quite likely to perform very poorly. This
again is an additional potential application of our work.

\subsection{Multi-label Image Segmentation}
\label{sec:multi-label-image}

Segmenting images into different regions is one of the more important
problems in computer vision.  Historically, the problem was addressed
by finding a MAP solution to to a Markov random field (MRF), where one constructs a distribution $p(y,\bar x) =
\frac{1}{Z}\exp( - E(y,\bar x))$. Here, $y = (y_1, y_2, \dots,
y_{|V|})$ is a set of pixel labels for a set of pixels $V$, $\bar x =
(\bar x_1, \bar x_2, \dots, \bar x_{|V|})$ is a set of pixel values, and
where $E(y, \bar x)$ is an energy function.  For binary image
segmentation, $y_i \in {0,1}$, and for multi-label instance, $y_i \in
\{0, 1, \dots, m - 1 \}$, and the image labeling problem computes
$\max_{y} p(y|\bar x)$. In many instances
\cite{Boykov2001,Boykov2004,boyVek06}, it was found that performing
this maximization can be performed by a graph cut algorithm, in
particular when the energy function takes the form $E(y,\bar x) =
\sum_{v \in V} \phi_v(y_v, \bar x_v) + \sum_{C \in \mathcal C}
\phi_C(y_C)$ where $\mathcal C$ are the set of cliques of a
graph. When the order of the cliques is two, and when the $\phi_C$
functions are submodular, the MAP inference problem can be solved via
a minimum graph cut algorithm, and if the order if $\phi_C$ is larger,
then general submodular function minimization
\cite{fujishige2005submodular} (which runs in polynomial time in
$|V|$) may be used. This corresponds, however, only to the binary
image segmentation problem. In order to perform multi-label image
segmentation, there have been various elaborate extensions of the
energy functions to support this (e.g.,
\cite{jegelka2011-nonsubmod-vision,shelhamer14,heng15,taniai15,kohliOJ13,silberman14}
to name just a few), but in all cases the energy function $E(y, \bar
x)$ has to be extended to, at the very least, a non-pseudo Boolean
function.

Supposing that we know the number of labels $m$ in advance, an
alternate strategy for performing multi-label image segmentation that
still utilizes a pseudo-Boolean function is to perform a partition of
the pixels. Each block of the partition corresponds to one segment,
and for a pixel partitioning to make sense in an image segmentation
context, it should be the case that each block consists of pixels that
are as non-diverse as possible. For example, the pixels consisting of
one object (e.g., a car, a tree, a person) tend to have much less
diversity than do a set pixels chosen randomly across the image.  A
natural set partitioning objective to address multi-label image
segmentation is then \minmaxprob{}, the reason being that submodular
functions, when minimized, tend to choose sets that are non-diverse
and, when forced also to be large, are redundant.  \minmaxprob{} finds
a partition that simultaneously minimizes the worst case diversity of
any block, and the average case diversity over all blocks. Note that
in this case, the energy function $E(y, \bar x)$ takes on a different
form and in order to define it, we need a bit of additional
information. The vector $y \in \{0, 1, \dots, m -1 \}^V$ corresponds
to the multi-label labeling of all the pixels in the image.
For $i \in \{1, 2, \dots, m-1\}$ define $A_i(y) = \{ v \in V : y_v = i\} \subseteq V$
be the subset of $V$ corresponding to the pixels with label $i$ in vector $y$. The
multi-label energy function then becomes:
\begin{align}
E(y, \bar x) = 
\sum_{v \in V} \phi_v(y_v, \bar x_v)
+ 
\Bigl[(1 - \lambda) \max_{i} f_i(A_i(y)) + \frac{\lambda}{m} \sum_{j = 1}^m f_j(A_j(y))\Bigr],
\end{align}
and \minmaxprob{} minimizes it.
After solving \minmaxprob{} 
and given the
result $\pi$ of such a partition procedure, the pixels corresponding to
block $A_i^\pi$ should, intuitively, correspond to an object. What is
interesting about this approach to multi-label segmentation, however,
is that the underlying submodular function (or set thereof) is still
binary (i.e., it operates on subsets of sets, or equivalently, on
vectors Boolean values corresponding to the characteristic vectors of
sets). Instead of changing the function being optimized, \minmaxprob{}
changes the way that the function is being used to be more appropriate
for the multi-label setting.  We evaluate \minmaxprob{} for image
segmentation in Section~\ref{sec:experiments}.

\subsection{Sub-categorizations and Related Previous Work}
\label{sec:subc-prev-work}

%\everyone{Jeff}{We need to reorganize this section. We should talk about all the problem 1 results first, and then move on to the problem 2 results (i.e., keep things in order).}

\arxivalt{

\paragraph{Problem~\ref{prob:maxmin-genmix}: } 
Special cases of Problem~\ref{prob:maxmin-genmix} have appeared previously in the literature. \jeff{Kai}{I removed the matroid example in this case, since it is unclear how it can be formulated as an instance of either Problem 1 or 2.}  Problem~\ref{prob:maxmin-genmix} with $\lambda=0$ is called \emph{submodular fair allocation} (SFA), which has been studied mostly in the heterogeneous setting. When $f_i$'s are all modular, the tightest algorithm, so far, is to iteratively round an LP solution achieving $O(1/(\sqrt{m}\log^3 m))$ approximation~\cite{asadpour2010approximation}, whereas the problem is NP-hard to $1/2+\epsilon$ approximate for any $\epsilon > 0$~\cite{golovin2005max}.  When $f_i$'s are submodular,~\cite{golovin2005max} gives a matching-based algorithm with a factor $1/(n-m+1)$ approximation that performs poorly when $m\ll n$. \cite{khot2007approximation} propose a binary search algorithm yielding an improved factor of $1/(2m-1)$.  Another approach approximates each submodular function by its ellipsoid approximation (non-scalable) and reduces SFA to its modular version leading to an approximation factor of $O(\sqrt{n} m^{1/4} \log n \log^{3/2} m)$.  These approaches are theoretically interesting, but they either do not fully exploit the problem structure or cannot scale to large problems.  On the other hand, Problem~\ref{prob:maxmin-genmix} for $\lambda=1$ is called \emph{submodular welfare}. This problem has been extensively studied in the literature and can be equivalently formulated as submodular maximization under a partition matroid constraint~\cite{vondrak2008optimal}.  It admits a scalable greedy algorithm that achieves a $1/2$ approximation~\cite{fisher1978analysis}. More recently a multi-linear extension based algorithm nicely solves the submodular welfare problem with a factor of $(1-1/e)$ matching the hardness of this problem~\cite{vondrak2008optimal}.  It is worth noting that the homogeneous instance of the submodular welfare with $m=2$ generalizes the well-known max-cut problem~\citep{goemans1995improved}, where the submodular objective is defined as the submodular neighborhood function~\citep{nipssubcons2013,schrijver2003combinatorial}.  Given a set of vertices $X \subset V$, of a graph $G=(V,E)$, the neighborhood function $f(X)$ counts the number of edges that are incident to at least one vertex in $X$, a monotone non-decreasing submodular function.  The function $h(X) = f(X) + f(V \setminus X) - f(V)$ is then the cut function, and since the last term is a constant, maximizing $h$ corresponds to the submodular welfare problem.  As far as we know, Problem~\ref{prob:maxmin-genmix} for general $0<\lambda<1$ has not been studied in the literature.

\begin{table}[t]
{
\centerline{\begin{tabular}{|c|c|}
\hline
\multicolumn{2}{|c|}{{\color[HTML]{000000} \maxminprob{} (Max-(Min$+$Avg))}}             \\ \hline
                     & Approximation factor      \\ \hline					 
 $\lambda=0$, \textsc{BinSrch}~\cite{khot2007approximation}  & $1/(2m-1)$    \\ \hline
$\lambda=0$, \textsc{Matching}~\cite{golovin2005max}  & $1/(n-m+1)$    \\ \hline
$\lambda=0$, \textsc{Ellipsoid}~\cite{goemans2009approximating} &   $O(\sqrt{n} m^{1/4} \log n \log^{3/2} m)$   \\ \hline
$\lambda = 1$, \textsc{GreedWelfare}~\cite{fisher1978analysis}  &           $1/2$              \\ \hline
$\lambda=0$, \textsc{GreedSat}$^*$ &          $(1/2-\delta,\frac{\delta}{1/2+\delta})$   \\ \hline
$\lambda=0$, \textsc{MMax}$^*$ & $O( \underset{i}{\min} \frac{1+(|{A}^{\hat{\pi}}_i|-1)(1-\kappa_{f_i}({A}^{\hat{\pi}}_i))}{|A^{\hat{\pi}}_i| \sqrt{m} \log^3 m})$       \\ \hline
$\lambda=0$, \textsc{GreedMax}$^{\dagger *}$       &           $1/m$   \\ \hline
$0<\lambda<1$, \textsc{GeneralGreedSat}$^{*}$       &           $\lambda/2$   \\ \hline
$0 < \lambda < 1$, \textsc{CombSfaSwp}$^*$  &    $\max\{\frac{\beta \alpha}{\bar{\lambda}\beta+\alpha},\lambda\beta \}$            \\ \hline
$0 < \lambda < 1$, \textsc{CombSfaSwp}$^{\dagger *}$  &  $\max\{\min\{\alpha,\frac{1}{m}\}, \frac{\beta \alpha}{\bar{\lambda}\beta + \alpha}, \lambda \beta\}$ \\ \hline
\hline
 $\lambda = 0$, Hardness &       $1/2$~\cite{golovin2005max}  \\ \hline
 $\lambda = 1$, Hardness  &   $1 - 1/e$~\cite{vondrak2008optimal}                \\ \hline
\end{tabular}
}
\caption[foo]{Summary of our contributions and existing work 
on Problem~\ref{prob:maxmin-genmix}.\footnotemark{} See text for details.\footnotetext{Results obtained in this paper are marked as $*$. Methods for only the homogeneous setting are marked as $\dagger$.}
}}
\label{tab:summary_theoretical_results_max_case}
\end{table}

% \footnotetext{Results obtained in this paper are marked as $*$. Methods for only the homogeneous setting are marked as $\dagger$.}

\paragraph{Problem~\ref{prob:minmax-genmix}: }
When $\lambda=0$, Problem~\ref{prob:minmax-genmix} is studied as \emph{submodular load balancing} (SLB). When $f_i$'s are all modular, SLB is called {\em minimum makespan
  scheduling}. \arxivalt{In the homogeneous setting,
~\cite{hochbaum1988polynomial} give a PTAS scheme\arxiv{
  ($(1+\epsilon)$-approximation algorithm which runs in polynomial
  time for any fixed $\epsilon$)}, while an LP relaxation algorithm
provides a $2$-approximation for the heterogeneous
setting~\cite{lenstra1990approximation}.}{An LP relaxation algorithm provides a $2$-approximation for the heterogeneous setting~\cite{lenstra1990approximation}.}  When the objectives are
submodular, the problem becomes much harder.  Even in the homogeneous
setting,~\cite{svitkina2008submodular} show that the problem is
information theoretically hard to approximate within $o(\sqrt{n/\log
  n})$. They provide a balanced partitioning algorithm yielding a
factor of $\min\{m,n/m\}$ under the homogeneous setting. They also
give a sampling-based algorithm achieving $O(\sqrt{n/\log n})$ for the homogeneous setting. However, the sampling-based algorithm is not practical
and scalable since it involves solving, in the worst-case,
$O(n^3\log n)$ instances of submodular function
minimization each of which in general currently requires $O(n^5\gamma+n^6)$
computation~\cite{orlin2009faster}, where $\gamma$ is the cost of
a function valuation.
Similar to Submodular Fair Allocation,~\cite{goemans2009approximating} applies the same ellipsoid approximation techniques 
leading to a factor of $O(\sqrt{n }\log
n)$~\cite{goemans2009approximating}. 
When $\lambda=1$, Problem~\ref{prob:minmax-genmix} becomes the \emph{submodular
  multiway partition} (SMP) for which one can obtain a relaxation based
$2$-approximation~\cite{chekuri2011approximation} in the homogeneous case. In the non-homogeneous case, the guarantee is $O(\log n)$~\cite{chekuri2011submodularSCA}. Similarly, \cite{zhao2004generalized, narasimhan2005q} propose a greedy splitting $2$-approximation algorithm for the homogeneous setting. 
To the best of our knowledge, there does not exist any work on Problem~\ref{prob:minmax-genmix} with general $0<\lambda<1$.

We illustrate both the special and the general cases of Problem~\ref{prob:maxmin-genmix} and~\ref{prob:minmax-genmix} in Figure~\ref{fig:illustration_problem_in_table}.

\arxiv{
\begin{table}[t]
{
\centerline{\begin{tabular}{|c|c|}
\hline
 \multicolumn{2}{|c|}{\minmaxprob{} (Min-(Max$+$Avg))}              \\ \hline                                     & Approximation factor \\ \hline					
 $\lambda=0$, \textsc{Balanced}$^{\dagger}$~\cite{svitkina2008submodular}             &                           $\min\{m,n/m\}$       \\ \hline
 $\lambda=0$, \textsc{Sampling}~\cite{svitkina2008submodular}  &  $O(\sqrt{n\log n})$        \\ \hline
$\lambda=0$, \textsc{Ellipsoid}~\cite{goemans2009approximating}       &            $O(\sqrt{n}\log n)$  \\ \hline
$\lambda = 1$, \textsc{GreedSplit}$^{\dagger}$~\cite{zhao2004generalized, narasimhan2005q}   &      $2$   \\ \hline
$\lambda = 1$, \textsc{Relax}~\cite{chekuri2011approximation}    &  $O(\log n$)  \\ \hline
$\lambda=0$, \textsc{MMin}$^*$ & $\underset{i}{\max} \frac{2|A_i^{\pi^*}|}{1+(|A_i^{\pi^*}|-1)(1-\kappa_{f_i}(A_i^{\pi^*}))}$ \\ \hline
$\lambda=0$, \textsc{\lovasz{}\!Round}$^{*}$ & $m$     \\ \hline
$0<\lambda<1$ \textsc{General\lovasz{}\!Round}$^{*}$ & $m$     \\ \hline
$0 < \lambda < 1$, \textsc{CombSlbSmp}$^*$   & $\min\{\frac{m\alpha}{m\bar{\lambda}+\lambda},\beta(m\bar{\lambda}+\lambda) \}$   \\ \hline
$0 < \lambda < 1$, \textsc{CombSlbSmp}$^{\dagger *}$   &  $\min\{m, \frac{m\alpha}{m\bar{\lambda}+\lambda}, \beta(m\bar{\lambda}+\lambda)\}$  \\ \hline
\hline
  $\lambda = 0$, Hardness$^*$              &           $m$ \\ \hline
 $\lambda = 1$, Hardness &       $2 - 2/m$~\cite{ene2013local}     \\ \hline
\end{tabular}
}
\caption[foo]{Summary of our contributions and existing work 
on Problem~\ref{prob:minmax-genmix}\footnotemark{}.}
\label{tab:summary_theoretical_results_min_case}

}
\vspace{-1.3\baselineskip}
%\vspace{-0.3em}
\end{table}
\footnotetext{Results obtained in this paper are marked as $*$. Methods for only the homogeneous setting are marked as $\dagger$.}
}

}{}
\arxivalt{}{
\textbf{Previous work: }{Special cases of \theproblems{} have appeared previously\arxiv{ in the literature}. \arxiv{Perhaps the first instance of this kind of problem 
is Edmonds's matroid partition problem \cite{edmonds1968matroid}. Here, 
if $\{ M_i \}_i$ is a set of matroids, 
then the ground set $V$ can be partitioned into a set
of independent sets $\{ I_i \}_i$ with $I_i \in \mathcal I(M_i)$ iff
there is no $X \subseteq V$ such that $|X| > \sum_{i} r_i(X)$.
If all matroids are the same, and there are $k$ of them, this becomes $|X|/k > r(X)$.
Edmonds both proved these conditions and gave a polynomial time
algorithm (a special case of submodular function minimization) for finding the partition when the conditions are true. Little additional work on this sort of problem occurred until many years later when submodularity was further developed. 
\jeff{Kai}{I am not sure how the matroid example fit into the problems we studied in this work. In particular, it is unclear if we can formulate matroid example as an instance of Problem 1 or 2? }
In this work we study \theproblems{} for general monotone submodular functions. 
}%
Problem~\ref{prob:maxmin-genmix} with $\lambda=0$ is called
\emph{submodular fair allocation} (SFA), and
Problem~\ref{prob:minmax-genmix} with $\lambda=0$ is called
\emph{submodular load balancing} (SLB), robust optimization problems
both of which previously have been studied.  When $f_i$'s are all modular, SLB is called {\em minimum makespan
  scheduling}. \arxivalt{In the homogeneous setting,
~\cite{hochbaum1988polynomial} give a PTAS scheme\arxiv{
  ($(1+\epsilon)$-approximation algorithm which runs in polynomial
  time for any fixed $\epsilon$)}, while an LP relaxation algorithm
provides a $2$-approximation for the heterogeneous
setting~\cite{lenstra1990approximation}.}{An LP relaxation algorithm provides a $2$-approximation for the heterogeneous setting~\cite{lenstra1990approximation}.} When the objectives are
submodular, the problem becomes much harder.  Even in the homogeneous
setting,~\cite{svitkina2008submodular} show that the problem is
information theoretically hard to approximate within $o(\sqrt{n/\log
  n})$. They provide a balanced partitioning algorithm yielding a
factor of $\min\{m,n/m\}$ under the homogeneous setting. They also
give a sampling-based algorithm achieving $O(\sqrt{n/\log n})$ for the homogeneous setting. However, the sampling-based algorithm is not practical
and scalable since it involves solving, in the worst-case,
$O(n^3\log n)$ instances of submodular function
minimization each of which requires $O(n^5\gamma+n^6)$
computation~\cite{orlin2009faster}, where $\gamma$ is the cost of
a function valuation.
% \kai{Jeff}{give SFM hardness and cite relevant
%   folks, Iwata, Orlin, etc.}\jeff{Kai}{done.}
%Under the heterogeneous setting,
Another approach approximates each submodular function by its
ellipsoid approximation (again non-scalable) and reduces SLB to its
modular version (minimum makespan scheduling) leading to an
approximation factor of $O(\sqrt{n }\log
n)$~\cite{goemans2009approximating}.
% provides an algorithm leading to a factor of $O(\sqrt{n}\log n)$. Their idea is to apply the ellipsoid approximation technique to construct a separate proxy function that approximates each submodular function within a factor of $O(\sqrt{n }\log n)$. SLB in terms of the proxy functions are reduced to the minimum makespan scheduling problem, which can be approximated up to a constant factor. 
SFA, on the other hand, has been studied mostly in the heterogeneous
setting.  When $f_i$'s are all modular, the tightest algorithm, so
far, is to iteratively round an LP solution
% \kai{Jeff}{Not sure what ``round a
%   configuration LP'' means. Fix wording.}\jeff{Kai}{I removed the
%   ``configuration'' and make it simply round an LP to avoid
%   confusion. } 
% \kai{Jeff}{I've further reworded here.}
achieving $O(1/(\sqrt{m} \log^3 m ))$
approximation~\cite{asadpour2010approximation}, whereas 
% \kai{Jeff}{
% The original text was ``the best
% NP-hardness result
% stands at $1/2$'', 
% What do you mean ``best known
%   NP-hardness''? Fix wording.}\jeff{Kai}{What I mean is ``the best
%   NP-hardness result shown, so far, for this problem is only $1/2$''.}
% \kai{Jeff}{I still don't think this is well defined. 
% An ``NP-hardness result'' can't be best since a problem is either
% NP-hard or it is not, and a result that shows it is NP-hard is all that is needed.
% Factors involving constants can either be an approximation algorithm,
% it can, say be a lower bound. Mentioning an approximation algorithm would
% be worded ``The best known approximation algorithm would be ... ``
% and the hardness would be worded ``The known hardness of this problem stands at ...'',
% you don't say ``the best NP-hardness stands at ...'' ,
% what you have now seems to be conflating several things together, i.e.,
% that the problem is NP-hard, and that it has a known hardness. Can
% you please completely re-word to achieve clarity? Also, please let me know
% if you know what I'm asking for. thanks! I've now changed the wording.} 
the problem is NP-hard to $1/2+\epsilon$
approximate for any $\epsilon > 0$~\cite{golovin2005max}.
%~\cite{golovin2005max} shows that it is NP-hard to approximate better than $1/2$, however, 
%the best approximation factor is only $O(1/(\sqrt{m} \log^3 m ))$, which is achieved by rounding a configuration LP formulation~\cite{asadpour2010approximation}. 
When $f_i$'s are submodular,~\cite{golovin2005max} gives a matching-based algorithm with a factor $1/(n-m+1)$ approximation that performs poorly when $m\ll n$. \cite{khot2007approximation} proposes a binary search algorithm yielding an improved factor of $1/(2m-1)$.
%When the objectives are sub-additive, which generalizes the notion of submodularity,~\cite{khot2007approximation} gives an algorithm with a factor of $1/(2m-1)$. When the objectives are submodular, ~\cite{golovin2005max} gives a different algorithm yielding a factor of $\frac{1}{n-m+1}$. This algorithm does not perform well in practice, where we always have $m\ll n$. 
Similar to SLB,~\cite{goemans2009approximating} applies the same ellipsoid approximation techniques 
%to approximate each submodular function with a simple proxy function and then reduce the original submodular fair allocation problem to a simpler one where the objectives are modular, 
leading to a factor of $O(\sqrt{n} m^{1/4} \log n \log^{3/2} m)$. 
These approaches are theoretically interesting, but they do not scale to large problems. \theproblems{}, when $\lambda=1$, have also been previously studied.
Problem~\ref{prob:minmax-genmix} becomes the \emph{submodular
  multiway partition} (SMP) for which one can obtain a relaxation based
$2$-approximation~\cite{chekuri2011approximation} in the homogeneous case. In the heterogeneous case, the guarantee is $O(\log n)$~\cite{chekuri2011submodularSCA}. Similarly, \cite{zhao2004generalized, narasimhan2005q} propose a greedy splitting $2$-approximation algorithm for the homogeneous setting.
Problem~\ref{prob:maxmin-genmix} becomes the \emph{submodular
  welfare}~\cite{vondrak2008optimal} for which a scalable greedy
algorithm achieves a $1/2$ approximation~\cite{fisher1978analysis}. Unlike the worst case ($\lambda=0$), many of the algorithms proposed for these problems are scalable.
The general case ($0 < \lambda < 1$) of \theproblems{} differs from either of these extreme
cases since we wish both for a \emph{robust} (worst-case) and average
case partitioning, and controlling $\lambda$ allows one to trade off
between the two.
% Intuitively, a more worst-case partitioning scheme
% achieves balanced solutions since it tries to optimize all the
% submodular valuations over the blocks. 
As we shall see, the flexibility of a mixture can be more natural
in certain applications.
}
}

% see also \cite{fujishige1988fair}

% quick 1-page summary of previous average case instances
% \url{http://research.microsoft.com/en-us/um/redmond/groups/theory/theorylunch/chekuri_2012_5_17.pdf}
% 

%We also address in Section~\ref{sec:gener-subm-part} objectives that mix between worst- and average-case. Next we introduce several applications of submodular partitioning

%Following the intuition shown in~\cite{wei2015-submodular-data-active} that  the class of facility location functions $f_{\mbox{fac}}$ naturally model the utility of a data set for training Nearest Neighbor classifiers, we heuristically apply $f_{\mbox{fac}}$ to model the efficiency of k-d tree nearest neighbor search on each data subset and solve Problem 1 with $f_{\mbox{fac}}$ to achieve a partition the training data such that each block of data in the partition leads to efficient k-d tree nearest neighbor search. Empirically, consistent speed-up has been observed by the proposed partition over the random partition.       

%\everyone{Kai}{In the Journal version, we should have the applications as a separate section, where we explain how to formulate the applications as submodular partitioning.}
\notarxiv{
\arxivalt{
\subsection{Applications of Submodular Partitionings in Machine
  Learning and Data Science}
\label{sec:appl-subm-part}
}{
{\bf Applications: }%
}%
There are a number of applications of submodular partitioning in ML
as outlined below. Some of these we evaluate in
Section~\ref{sec:experiments}.
% As we argue below, when $V$ is a finite data set
% and the $f_i$'s model data utility, this problem naturally addresses
% partitioning a data set for good distributed machine learning. On the
% other hand, when the $f_i$'s model workload, cost, or data subset
% complexity, Problem 2 (Submodular Load Balancing) is more
% natural. This is useful in applications such as
% clustering,\kai{Jeff}{I don't think the term ``working load
%   scheduling'' is ever used, but if it is can you point to some papers
%   that use this term as stated?}\jeff{Kai}{I agree. I don't think the
%   term ``working load scheduling'' is ever used in the literature. I
%   think ``load balancing'' is good.} load balancing, and image
% segmentation. 
%  \arxiv{Moreover, a number of data
%   partitioning (and clustering) applications can be posed as SFA or
%   SLB under the homogeneous setting.}\looseness-1
% 
% \textbf{Balanced Data Clustering and Image Segmentation: } 
Submodular
functions naturally capture notions of interacting cooperative costs
and homogeneity and thus are useful for {\bf clustering and
image segmentation}\arxivalt{~\cite{narasimhan2005q,
  boykovJolly01,kolmogorov2004energy,kohli2013principled}}{~\cite{narasimhan2005q,kolmogorov2004energy}}.
% \kai{Jeff}{add more cites, esp.\ to image segmentation, lots of the
%   Kolmogorov stuff, Rother, Kohli, etc.}.\jeff{Kai}{Done. We have to
%   be aware that only one page of references is allowed. So far, we
%   have already more than one page of references.  Later we may need to
%   prune away some references in case that we go over the space on
%   citations.}  
While the average case instance has been used before, a more worst-case
variant (i.e., Problem~\ref{prob:minmax-genmix} with $\lambda \approx
0$) is useful to produce {\bf balanced clusterings} (i.e.,
the submodular valuations of all the blocks should be similar to each other).
% \kai{Jeff}{6/4: it would be useful to add a sentence on how precisely
% the clusters are balanced, since it depends entirely on the submodular functions.
% I.e., they might be still very imbalanced in terms of cardinality, but
% they are balanced in terms of submodular valuation. I added a few
% words about ``submodualr valuation'' please read it over.}
Problem~\ref{prob:minmax-genmix} also addresses a problem in image
segmentation, namely how to use only submodular functions (which are
instances of pseudo-Boolean functions) for multi-label (i.e.,
non-Boolean) image segmentation. Problem~\ref{prob:minmax-genmix}
addresses this problem by allowing each segment $j$ to have its own
submodular function $f_j$, and the objective measures the homogeneity
$f_j(A^\pi_j)$ of segment $j$ based on the image pixels $A^\pi_j$
assigned to it. Moreover, by combining the average case and the worst case objectives, one can achieve a tradeoff between the two. Empirically, we evaluate our algorithms on unsupervised image segmentation (Section~\ref{sec:experiments}) and find that it outperforms other clustering methods including 
$k$-means, $k$-medoids, spectral clustering, and graph cuts.\looseness-1
% \textbf{Memory and Communications Cost Load Balancing: } 

Submodularity also accurately represents computational costs in
{\bf distributed systems}, as shown in~\cite{smola2015graphpartitioning}.
% \kai{Jeff}{I think we should still cite the recent arxiv paper, by
% Smola, and describe it briefly.}  \jeff{Kai}{I think the recent
% arxiv paper has already been cited here, right?}  \kai{Jeff}{Yes,
% but there was a problem with the bibtex file which was that the last
% and first names were reversed, and the last names were abbreviated
% which is why I missed it. Make sure to double check the bibtex
% entries to ensure that the last/first names are not reversed.
In fact, \cite{smola2015graphpartitioning} considers two separate
problems: 1) text data partitioning for balancing memory demands; and
2) parameter partitioning for balancing communication costs.  Both are
treated by solving an instance of SLB
(Problem~\ref{prob:minmax-genmix}, $\lambda=0$) where memory costs are
modeled using a set-cover submodular function and the communication
costs are modeled using a modular (additive) function.
%They model the memory usage for a set of documents as the set cover function and formulate the data partitioning for memory requirement as an instance of Submodular Load Balancing. Similarly, they model the communication costs associated with each key as a modular function, and solve the key partitioning for communication cost minimization as an instance of SLB. 
%They consider two separate problems: (1) text data partitioning for memory requirements, and (2) parameter partitioning for communication costs balancing. They treat both problems as an instance of Submodular Load Balancing where they model the memory usage as a set-cover function and the communication cost as a modular set function.
% 
% \textbf{Distributed Training: }
%

Another important ML application, evaluated in
Section~\ref{sec:experiments}, is {\bf distributed training} of
statistical models. As data set sizes grow, the need for statistical
training procedures tolerant of the {\bf distributed data
  partitioning} becomes more important. Existing schemes are often
developed and performed assuming data samples are distributed
\arxiv{to their computational clients }in an arbitrary or random
fashion. As an alternate strategy, if the data is intelligently
partitioned such that each block of samples can itself lead to a good
approximate solution, a consensus amongst the distributed results
could be reached more quickly than when under a poor partitioning.
Submodular functions can in fact express the value of a subset of
training data for certain machine learning risk functions,
e.g.,~\cite{wei2015-submodular-data-active}\arxiv{in the case of
  classification}.  Using these functions within
Problem~\ref{prob:maxmin-genmix}, one can expect a partitioning (by formulating the problem as an instance of Problem~\ref{prob:maxmin-genmix}, $\lambda \approx 0$) where
each block is a good \emph{representative} of the entire set, thereby
achieving faster convergence in distributed settings.  We demonstrate
empirically, in Section~\ref{sec:experiments}, that this provides
better results on several machine learning tasks, including the
training of deep neural networks.
\looseness-1}

\arxivalt{
\subsection{Our Contributions}
\label{sec:our-contributions}
}{
{}%
}%

{\notarxiv{\textbf{Our Contributions: }}In contrast to \theproblems{} in the average case (i.e., $\lambda = 1$), existing algorithms 
for the worst case ($\lambda = 0$) are not scalable. This paper closes this gap, by proposing three new classes of algorithmic frameworks to solve SFA
and SLB: (1) greedy algorithms; (2) semigradient-based algorithms; and
(3) a \lovasz{} extension based relaxation algorithm. 
\arxiv{

}
For SFA, when
$m=2$, we formulate the problem as non-monotone submodular
maximization, which can be approximated up to a factor of $1/2$ with
$O(n)$ function evaluations~\cite{buchbinder2012tight}. For
general $m$, we give a simple and scalable greedy algorithm
(\textsc{GreedMax}), and show a factor of $1/m$ in the homogeneous
setting, improving the state-of-the-art factor of $1/(2m-1)$ under the
heterogeneous setting~\cite{khot2007approximation}. 
For the heterogeneous setting, we propose a ``saturate'' greedy
algorithm (\textsc{GreedSat}) that iteratively solves instances of
submodular welfare problems. We show \textsc{GreedSat} has a
bi-criterion guarantee of $(1/2-\delta, {\delta}/{(1/2+\delta)})$,
which ensures at least $\ceil{m(1/2 - \delta)}$ blocks receive utility at
least $\delta/(1/2 +\delta)OPT$ for any $0 < \delta < 1/2$.
% (in the worst-case $\alpha=1-1/e$, but in practice, $\alpha$ is often close to 1). 
% yielding a bi-criterion factor of $(1/2-\delta, \frac{\delta}{1/2 + \delta})$. The notion of bi-criterion guarantee 
%Sharing similar spirit with the algorithm proposed in~\cite{krause2008robust}, \textsc{GreedSat} iteratively solves a submodular welfare problem. 
For SLB, we first generalize the hardness result in~\cite{svitkina2008submodular} and show that it is hard to approximate better than $m$ for any $m=o(\sqrt{n/\log n})$ even in the homogeneous setting.
%, which generalizes the hardness result in~\cite{svitkina2008submodular}. 
We then give a \lovasz{}extension based relaxation algorithm (\textsc{\lovasz{}\!Round}) yielding a tight factor of $m$ for the heterogeneous setting. As far as we know, this is the first algorithm achieving a factor of $m$ for SLB in this setting. 
%Though providing tight analysis, \textsc{\lovasz{}\!Round} requires many evaluations of the \lovasz extension, which can be computationally unfeasible for large data set. 
For both SFA and SLB, we also obtain more efficient algorithms with bounded approximation factors, which we call majorization-minimization (\textsc{MMin}) and minorization-maximization (\textsc{MMax}). 
\looseness-1
%Similar in flavor to~\cite{iyer2013fast}, the idea is to iteratively minimize the modular upper bound of each submodular function obtained by its super-gradient. 

Next we show algorithms to handle \theproblems{} with general $0<\lambda<1$. We first give two simple and generic schemes (\textsc{CombSfaSwp} and \textsc{CombSlbSmp}), both of which 
efficiently combines
an algorithm for the worst-case problem (special case with $\lambda = 0$), 
and an algorithm for the average case (special case with $\lambda = 1$)  to provide a guarantee  interpolating between the two bounds. \arxivalt{Given the efficient algorithms proposed in this paper for the robust (worst case) problems (with guarantee $\alpha$), and an existing algorithm for the average case (say, with a guarantee $\beta$), we can obtain a combined guarantee in terms of $\alpha, \beta$ and $\lambda$.
We then generalize the proposed algorithms for SLB and SFA to give more practical algorithmic frameworks to solve \theproblems{} for general $\lambda$.
In particular we generalize \textsc{GreedSat} leading to \textsc{GeneralGreedSat}, whose guarantee smoothly interpolates in terms of $\lambda$ between the bi-criterion factor by \textsc{GreedSat} in the case of $\lambda=0$ and the constant factor of $1/2$ by the greedy algorithm in the case of $\lambda=1$. For Problem~\ref{prob:minmax-genmix} we generalize \textsc{\lovasz{}\!Round} to obtain a relaxation algorithm (\textsc{General\lovasz{}\!Round}) that achieves an $m$-approximation for general $\lambda$. 
Motivated by the computational limitation of \textsc{General\lovasz{}\!Round} we also give a simple and efficient greedy heuristic called \textsc{GeneralGreedMin} that works for the homogeneous setting of Problem~\ref{prob:minmax-genmix}.

Lastly we demonstrate a number of applications of submodular partitioning in real-world machine learning problems. 
In particular,
corresponding to 
Sections~\ref{sec:data-part-parall}
and~\ref{sec:multi-label-image},
we show Problem~\ref{prob:maxmin-genmix} is applicable in distributed training of statistical models. Problem~\ref{prob:minmax-genmix} is useful for data clustering, image segmentation, and computational load balancing. In the experiments we empirically evaluate Problem~\ref{prob:maxmin-genmix} on data partitioning for ADMM and distributed deep neural network training.
The efficacy of Problem~\ref{prob:minmax-genmix} is tested on an unsupervised image segmentation task. 
} 
{%Given the efficient algorithms proposed in this paper for the robust (worst case) problems (with guarantee $\alpha$), and an existing algorithm for the average case (say, with a guarantee $\beta$), we can obtain a combined guarantee in terms of $\alpha, \beta$ and $\lambda$.
For Problem~\ref{prob:maxmin-genmix}  we generalize \textsc{GreedSat} leading to \textsc{GeneralGreedSat}, whose guarantee smoothly interpolates in terms of $\lambda$ between the bi-criterion factor by \textsc{GreedSat} in the case of $\lambda=0$ and the constant factor of $1/2$ by the greedy algorithm in the case of $\lambda=1$. For Problem~\ref{prob:minmax-genmix} we generalize \textsc{\lovasz{}\!Round} to obtain a relaxation algorithm (\textsc{General\lovasz{}\!Round}) that achieves an $m$-approximation for general $\lambda$. 
%Motivated by the computational limitation of \textsc{General\lovasz{}\!Round} we also give a simple and efficient greedy heuristic called \textsc{GeneralGreedMin} that works for the homogeneous setting of Problem~\ref{prob:minmax-genmix}.
The theoretical contributions and the existing work for \theproblems{} are summarized in Table~\ref{tab:summary_theoretical_results_both_case}.\looseness-1} 

\arxiv{

}

\notarxiv{Lastly, we
demonstrate the efficacy of Problem~\ref{prob:minmax-genmix} on unsupervised image
segmentation, and the success of Problem~\ref{prob:maxmin-genmix} to distributed machine
learning, including ADMM and neural network training.\looseness-1}
}
% \kai{Jeff}{add section on empirical contributions.}
% \jeff{Kai}{Done.} 
% \subsection{Outline} 

% Section~\ref{sec:gener-subm-part} studies the case where $\lambda > 0$.

% Problem~\ref{prob:minmax-genmix} and
% Problem~\ref{prob:maxmin-genmix}.

\begin{figure}[t]
\centering
% \caption*{\Large Problem~\ref{prob:maxmin-genmix}: $\max_{\pi \in \Pi} \Bigl[\bar\lambda \min_{i} f_i(A_i^{\pi}) + \frac{\lambda}{m} \sum_{j = 1}^m f_j(A^{\pi}_j)\Bigr]$;\\ 
% Problem~\ref{prob:minmax-genmix}: $\min_{\pi \in \Pi} \Bigl[\bar\lambda \max_{i} f_i(A^\pi_i) + \frac{\lambda}{m} \sum_{j = 1}^m f_j(A^\pi_j)\Bigr]$.}}
% \includegraphics[page=3,trim=0 0 0 2cm, clip=true, width=1\textwidth]{submodular_partitioning_prob_def_table.pdf}
\includegraphics[page=3,width=1.0\textwidth]{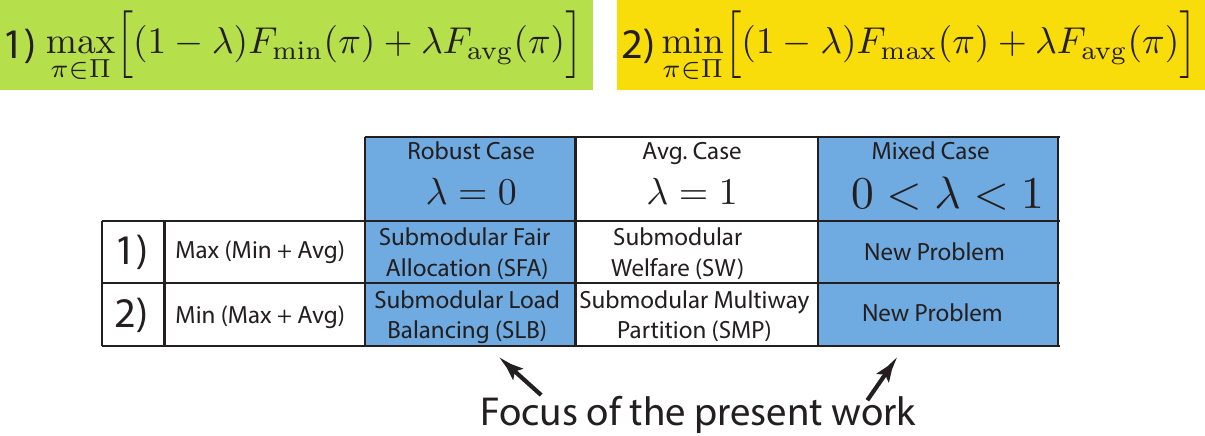}
\caption{A reminder of the definitions of
Problem~\ref{prob:maxmin-genmix} and~\ref{prob:minmax-genmix} 
(above) and a table that illustrates the cases we focus
on in the present work.}
\label{fig:illustration_problem_in_table}
\end{figure}

%\refstepcounter{problemnumber}\label{prob:maxmin-genmix}
%\abovedisplayshortskip=.2ex\belowdisplayshortskip=.2ex\abovedisplayskip=.2ex\belowdisplayskip=.2ex
%\begin{align}
%\text{Problem \theproblemnumber: }\qquad
%\max_{\pi \in \Pi} \Bigl[\bar\lambda \min_{i} f_i(A_i^{\pi}) + \frac{\lambda}{m} \sum_{j = 1}^m f_j(A^{\pi}_j)\Bigr],
%\end{align}
%
%\refstepcounter{problemnumber}\label{prob:minmax-genmix}
%\abovedisplayshortskip=.2ex\belowdisplayshortskip=.2ex\abovedisplayskip=.2ex\belowdisplayskip=.2ex
%\begin{align}
%\text{Problem \theproblemnumber: }\qquad
%\min_{\pi \in \Pi} \Bigl[\bar\lambda \max_{i} f_i(A^\pi_i) + \frac{\lambda}{m} \sum_{j = 1}^m f_j(A^\pi_j)\Bigr],
%\end{align}

\section{Robust Submodular Partitioning (\theproblems{} when $\lambda = 0$)}
\label{sec:algorithms_for_SFA_SLB}
\notarxiv{\vspace{-0.5\baselineskip}}

Notation: we define $f(j | S) \triangleq f(S \cup j) - f(S)$ as the gain of
$j\in V$ in the context of $S \subseteq V$.
\arxiv{Then, $f$ is submodular if
and only if %\emph{Iff}
$f(j | S) \geq f(j | T)$ for all $S \subseteq T$ and $j \notin
T$. Also, $f$ is monotone iff $f(j | S) \geq 0, \forall j \notin S, S
\subseteq V$.}  We assume w.l.o.g.\ that the ground set is $V
= \{1, 2, \cdots, n\}$.\notarxiv{\\[-2\baselineskip]}

\subsection{Approximation Algorithms for SFA (Problem~\ref{prob:maxmin-genmix} with $\lambda=0$)}
\label{sfaapproxsec}
\arxivalt{We first investigate a special case of SFA with $m=2$.%We next propose two variants of a greedy algorithm that approximates SFA for general $m$. Lastly we describe a subgradient-based algorithm.  
}
{We first study approximation algorithms for SFA.} 
\arxivalt{When $m=2$, the problem becomes 
\begin{align}
\max_{A\subseteq V} g(A),
\label{eqn:non-monotone_submod_max}
\end{align}
where $g(A) = \min\{f_1(A), f_2(V\setminus A)\}$ and is
submodular thanks to Theorem~\ref{thm:submod_two_case}.
}
{When $m=2$, the problem becomes $\max_{A\subseteq V}
g(A)$ where $g(A) = \min\{f_1(A), f_2(V\setminus A)\}$ and is
submodular thanks to Theorem~\ref{thm:submod_two_case}.}
\begin{theorem}
\label{thm:submod_two_case}
If $f_1$ and $f_2$ are monotone submodular, 
$\min\{f_1(A), f_2(V\setminus A)\}$ is also submodular.
\end{theorem}
\arxivalt{All proofs for the theoretical results are given in Appendix. Interestingly SFA for $m=2$ can be equivalently formulated as unconstrained submodular maximization. This problem has been well studied in the literature~\cite{buchbinder2012tight, dobzinski2015deterministic, fiege2011submodmax}.}{Proofs for all theorems
in this paper are given in~\cite{datapartitionextend}.} \arxivalt{A simple bi-directional randomized greedy algorithm~\cite{buchbinder2012tight} solves Eqn~\ref{eqn:non-monotone_submod_max} with a tight factor of $1/2$. Applying this randomized algorithm to solve SFA then achieves a guarantee of $1/2$ matching the problem's hardness.}{The simple bi-directional randomized greedy algorithm~\cite{buchbinder2012tight} therefore approximates SFA with $m=2$ to a factor of $1/2$ matching the problem's hardness.} \arxiv{However, the same idea does not apply to the general case of $m>2$.}
% \kai{Jeff}{why is the bit about $k$-submodularity commented out? This seems worthy of
% inclusing in the extended version.}\jeff{Kai}{That's a good idea. I put it in the extended version. } 
\arxiv{

}
For general $m$, we approach SFA from the perspective of the greedy algorithms.  
\arxiv{Greedy is often the algorithm that practitioners use for combinatorial optimization problems since they are intuitive,
simple to implement, and often lead to very good solutions.} In this work  we introduce two variants of a greedy algorithm -- \textsc{GreedMax} (Alg.~\ref{alg:greedMax}) and \textsc{GreedSat} (Alg.~\ref{alg:greedSat}), suited to the homogeneous and heterogeneous settings, respectively.\looseness-1
%Theoretical analysis is given in Section 3 -- the current section defines and offers intuition for the methods. 
%Initialized with $A_i=\emptyset$ for each block $i$, \textsc{GreedMax}, in each iteration, proceeds as follows: (1) finds a block $j$ whose current function value is smallest (i.e., $j\in \argmin_{i=1,\dots,m}f_i(A_i)$), (2) greedily chooses an item $v\in V\setminus \cup_{i=1,\dots,m} A_i$ that maximizes the marginal gain on block $j$ ($v\in \argmax_{v^\prime \in V\setminus \cup_{i=1,\dots,m} A_i} f_j(v^\prime | A_{j})$), and (3) updates as $A_{j} \leftarrow A_{j} \cup v$. \textsc{GreedMax} terminates when all items in $V$ have been assigned, or equivalently, $\{A_i\}_{i=1}^m$ forms a partition of the ground set $V$.
%The key idea of \textsc{GreedMax} is to greedily adds item with the maximum marginal gain to the block whose current solution is minimum. 
%Since the overall objective $\min_{i=1,\dots,m}f_i(A_i)$ can only be improved in each iteration by assigning item to a block with the smallest function value, 

\notarxiv{
\begin{figure}[!t]
\begin{minipage}{\textwidth}
\begin{minipage}{0.48\textwidth}
\begin{algorithm}[H]
{\fontsize{8}{8}\selectfont	
\begin{algorithmic}[1]
\caption{\textsc{GreedMax}}
\label{alg:greedMax}
%\begin{algorithmic}
\STATE Input: $f$, $m$, $V$.\\
\STATE Let $A_1=,\dots, =A_m = \emptyset$; $R = V$. \\
\label{line:greedMax_initialization}
\WHILE {$R\neq \emptyset$}
\STATE $j^*\in \argmin_j f(A_j)$; \\
\STATE $a^* \in \argmax_{a\in R} f(a|A_{j^*})$\\	
    \label{line:greedy_step_GreedMax}
  \STATE  $A_{j^*} \leftarrow A_{j^*} \cup \{a^*\}$; $R \leftarrow R \setminus a^*$ \\
	\ENDWHILE
\STATE Output $\{A_i\}_{i=1}^m$.
\end{algorithmic}}
\end{algorithm} 

\vspace{-0.5em}

\begin{algorithm}[H]
\caption{\textsc{GreedSat}}
\label{alg:greedSat}
{\fontsize{8}{8}\selectfont	
\begin{algorithmic}[1]
\STATE Input: \arxiv{$\epsilon$, }$\{f_i\}_{i=1}^m$, $m$, $V$, $\alpha$.\\
\STATE Let $\bar{F}^c(\pi) = \frac{1}{m}\sum_{i=1}^m \min\{f_i(A^\pi_i),c\}$.\\
\label{line:def_F_c_function}
\STATE Let $c_{\min} = 0$,
$c_{\max} = \min_i f_{i}(V)$\\
\WHILE {$ c_{\max} - c_{\min}   \geq \epsilon $}
\STATE $c = \frac{1}{2}(c_{\max} + c_{\min})$\\
\STATE $\hat{\pi}^c \in \argmax_{\pi \in \Pi} \bar{F}^c(\pi)$\\
\label{line:swp_greedy_sat}
\IF{$\bar{F}^c(\hat{\pi}^c) < \alpha c$}
\STATE $c_{\max} = c$ \\	
\ELSE 
\STATE $c_{\min} = c$;  $\hat{\pi} \leftarrow \hat{\pi}^c$\\
\ENDIF
\ENDWHILE    
\STATE Output: $\hat{\pi}$.
\end{algorithmic}}
\end{algorithm} 

\vspace{-0.5em}

\begin{algorithm}[H]
{\fontsize{8}{8}\selectfont		
\begin{algorithmic}[1]
\caption{\textsc{\lovasz{}\!Round}}
\label{alg:lovasz_ext_alg}
\STATE Input: $\{f_i\}_{i=1}^m$, $\{\tilde{f}_i\}_{i=1}^m$, $m$, $V$.\\
{
\STATE Solve for $\{x^*_i\}_{i=1}^m$ via convex relaxation. \\
\label{line:solve_relaxation_lovasz}
}
\STATE Rounding: Let $A_1=,\dots,=A_m=\emptyset$.\\
\label{line:start_rounding}
\FOR{$j=1,\dots, n$}
\STATE $\hat{i}\in \argmax_{i} x^*_i(j)$; $A_{\hat{i}} = A_{\hat{i}} \cup j$\\
\ENDFOR
\label{line:end_rounding}
\STATE Output $\{A_i\}_{i=1}^m$.
\end{algorithmic}}
\end{algorithm}
\end{minipage} 
\hspace{1em}
\begin{minipage}{0.48\textwidth}

\begin{algorithm}[H]
{\fontsize{8}{8}\selectfont		
\begin{algorithmic}[1]
\caption{\textsc{GreedMin}}
\label{alg:greed_min}
\STATE Input: $f$, $m$, $V$; \\ 
\STATE Let $A_1=,\dots,=A_m=\emptyset$; $R=V$.\\
\WHILE{$R\neq \emptyset$}
\STATE $j^*\in \argmin_{j} f(A_j)$ \\ 
\STATE $a^* \in \min_{a\in R} f(a|A_{j^*})$\\
\label{line:greedy_step_greed_min}
\STATE $A_{j^*}\leftarrow A_{j^*} \cup {a^*}; R\leftarrow R\setminus a^*$
\ENDWHILE
\STATE Output $\{A_i\}_{i=1}^m$.
\end{algorithmic}}
\end{algorithm}

\begin{algorithm}[H]
{\fontsize{8}{8}\selectfont		
\begin{algorithmic}[1]
\caption{\textsc{MMin}}
\label{alg:MMin}
\STATE Input: $\{f_i\}_{i=1}^m$, $m$, $V$, partition $\pi^0$.\\
\STATE Let  $t=0$ \\
\REPEAT
\FOR{$i=1,\dots,m$}
\STATE   Pick a supergradient $m_i$ at $A_i^{\pi^t}$ for $f_i$.\\
\ENDFOR
\STATE $\pi^{t+1} \in \argmin_{\pi\in \Pi} \max_{i} m_i (A^{\pi}_i)$
\label{line:MMin_modular_version}	\\
\STATE	$t = t+1$;\\
\UNTIL{$\pi^t = \pi^{t-1}$}
\STATE Output: $\pi^t$.
\end{algorithmic}}
\end{algorithm} 

\begin{algorithm}[H]
{\fontsize{8}{8}\selectfont	
\begin{algorithmic}[1]
\caption{\textsc{MMax}}
\label{alg:MMax}
\STATE Input: $\{f_i\}_{i=1}^m$, $m$, $V$, partition $\pi^0$.\\
\STATE Let $t=0$.\\
\REPEAT
\FOR{$i=1,\dots,m$}
\STATE   Pick a subgradient $h_i$ at $A_i^{\pi^t}$ for $f_i$.\\
\ENDFOR
\STATE $\pi^{t+1} \in \argmax_{\pi\in \Pi} \min_{i} h_i(A^{\pi}_i)$
\label{line:MMax_modular_version}	\\
\STATE $t = t+1$; \\
\UNTIL{$\pi^t = \pi^{t-1}$}
\STATE Output: $\pi^t$.
\end{algorithmic}}
\end{algorithm} 
\end{minipage}
%\caption{Algorithms }
\end{minipage}
\vspace{-2.1\baselineskip}
\end{figure}
}
\arxiv{
\begin{algorithm}[]
{
\begin{algorithmic}[1]
\caption{\textsc{GreedMax}}
\label{alg:greedMax}
%\begin{algorithmic}
\STATE Input: $f$, $m$, $V$.\\
\STATE Let $A_1=,\dots, =A_m = \emptyset$; $R = V$. \\
\label{line:greedMax_initialization}
\WHILE {$R\neq \emptyset$}
\STATE $j^*\in \argmin_j f(A_j)$; \\
\STATE $a^* \in \argmax_{a\in R} f(a|A_{j^*})$\\	
    \label{line:greedy_step_GreedMax}
  \STATE  $A_{j^*} \leftarrow A_{j^*} \cup \{a^*\}$; $R \leftarrow R \setminus a^*$ \\
	\ENDWHILE
\STATE Output $\{A_i\}_{i=1}^m$.
\end{algorithmic}}
\end{algorithm} 
}

\textbf{\textsc{GreedMax}: } 
%Initialized with $A_i=\emptyset$ for each block $i$, \textsc{GreedMax}, in each iteration, proceeds as follows: (1) finds a block $j$ whose current function value is smallest (i.e., $j\in \argmin_{i=1,\dots,m}f_i(A_i)$), (2) greedily chooses an item $v\in V\setminus \cup_{i=1,\dots,m} A_i$ that maximizes the marginal gain on block $j$ ($v\in \argmax_{v^\prime \in V\setminus \cup_{i=1,\dots,m} A_i} f_j(v^\prime | A_{j})$), and (3) updates as $A_{j} \leftarrow A_{j} \cup v$. \textsc{GreedMax} terminates when all items in $V$ have been assigned, or equivalently, $\{A_i\}_{i=1}^m$ forms a partition of the ground set $V$.
The key idea of \textsc{GreedMax} (see Alg.~\ref{alg:greedMax}) is to greedily add an item with the maximum marginal gain to the block whose current solution is minimum. 
%Since the overall objective $\min_{i=1,\dots,m}f_i(A_i)$ can only be improved in each iteration by assigning item to a block with the smallest function value, 
Initializing $\{A_i\}_{i=1}^m$ with the empty sets, the greedy flavor also comes from that it incrementally grows the solution by greedily improving the overall objective 
$\min_{i=1,\dots,m} f_i(A_i)$ until $\{A_i\}_{i=1}^m$ forms a partition.
Besides its simplicity, Theorem~\ref{thm:greedy_max_guarantee} offers the optimality guarantee.\looseness-1
 %Thanks to submodularity, the number of marginal gain evaluations needed in step (2) can be significantly reduced using the lazy evaluation trick as described in~\cite{minoux1978accelerated}, leading \textsc{GreedMax} to scale to large data sets. 
\begin{theorem}
\label{thm:greedy_max_guarantee}
\arxivalt{
Under the homogeneous setting ($f_i=f$ for all $i$), \textsc{GreedMax} is guaranteed to find a partition $\hat{\pi}$ such that 
\arxivalt{
\begin{align}
\min_{i=1,\dots,m} f(A_i^{\hat{\pi}}) \geq \frac{1}{m} \max_{\pi \in \Pi }\min_{i=1,\dots,m}  f(A_i^\pi).
\end{align}
}
{$\min_{i=1,\dots,m} f(A_i^{\hat{\pi}}) \geq \frac{1}{m} \max_{\pi \in \Pi }\min_{i=1,\dots,m}  f(A_i^\pi)$.} 
}
{
\textsc{GreedMax} achieves a guarantee of $1/m$ under the homogeneous setting.\looseness-1
}
%Conversely, for any $m\geq 2$, there exists a submodualr function $f$, on which an instance of \textsc{GreedMax} achieves an approximation factor $1/2$.
\end{theorem}
%\arxiv{In fact, the proof of Theorem~\ref{thm:greedy_max_guarantee} only requires $f$ to be sub-additive. Moreover, the $1/m$-approximation also holds for a much weaker version of \textsc{GreedMax}, which is a streaming algorithm. We hypothesize that $1/m$-approximation guarantee may not be tight and the analysis may be further improved. }
By assuming the homogeneity of the $f_i$'s, we obtain a very simple $1/m$-approximation algorithm improving upon the state-of-the-art factor $1/(2m-1)$~\cite{khot2007approximation}. 
Thanks to the lazy evaluation trick as described in~\cite{minoux1978accelerated}, Line~\ref{line:greedy_step_GreedMax} in Alg.~\ref{alg:greedMax} need not to recompute the marginal gain for every item in each round, leading \textsc{GreedMax} to scale to large data sets. 

\arxiv{

\begin{algorithm}[htb]
\caption{\textsc{GreedSat}}
\label{alg:greedSat}
{
\begin{algorithmic}[1]
\STATE Input: \arxiv{$\epsilon$, }$\{f_i\}_{i=1}^m$, $m$, $V$, $\alpha$.\\
\STATE Let $\bar{F}^c(\pi) = \frac{1}{m}\sum_{i=1}^m \min\{f_i(A^\pi_i),c\}$.\\
\label{line:def_F_c_function_greed_sat}
\STATE Let $c_{\min} = 0$,
$c_{\max} = \min_i f_{i}(V)$\\
\WHILE {$ c_{\max} - c_{\min}   \geq \epsilon $}
\STATE $c = \frac{1}{2}(c_{\max} + c_{\min})$\\
\STATE $\hat{\pi}^c \in \argmax_{\pi \in \Pi} \bar{F}^c(\pi)$ // solved by \textsc{GreedSWP} (Alg~\ref{alg:greedSWP})\\
\label{line:swp_greedy_sat}
\IF{$\bar{F}^c(\hat{\pi}^c) < \alpha c$}
\STATE $c_{\max} = c$ \\	
\ELSE 
\STATE $c_{\min} = c$;  $\hat{\pi} \leftarrow \hat{\pi}^c$\\
\ENDIF
\ENDWHILE    
\STATE Output: $\hat{\pi}$.
\end{algorithmic}}
\end{algorithm} 

\begin{algorithm}[htb]
\caption{\textsc{GreedSWP}}
\label{alg:greedSWP}
\begin{algorithmic}
\STATE Input: $\{f_i\}_{i=1}^m$, $c$, $V$\\
\STATE Initialize: $A_1=,\dots,=A_m = \emptyset$, and $R \leftarrow V$ \\
\While {$R \neq \emptyset$}{
\For{$i=1,\dots,m$}{
$\delta_i = \max_{r\in R} \min\{f_i(A_i\cup r),c\} - \min\{f_i(A_i),c\} $ \\	
$a_i \in \argmax_{r\in R} \min\{f_{i}(A_{i}\cup r),c\} - \min\{f_{i}(A_{i}),c\}$\\	
}
	$j \in \argmax_{i} \delta (i)$\\    
    
	$A_{j} \leftarrow A_{j} \cup \{a_j\}$ \\
	$R \leftarrow R \setminus a_j$ \\
	}
	Output $\hat{\pi}^c = (A_1,\dots,A_m)$.
\end{algorithmic}
\end{algorithm}

}

\textbf{\textsc{GreedSat}: } Though simple and effective in the homogeneous setting, \textsc{GreedMax} performs arbitrarily poorly under the heterogeneous setting.\arxiv{Consider the following example: $V=\{v_1, v_2\}$, $f_1(v_1)=1, f_1(v_2)=0, f_1(\{v_1,v_2\})=1, f_2(v_1) = 1+\epsilon, 
f_2(v_2)=1$, $f_2(\{v_1,v_2\}) = 2+\epsilon$. $f_1$ and $f_2$ are monotone submodular. The optimal partition is to assign $v_1$ to $f_1$ and $v_2$ to $f_2$ leading to a solution of value $1$. However, \textsc{GreedMax} may assign $v_1$ to $f_2$ and $v_2$ to $f_1$ leading to a solution of value $0$.
Therefore, \textsc{GreedMax} performs arbitrarily poorly on this example.
	
}
To this end we provide another algorithm -- ``Saturate'' Greedy (\textsc{GreedSat}, see Alg.~\ref{alg:greedSat}). 
The key idea of \textsc{GreedSat} is to relax SFA to a much simpler problem -- Submodular Welfare (SWP), i.e., Problem~\ref{prob:maxmin-genmix} with $\lambda=0$.
Similar in flavor to the one proposed in~\cite{krause08robust} 
\textsc{GreedSat} 
defines an intermediate objective $\bar{F}^c(\pi) =\sum_{i=1}^m f_i^c(A_i^{\pi})$, where $f_i^{c}(A) = \frac{1}{m} \min\{f_i(A), c\}$ (Line~\ref{line:def_F_c_function_greed_sat}). The parameter $c$ controls the saturation in each block. \arxiv{It is easy to verify that }$f_i^c$ satisfies submodularity for each $i$. 
Unlike SFA, the combinatorial optimization problem $\max_{\pi \in \Pi} \bar{F}^c(\pi)$ (Line~\ref{line:swp_greedy_sat})
is much easier and is an instance of SWP. 
\arxivalt{In this work, we solve Line~\ref{line:swp_greedy_sat} using the greedy algorithm as described in Alg~\ref{alg:greedSWP}, which attains a constant 1/2-approximation~\cite{fisher1978analysis}. Moreover the lazy evaluation trick also applies for Alg~\ref{alg:greedSWP} enabling the wrapper algorithm \textsc{GreedSat} scalable to large data sets.}{In this work, we solve Line~\ref{line:swp_greedy_sat}
by 
the efficient greedy algorithm as described in~\cite{fisher1978analysis} with a factor $1/2$.} 
One can also use a more computationally expensive multi-linear relaxation algorithm as given in~\cite{vondrak2008optimal} to solve Line~\ref{line:swp_greedy_sat} with a tight factor $\alpha=(1-1/e)$. 
%Let SWP$(\bar{F}^c_{\min}(\pi))$ be an instance of the submodular welfare problem with an objective $\bar{F}^c_{\min}(\pi)$. 
%Suppose $\alpha \leq 1$ is the approximation factor of an algorithm for solving SWP$(\bar{F}_{\min}^c(\pi))$ and $\hat{\pi}^c\in $SWP$(\bar{F}_{\min}^c(\pi))$ is the corresponding solution ($\alpha = 1/2$ for \textsc{GreedSWP} and $\alpha = 1-1/e$ for the multi-linear extension algorithm). 
Setting the input argument $\alpha$ as the approximation factor for Line~\ref{line:swp_greedy_sat}, the essential idea of \textsc{GreedSat} is to 
perform a binary search over the parameter $c$ to find the largest $c^*$ such that the returned solution $\hat{\pi}^{c^*}$ for the instance of SWP
satisfies $\bar{F}^{c^*}(\hat{\pi}^{c^*})\geq \alpha
c^*$. 
\textsc{GreedSat} terminates after solving $O(\log (\frac{\min_i f_i(V)}{\epsilon}))$  instances of SWP.
Theorem~\ref{thm:greedy_sat_bound} gives a bi-criterion optimality guarantee.\looseness-1
\begin{theorem}
\label{thm:greedy_sat_bound}
Given $\epsilon>0$, $0\leq \alpha \leq 1$ and any $0<\delta<\alpha$, \textsc{GreedSat} finds a partition such that at least $\ceil{m(\alpha-\delta)}$ blocks receive utility at least $\frac{\delta}{1-\alpha + \delta} (\max_{\pi\in \Pi}\min_{i}f_i(A_i^{\pi})-\epsilon)$.
\end{theorem}
For any $0<\delta < \alpha$  Theorem~\ref{thm:greedy_sat_bound} ensures that the top $\ceil{m(\alpha - \delta)}$ valued blocks in the partition returned by 
\textsc{GreedSat} are $(\delta/(1-\alpha + \delta)-\epsilon)$-optimal. $\delta$ controls the trade-off between the number of top valued blocks to bound and the performance guarantee attained for these blocks. The smaller $\delta$ is, the more top blocks are bounded, but with a weaker guarantee.
%When $f_i$'s take only integral values and $\epsilon$ is set as $1/m$, the additive performance loss in terms of $\epsilon$ can be removed, hence we obtain a slightly improved guarantee in this scenario. 
 We set the input argument $\alpha=1/2$ (or $\alpha =1-1/e$) as the worst-case performance guarantee for solving SWP so that the above theoretical analysis follows. 
However, the worst-case is often achieved only by very contrived submodular functions. For the ones used in practice, the greedy algorithm often leads to near-optimal solution (\cite{krause08robust} and our own observations). Setting $\alpha$ as the actual performance guarantee for SWP (often very close to 1) can improve the empirical bound, and
%, which empirically often performs rather well. 
we, in practice, typically set $\alpha=1$ to good effect. \looseness-1

\arxiv{
\begin{algorithm}[htb]
{
\begin{algorithmic}[1]
\caption{\textsc{MMax}}
\label{alg:MMax}
\STATE Input: $\{f_i\}_{i=1}^m$, $m$, $V$, partition $\pi^0$.\\
\STATE Let $t=0$.\\
\REPEAT
\FOR{$i=1,\dots,m$}
\STATE   Pick a subgradient $h_i$ at $A_i^{\pi^t}$ for $f_i$.\\
\ENDFOR
\STATE $\pi^{t+1} \in \argmax_{\pi\in \Pi} \min_{i} h_i(A^{\pi}_i)$
\label{line:MMax_modular_version}	\\
\STATE $t = t+1$; \\
\UNTIL{$\pi^t = \pi^{t-1}$}
\STATE Output: $\hat{\pi} \in \argmax_{\pi=\pi^1,\dots, \pi^t} \min_i f_i(A_i^{\pi})$.
\end{algorithmic}}
\end{algorithm} 
}

\textbf{\textsc{MMax}}: \arxivalt{In parallel to \textsc{GreedSat}, we also introduce a semi-gradient based approach for solving SFA under the heterogeneous setting. We call this algorithm minorization-maximization (\textsc{MMax}, see Alg.~\ref{alg:MMax})}{Lastly, we introduce another algorithm for the heterogeneous setting, called minorization-maximization (\textsc{MMax}, see Alg.~\ref{alg:MMax})}.
Similar to the \arxivalt{ones}{one} proposed in\arxivalt{~\cite{rkiyersemiframework2013, nipssubcons2013, rkiyeruai2012}}{~\cite{rkiyersemiframework2013}}, %Similar to the \textsc{GreedSat}, this algorithm works in both the homogeneous and heterogeneous setting. 
the idea is to iteratively maximize tight lower bounds of the submodular functions. Submodular functions have tight modular lower bounds, which are related to the subdifferential $\partial_f(Y)$ of the submodular set function $f$ at a set $Y \subseteq V$\arxivalt{, which is defined \cite{fujishige2005submodular}
as:
\begin{align}
\partial_f(Y) = \{y \in \mathbb{R}^n: f(X) - y(X) \geq f(Y) - y(Y)\;\text{for all } X \subseteq V\}.
\end{align}
For a vector $x \in \mathbb{R}^V$ and $X \subseteq V$, we write $x(X)
= \sum_{j \in X} x(j)$.}{~\cite{fujishige2005submodular}.}  
Denote a subgradient at $Y$ by $h_Y \in \partial_f(Y)$, the extreme points of
$\partial_f(Y)$ may be computed via a greedy algorithm: Let $\sigma$
be a permutation of $V$ that assigns the elements in $Y$ to the first
$|Y|$ positions ($\sigma(i) \in Y$ if and only if $i \leq  |Y|$). Each
such permutation defines a chain with elements $S_0^\sigma =
\emptyset$, $S^{\sigma}_i = \{ \sigma(1), \sigma(2), \dots, \sigma(i)
\}$, and $S^{\sigma}_{|Y|} = Y$. An extreme point
$h^{\sigma}_Y$ of $\partial_f(Y)$ has each entry as \arxivalt{
\begin{align}
    h^{\sigma}_Y(\sigma(i)) = f(S^{\sigma}_i) - f(S^{\sigma}_{i-1}).
  \end{align}}
  {$h^{\sigma}_Y(\sigma(i)) = f(S^{\sigma}_i) -
  f(S^{\sigma}_{i-1})$.}  Defined as above, $h^{\sigma}_Y$
forms a lower bound of $f$, tight at $Y$ --- i.e.,
$h^{\sigma}_Y(X) = \sum_{j \in X} h^{\sigma}_Y(j) \leq f(X), \forall X
\subseteq V$ and $h^{\sigma}_Y(Y) = f(Y)$. 
The idea of \textsc{MMax} is to consider a modular lower bound tight at the set corresponding to each block of a partition. In other words, at iteration $t+1$, for each block $i$, we approximate $f_i$ with its modular lower bound tight at $A_i^{\pi^t}$ and solve a modular version of Problem 1 (Line~\ref{line:MMax_modular_version}), which admits efficient approximation algorithms~\cite{asadpour2010approximation}. \textsc{MMax} is initialized with a partition $\pi^0$, which is obtained by solving Problem 1, where each $f_i$ is replaced with a simple modular function $f^\prime_i(A) = \sum_{a \in A} f_i(a)$. 
%Note that in each iteration, we solve a modular version of Problem 1 (Line~\ref{line:MMax_modular_version}), for which there exists efficient approximation algorithms~\cite{asadpour2010approximation}. 
The following worst-case bound holds:\looseness-1
\begin{theorem}
\label{thm:MMax_bound}
\arxivalt{\textsc{MMax} achieves a worst-case guarantee of $$O( \min_i \frac{1+(|{A}^{\hat{\pi}}_i|-1)(1-\kappa_{f_i}({A}^{\hat{\pi}}_i))}{|{A}^{\hat{\pi}}_i| 
\sqrt{m} \log^3 m}),$$
where $\hat{\pi}=({A^{\hat{\pi}}_1},\cdots, 
A^{\hat{\pi}}_{m})$ is the partition obtained by the algorithm, and $$\kappa_f(A) = 1-\min_{v\in V} \frac{f(v|A\setminus v)}{f(v)}\in [0,1]$$  is the curvature of a submodular function $f$ at $A\subseteq V$.}{
\textsc{MMax} achieves a worst-case guarantee of $O( \min_i \frac{1+(|{A}^{\hat{\pi}}_i|-1)(1-\kappa_{f_i}({A}^{\hat{\pi}}_i))}{|{A}^{\hat{\pi}}_i| 
\sqrt{m} \log^3 m})$,
 where $\hat{\pi}=({A^{\hat{\pi}}_1},\cdots, 
A^{\hat{\pi}}_{m})$ is the partition obtained by the algorithm, and $\kappa_f(A) = 1-\min_{v\in V} \frac{f(v|A\setminus v)}{f(v)}\in [0,1]$  is the curvature of a submodular function $f$ at $A\subseteq V$.}
\end{theorem}
\arxiv{When each submodular function $f_i$ is modular, i.e., $\kappa_{f_i}(A) = 0,\forall A\subseteq V, i$, the approximation guarantee of \textsc{MMax} becomes $O(\frac{1}{\sqrt{m} \log^3 m})$, which matches the performance of the approximation algorithm for the modular problem. When each $f_i$ is fully curved, i.e., $\kappa_{f_i} = 1$, we still obtain a bounded guarantee of $O(\frac{1}{n \sqrt{m} \log^3 m})$. 
Theorem~\ref{thm:MMax_bound} suggests that the performance of \textsc{MMax} improves as the curvature $\kappa_{f_i}$ of each objective $f_i$ decreases. This is natural since \textsc{MMax} essentially uses the modular lower bounds as the proxy for each objective and optimizes with respect to the proxy functions. Lower the curvature of the objectives, the better the modular lower bounds approximate, hence better performance guarantee. 

Since the modular version of SFA is also NP-hard and cannot be exactly solved in polynomial time, we cannot guarantee that successive iterations of \textsc{MMax} improves upon the overall objective. However we still obtain the following Theorem giving a bounded performance gap between the successive iterations. 
\begin{theorem}
\label{thm:iterative_improvement_for_MMAX}
Suppose modular version of SFA is solved with an approximation factor $\alpha\leq 1$, we have for each iteration $t$ that
\begin{align}
\min_{i} f_i(A_i^{\pi_t}) \geq \alpha \min_{i} f_i(A_i^{\pi_{t-1}}).
\end{align}
\end{theorem}
}

%\everyone{Kai}{We may want to use subsection for each algorithm in the Journal version, instead of using paragraph in the current manuscript.}
\subsection{Approximation Algorithms for SLB (Problem~\ref{prob:minmax-genmix} with $\lambda=0$)}
\label{slbapproxsec}

\arxivalt{
In this section, we investigate the problem of submodular load balancing (SLB). It is a special case of Problem~\ref{prob:minmax-genmix} with $\lambda=0$. We first analyze the hardness of SLB. We then show a \lovasz extension-based algorithm with a guarantee matching the problem's hardness. Lastly we describe a more efficient supergradient based algorithm. 

Existing hardness for SLB is shown to be $o(\sqrt{n/\log n})$~\cite{svitkina2008submodular}. However it is independent of $m$, and~\cite{svitkina2008submodular} assumes  $m=\Theta(\sqrt{n/\log n})$ in their analysis. 
In most of the applications of SLB, we find
that the parameter $m$ is such that $m\ll n$ and can sometimes be treated as a constant w.r.t.\ $n$. To this end we offer a more general hardness analysis that is dependent directly on $m$.
}
{We next investigate SLB, where 
existing hardness results~\cite{svitkina2008submodular}
 are $o(\sqrt{n/\log n})$, which is independent of $m$ and implicitly assumes that $m=\Theta(\sqrt{n/\log n})$. 
 However, applications for SLB are often dependent on $m$ with $m\ll n$. We hence offer hardness analysis in terms of $m$ in the following.} %for general $m=O(\sqrt{n/\log n})$.
\begin{theorem}
\label{thm:hardness_SLB}
For any $\epsilon>0$, 
SLB cannot be approximated to a factor of $(1-\epsilon)m$ for any $m=o(\sqrt{n/\log n})$ with polynomial number of queries even under the homogeneous setting. 
\end{theorem}
\arxivalt{Though the proof technique for Theorem~\ref{thm:hardness_SLB} mostly carries over from~\cite{svitkina2008submodular}, the result strictly generalizes the analysis in~\cite{svitkina2008submodular}. For any choice of $m=o(\sqrt{n/\log n})$
Theorem~\ref{thm:hardness_SLB} implies that it is information theoretically hard to approximate SLB better than $m$ even for the homogeneous setting. For the rest of the paper, we assume $m=o(\sqrt{n/\log n})$ for SLB, unless stated otherwise.
It is worth pointing out that arbitrary partition $\pi \in \Pi$ already achieves the best approximation factor of $m$ that one can hope for under the homogeneous setting. Denote $\pi^*$ as the optimal partitioning for SLB, i.e., $\pi^*\in \argmin_{\pi\in \Pi} \max_i f(A_i^{\pi})$. 
This can be verified by considering the following:
\begin{align}
\max_{i}f(A_i^\pi) \leq f(V) \leq \sum_{i=1}^m f(A_i^{\pi^*}) \leq 
m \max_i f(A_i^{\pi^{*}}).
\end{align}
It is therefore theoretically interesting to consider only the heterogeneous setting. 

}
{For the rest of the paper, we assume $m=o(\sqrt{n/\log n})$ for SLB, unless stated otherwise.} 
\notarxiv{

\textbf{\textsc{GreedMin}: }
Theorem~\ref{thm:hardness_SLB} implies that SLB is hard to approximate better than $m$. However, an arbitrary partition $\pi \in \Pi$ already achieves the best approximation factor of $m$ that one can hope for under the homogeneous setting, since $\max_{i}f(A_i^\pi) \leq f(V) \leq \sum_{i}f(A_i^{\pi^\prime}) \leq 
m \max_i f(A_i^{\pi^{\prime}})$ for any $\pi^\prime \in \Pi$. 
%Arbitrary partition already gives the best theoretical guarantee one can hope for under the homogeneous setting. 
In practice, 
one can still implement a greedy style heuristic, which we refer to as \textsc{GreedMin} (Alg.~\ref{alg:greed_min}). Very similar to \textsc{GreedMax}, \textsc{GreedMin} only differs in Line~\ref{line:greedy_step_greed_min}, where the item with the smallest marginal gain is added. 
Since the functions are all monotone, any additions to a block can
(if anything) only increase its value, so we choose to add to the minimum valuation block
in Line 4 to attempt to keep the maximum valuation block from
growing further.
\arxiv{The lazy evaluation trick used for \textsc{GreedMax} does not apply in this case, but~\cite{smola2015graphpartitioning} give  efficient implementation of \textsc{GreedMin} for certain submodular function.} 
\arxiv{In general, \textsc{GreedMin} requires $O(|V|^2)$ function valuations, which is unfeasible for large-scale applications. In practice, one can relax the condition in Line~\ref{line:greedy_step_greed_min}. Instead of searching among all items in $R$, one can, in each round, randomly select a subset $\hat{R}\subset R$ and choose an item with the smallest marginal gain from only the subset $\hat{R}$. The resultant computational complexity is reduced to $O(|\hat{R}||V|)$ function valuations. Empirically we observe that \textsc{GreedMin} can be sped up more than 100 times by this trick without much performance loss.}
\looseness-1}

\arxiv{
\begin{algorithm}[htb]
{
\begin{algorithmic}[1]
\caption{\textsc{\lovasz{}\!Round}}
\label{alg:lovasz_ext_alg}
\STATE Input: $\{f_i\}_{i=1}^m$, $\{\tilde{f}_i\}_{i=1}^m$, $m$, $V$.\\
{
\STATE Solve for $\{x^*_i\}_{i=1}^m$ via convex relaxation. \\
\label{line:solve_relaxation_lovasz}
}
\STATE Rounding: Let $A_1=,\dots,=A_m=\emptyset$.\\
\label{line:start_rounding}
\FOR{$j=1,\dots, n$}
\STATE $\hat{i}\in \argmax_{i} x^*_i(j)$; $A_{\hat{i}} = A_{\hat{i}} \cup j$\\
\ENDFOR
\label{line:end_rounding}
\STATE Output $\hat{\pi} = \{A_i\}_{i=1}^m$.
\end{algorithmic}}
\end{algorithm}
}
\textbf{\textsc{\lovasz{}\!Round}: } 
\arxivalt{
Next we propose a tight algorithm -- \textsc{\lovasz{}\!Round} (see Alg.~\ref{alg:lovasz_ext_alg}) for the heterogeneous setting of SLB.
}{Next we consider the heterogeneous setting, for which we propose a tight algorithm -- \textsc{\lovasz{}\!Round} (see Alg.~\ref{alg:lovasz_ext_alg}).} The algorithm proceeds as follows: (1) apply the \lovasz extension of submodular functions to relax SLB to a convex program, which is exactly solved to a fractional solution (Line~\ref{line:solve_relaxation_lovasz});
%relax SLB to a convex program using \lovasz relaxation of submodular functions and globally solve the continuous relaxation for a fractional solution; 
(2) map the fractional solution to a partition using the $\theta$-rounding technique as proposed in~\cite{iyer2014uai-fast-cvx-relaxations} (Line~\ref{line:start_rounding} -~\ref{line:end_rounding}).
The \lovasz extension, which naturally connects a submodular function $f$ with its convex relaxation $\tilde{f}$, is defined as follows: given any $x\in [0,1]^n$, we obtain a permutation $\sigma_x$ by ordering its elements in non-increasing order, and thereby a chain of sets $S_0^{\sigma_x}\subset, \dots,\subset S_n^{\sigma_x}$ with $S_j^{\sigma_x} = \{\sigma_x(1),\dots, \sigma_x(j)\}$ for $j = 1,\dots, n$. The \lovasz extension $\tilde{f}$ for $f$ is the weighted sum of the ordered entries of $x$: 
\arxivalt{
\begin{align}
\tilde{f}(x) = \sum_{j=1}^n x(\sigma_x(j)) (f(S_j^{\sigma_x}) - f(S_{j-1}^{\sigma_x})).
\end{align}
}
{$\tilde{f}(x) = \sum_{j=1}^n x(\sigma_x(j)) (f(S_j^{\sigma_x}) - f(S_{j-1}^{\sigma_x}))$.} Given the convexity of the $\tilde{f}_i$'s , SLB is relaxed to the following convex program:\looseness-1
\notarxiv{\abovedisplayshortskip=.5ex\belowdisplayshortskip=.5ex\abovedisplayskip=.5ex\belowdisplayskip=.5ex}
\begin{align}
&\min_{x_1,\dots,x_m \in [0,1]^n} \max_{i} \tilde{f}_i(x_i), \mbox{ s.t }  \sum_{i=1}^m x_i(j) \geq 1, \text{ for } j=1,\dots,n
\label{eqn:convex_relaxation_SLB}
\end{align}
Denoting the optimal solution for Eqn~\ref{eqn:convex_relaxation_SLB} as $\{x_1^*,\dots,x_m^*\}$, the $\theta$-rounding step simply maps each item $j\in V$ to a block $\hat{i}$ such that $\hat{i} \in \argmax_{i}x^*_i(j)$ \arxiv{(ties broken arbitrarily)}. The bound for \textsc{\lovasz{}\!Round} is as follows:\looseness-1
\begin{theorem}
\label{thm:lovaszRound_bound}
\arxivalt{\textsc{\lovasz{}\!Round} is guaranteed to find a partition $\hat{\pi} \in \Pi$ such that $$\max_{i}f_i(A_i^{\hat{\pi}}) \leq m \min_{\pi \in \Pi} \max_{i} f_i(A_i^{\pi})$$.}
{\textsc{\lovaszRound} achieves a worst-case approximation factor $m$.}
\end{theorem}
We remark that, to the best of our knowledge, \textsc{\lovaszRound} is the first algorithm that is tight and that gives an approximation in terms of $m$ for the heterogeneous setting. 
%is the first algorithm achieving the tight approximation guarantee for SLB. 

\arxiv{
\begin{algorithm}[htb]
{
\begin{algorithmic}[1]
\caption{\textsc{MMin}}
\label{alg:MMin}
\STATE Input: $\{f_i\}_{i=1}^m$, $m$, $V$, partition $\pi^0$.\\
\STATE Let  $t=0$ \\
\REPEAT
\FOR{$i=1,\dots,m$}
\STATE   Pick a supergradient $m_i$ at $A_i^{\pi^t}$ for $f_i$.\\
\ENDFOR
\STATE $\pi^{t+1} \in \argmin_{\pi\in \Pi} \max_{i} m_i (A^{\pi}_i)$
\label{line:MMin_modular_version}	\\
\STATE	$t = t+1$;\\
\UNTIL{$\pi^t = \pi^{t-1}$}
\STATE Output: $\pi^t$.
\end{algorithmic}}
\end{algorithm} 
}

\textbf{\textsc{MMin}}: Similar to \textsc{MMax} for SFA, we propose Majorization-Minimization (\textsc{MMin}, see Alg.~\ref{alg:MMin}) for SLB. Here, we iteratively choose modular upper bounds, which are defined via superdifferentials $\partial^f(Y)$ of a submodular
function 
\cite{jegelka2011-nonsubmod-vision}
at
$Y$\arxivalt{: %Intuitively, it consists of tight linear upper bounds:
\begin{align}\label{supdiff-def}
\partial^f(Y) = \{y \in \mathbb{R}^n: f(X) - y(X) \leq f(Y) - y(Y);\text{for all } X \subseteq V\}.
\end{align}}{.}
Moreover, there are specific supergradients\arxivalt{~\cite{rkiyersubmodBregman2012, rkiyersemiframework2013}}{~\cite{rkiyersemiframework2013}} that define the following two modular upper bounds (when referring to either one, we use $m^f_X$):\looseness-1
\notarxiv{\small}
\begin{align*}
m^f_{X, 1}(Y) \triangleq f(X) - \!\!\!\! \sum_{j \in X \backslash Y } f(j| X \backslash j) + \!\!\!\! \sum_{j \in Y \backslash X} f(j| \emptyset)\scalebox{1.3}{,}\;\;\;\arxiv{\\}
m^f_{X, 2}(Y) \triangleq f(X) - \!\!\! \sum_{j \in X \backslash Y } f(j| V \backslash j) + \!\!\!\! \sum_{j \in Y \backslash X} f(j| X). \nonumber
\end{align*}
\normalsize
Then $m^f_{X, 1}(Y) \geq f(Y)$ and $m^f_{X, 2}(Y) \geq f(Y), \forall Y \subseteq V$ and $m^f_{X, 1}(X) = m^f_{X, 2}(X) = f(X)$. 
%The algorithm is shown in Alg.~\ref{alg:MMin}. 
%In this case, we choose tight modular upper bound at every iteration. 
At iteration $t+1$, for each block $i$, \textsc{MMin} replaces $f_i$ with a choice of its modular upper bound $m_i$ tight at $A_i^{\pi^t}$ and solves a modular version of Problem 2 (Line~\ref{line:MMin_modular_version}), for which there exists an efficient LP relaxation based algorithm~\cite{lenstra1990approximation}.
Similar to \textsc{MMax}, the initial partition $\pi^0$ is obtained by solving Problem 2, where each $f_i$ is substituted with $f_i^\prime(A) = \sum_{a \in A} f_i(a)$. 
% At each iteration, we solve a modular version of Problem 2, for which there exist LP relaxation based algorithms~\cite{lenstra1990approximation}. 
 The following worst-case bound holds:\looseness-1
\begin{theorem}
\label{thm:MMin_bound}
\arxivalt{\textsc{MMin} achieves a worst-case guarantee of $(2\max_i \frac{|A_i^{\pi^*}|}{1+(|A_i^{\pi^*}|-1)(1-\kappa_{f_i}(A_i^{\pi^*}))}),$
where $\pi^* = (A_1^{\pi^*}, \cdots, A_m^{\pi^*})$ denotes the optimal partition.}
{
\textsc{MMin} achieves a worst-case guarantee of $(2\max_i \frac{|A_i^{\pi^*}|}{1+(|A_i^{\pi^*}|-1)(1-\kappa_{f_i}(A_i^{\pi^*}))})$, where $\pi^* = (A_1^{\pi^*}, \cdots, A_m^{\pi^*})$ denotes the optimal partition.}
\end{theorem}
\arxiv{Similar to MMax, we can show that MMin has bounded performance gaps in successive iterations.
\begin{theorem}
\label{thm:iterative_improvement_for_MMIN}
Suppose the modular version of SLB can be solved with an approximation factor $\alpha\geq 1$, we have for each iteration $t$ that
\begin{align}
\max_{i} f_i(A_i^{\pi_t}) \leq \alpha \max_{i} f_i(A_i^{\pi_{t-1}}).
\end{align}
\end{theorem}
}

\notarxiv{
\begin{figure}
\centering
\begin{subfigure}{0.3\linewidth}
  \includegraphics[width=\linewidth, height=0.8\linewidth, trim=4.1cm 8.5cm 4.1cm 8.7cm, clip=true]{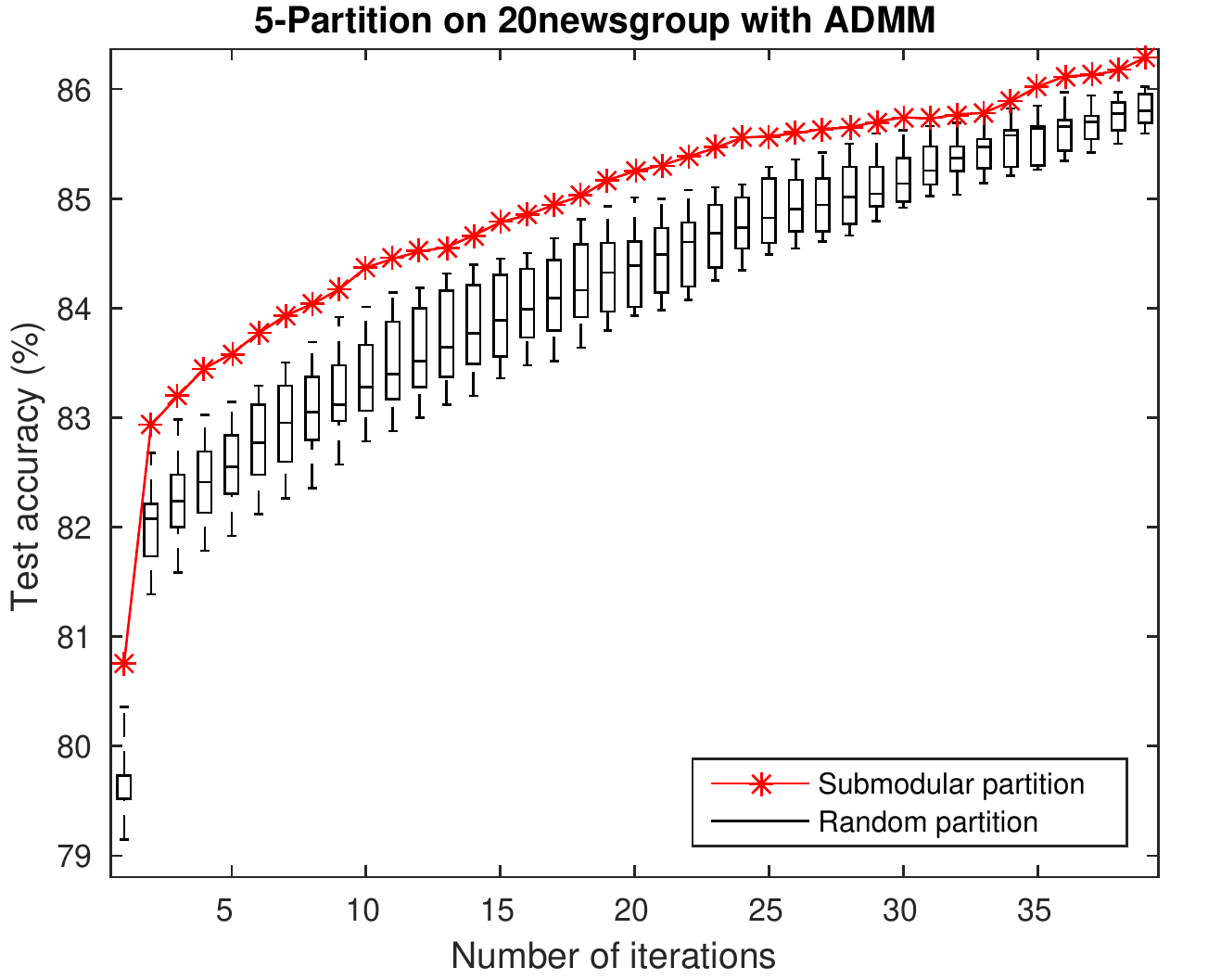}
  \includegraphics[width=\linewidth, height=0.8\linewidth, trim=4.1cm 8.5cm 4.1cm 8.7cm, clip=true]{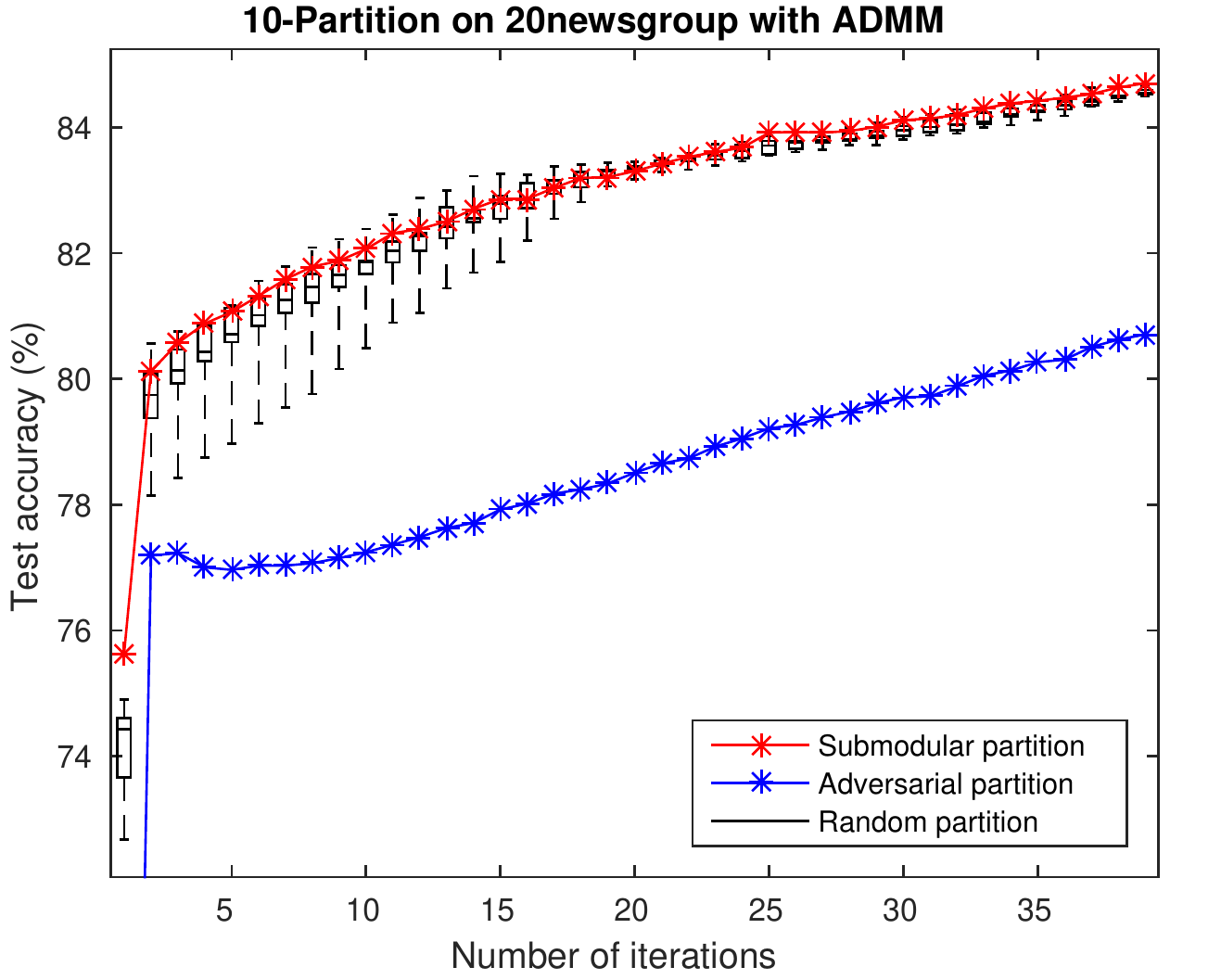}
%  \captionsetup{labelformat=empty}
  \vspace{-1.2\baselineskip}
  \caption{\hspace{-.2em}20Newsgroups} \label{fig:ADMM_20News_res}
\end{subfigure}
\begin{subfigure}{0.3\linewidth}
  \includegraphics[width=\linewidth,  height=0.8\linewidth, trim=3.8cm 8.5cm 4.1cm 8.7cm, clip=true]{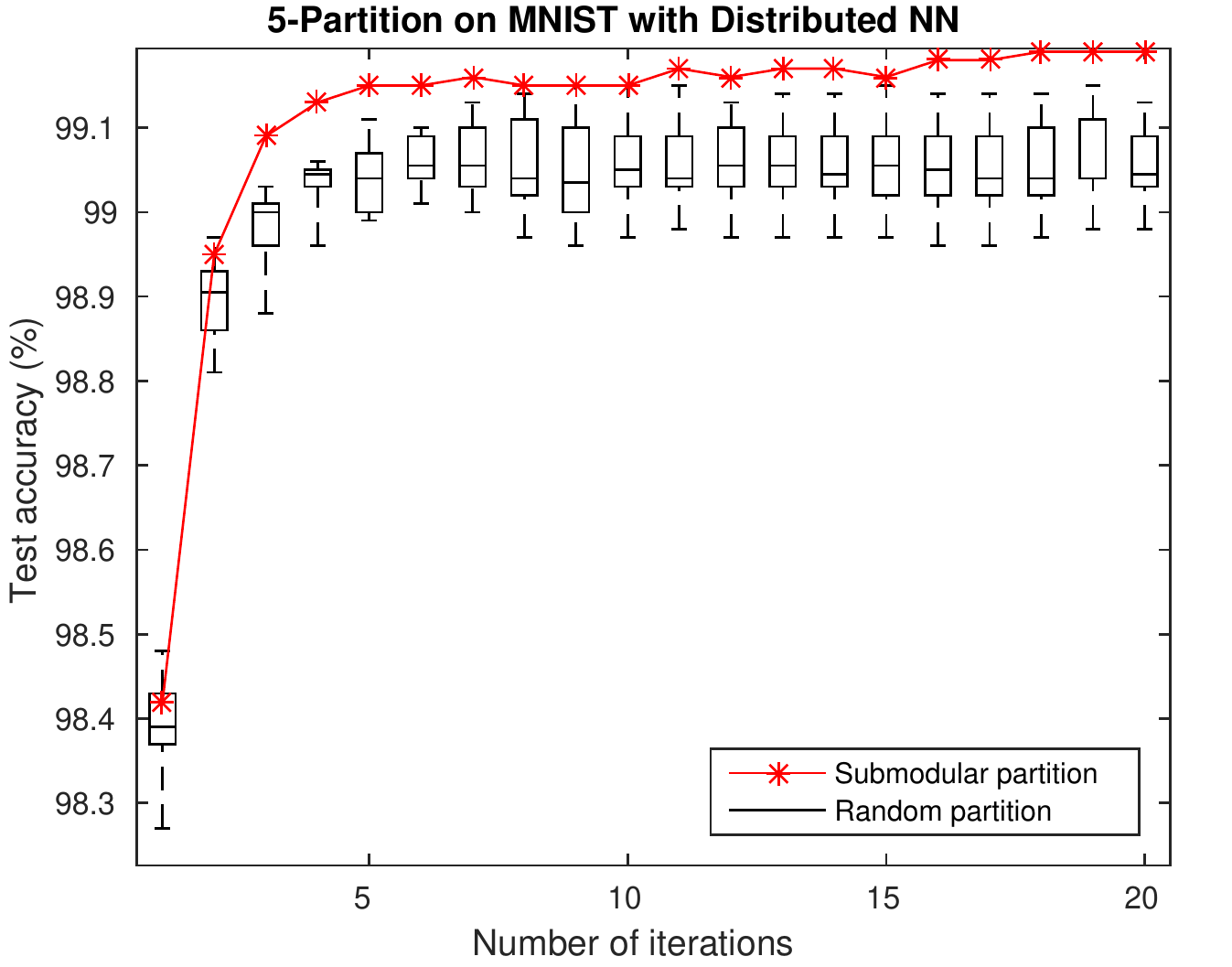}
%\shengjie{Jeff}{fix this file/missing figure}
    \includegraphics[width=\linewidth,  height=0.8\linewidth, trim=3.8cm 8.5cm 4.1cm 8.7cm, clip=true]{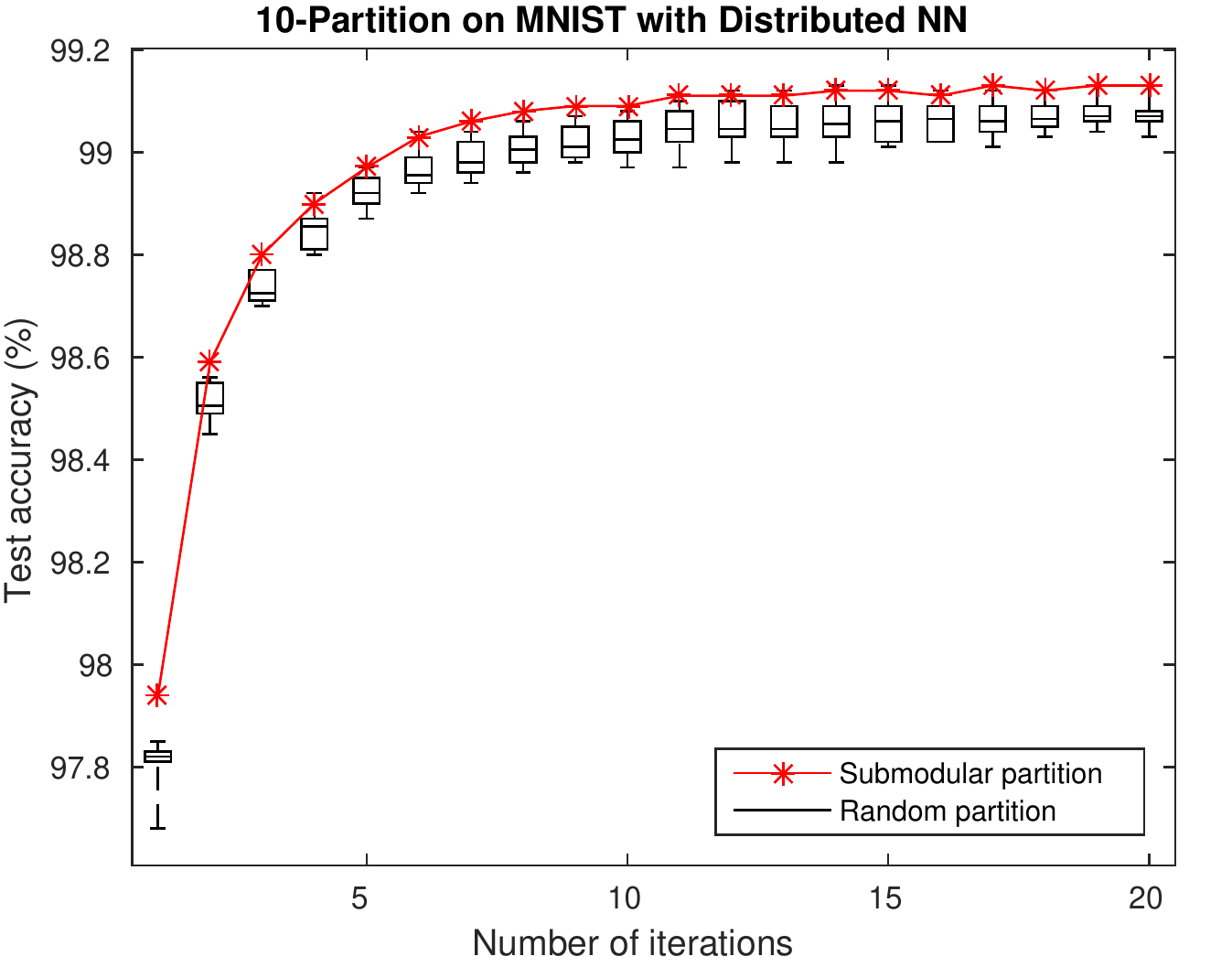}
   % \captionsetup{labelformat=empty}
  \vspace{-1.2\baselineskip}
  \caption{MNIST}\label{fig:DNN_MNIST_res}
\end{subfigure}
\begin{subfigure}{0.3\linewidth}
  \includegraphics[width=\linewidth,  height=0.8\linewidth, trim=3.8cm 8.5cm 4.1cm 8.7cm, clip=true]{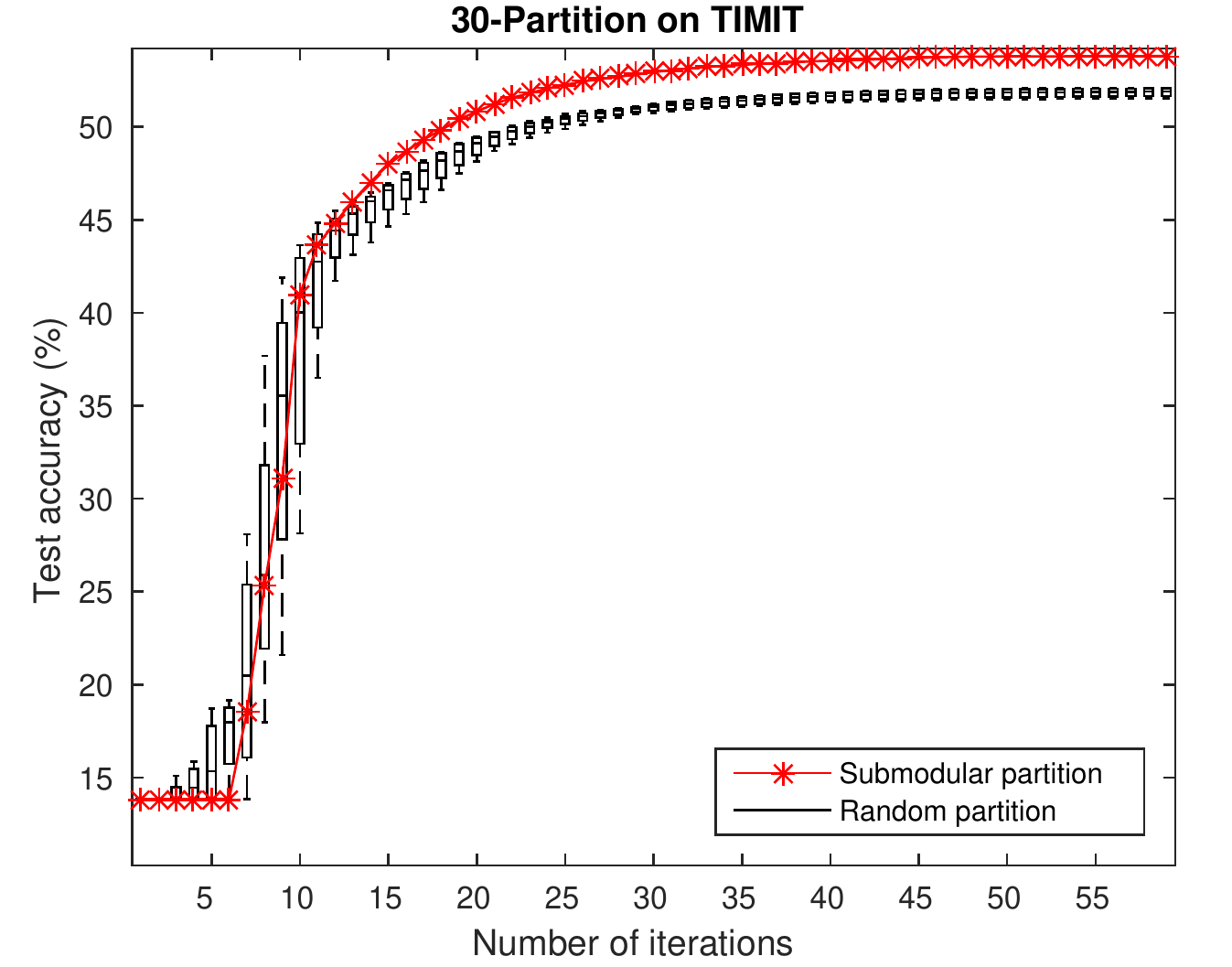}
%\shengjie{Jeff}{fix this file/missing figure}
    \includegraphics[width=\linewidth,  height=0.8\linewidth, trim=3.8cm 8.5cm 4.1cm 8.7cm, clip=true]{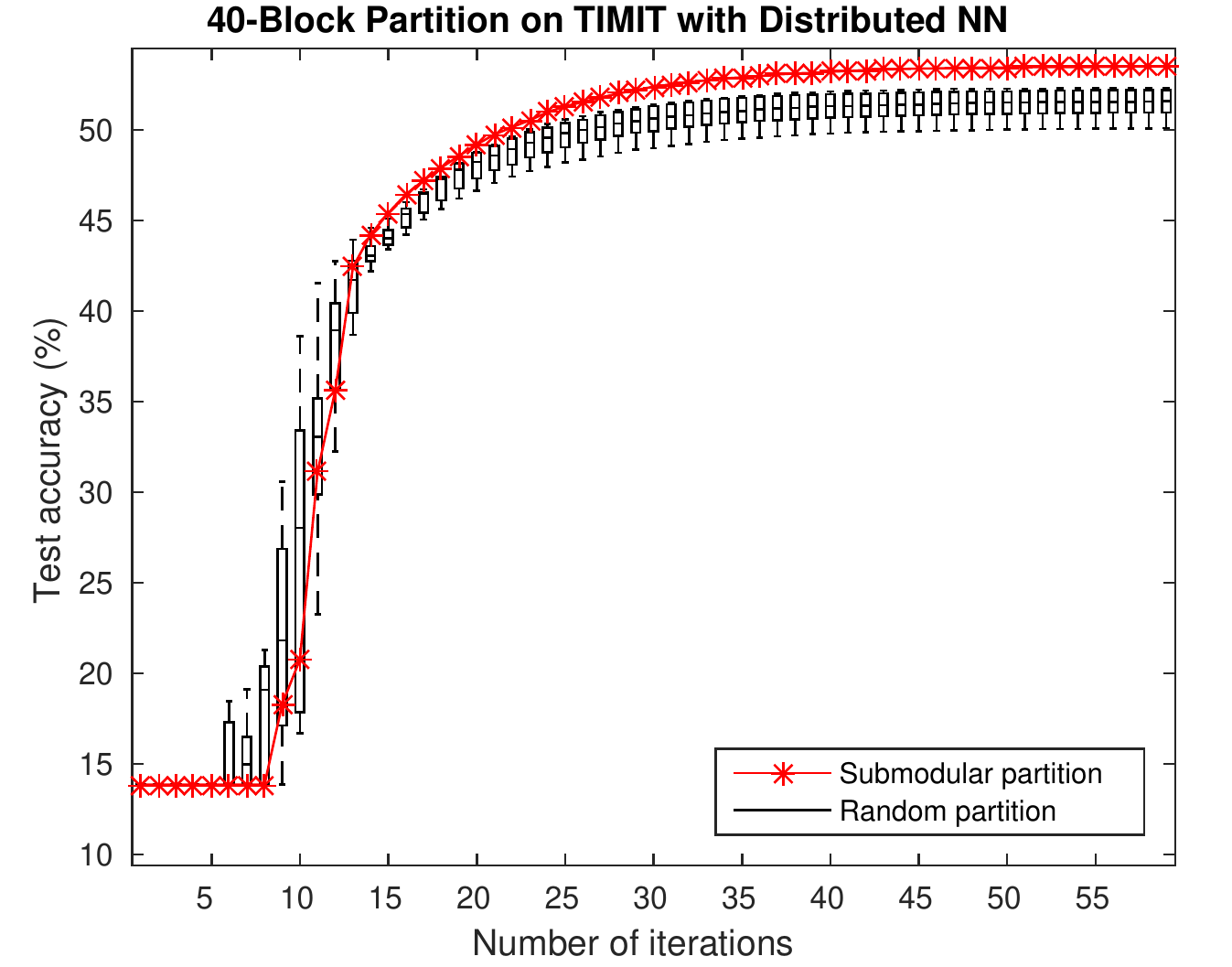}
   % \captionsetup{labelformat=empty}
  \vspace{-1.2\baselineskip}
  \caption{TIMIT}\label{fig:DNN_timit_res}
\end{subfigure}
\vspace{-.5\baselineskip}
\caption{Comparison between submodular and random partitions for
  distributed ML, including ADMM \arxiv{on 20Newsgroup }
  (Fig~\ref{fig:ADMM_20News_res}) and distributed neural nets\arxiv{ on MNIST }
  (Fig~\ref{fig:DNN_MNIST_res}) and\arxiv{ TIMIT }
  (Fig~\ref{fig:DNN_timit_res}).  For the
  box plots, the central mark is the median, the box edges are
  25th and 75th percentiles, and the bars denote the best and worst
  cases.\looseness-1 
  %In the right two columns, the adversarial partitions are so bad that they are off the plots. 
  }
\notarxiv{\vspace{-1\baselineskip}} 
\end{figure}

}

%\kai{Kai}{Think about moving it somewhere else.}
%\subsection{Generalization to $k$-cover Partitioning}

%\jeff{Kai}{Following section gives the discussion of how to extend the algorithms for SFA and SLB to the k-cover constraint. Since it is unclear how to make it fit for the entire paper, I am thinking about remove this small paragraph.}

\section{General Submodular Partitioning (\theproblems{} when $0 < \lambda < 1$)}
\label{sec:gener-subm-part}
\notarxiv{\vspace{-0.5\baselineskip}}
In this section we study Problem~\ref{prob:maxmin-genmix} and Problem~\ref{prob:minmax-genmix}, in the most general case, i.e., $0 < \lambda < 1$. \arxiv{We use the proposed algorithms for the special cases of \theproblems{} as the building blocks to design algorithms for the general scenarios $(0<\lambda<1)$. We first propose a simple and generic scheme that provides performance guarantee in terms of $\lambda$ for both problems. 
We then generalize the proposed \textsc{GreedSat} to obtain a more practically interesting algorithm for Problem~\ref{prob:maxmin-genmix}. 
For Problem~\ref{prob:minmax-genmix} we generalize \textsc{\lovasz Round} to obtain a relaxation based algorithm. 

}
\arxivalt{\textbf{Extremal Combination Scheme: }First we describe the scheme that works for both 
problem~\ref{prob:maxmin-genmix} and~\ref{prob:minmax-genmix}.}{We first propose a simple and general
``extremal combination'' scheme that works both for problem~\ref{prob:maxmin-genmix} and~\ref{prob:minmax-genmix}.} 
It 
naturally combines an algorithm for solving the worst-case problem ($\lambda=0$) with an algorithm for solving the average case ($\lambda=1$). 
We use Problem~\ref{prob:maxmin-genmix} as an example, but the same scheme easily works for Problem~\ref{prob:minmax-genmix}.
Denote \textsc{AlgWC} as the algorithm for the worst-case problem (i.e. SFA), and \textsc{AlgAC} as the algorithm for the  average case (i.e., SWP). The scheme is to first obtain a partition $\hat{\pi}_1$ by
running \textsc{AlgWC} on the instance of Problem~\ref{prob:maxmin-genmix} with $\lambda=0$ and a second partition $\hat{\pi}_2$ by running \textsc{AlgAC} with $\lambda=1$. 
Then we output one of $\hat{\pi}_1$ and $\hat{\pi}_2$, with which the higher valuation for Problem~\ref{prob:maxmin-genmix} is achieved.  
%We show that this simple scheme provides an approximation guarantee in terms of $\lambda$. 
We call this scheme \textsc{CombSfaSwp}. 
%and \textsc{CombSlbSmp} for Problem~\ref{prob:minmax-genmix}.
Suppose \textsc{AlgWC} solves the worst-case problem with a factor $\alpha\leq 1$ and \textsc{AlgAC} for the average case with $\beta\leq 1$. 
%It holds that \textsc{CombSfaSwp} solves the extreme cases with a factor $\alpha$ when $\lambda=0$ and with a factor $\beta$ when $\lambda=1$.
When applied to Problem~\ref{prob:minmax-genmix} we refer to this scheme as  \textsc{CombSlbSmp} ($\alpha\geq 1$ and $\beta\geq 1$). 
The following guarantee holds for both schemes:\looseness-1
%We show the bound for both schemes in the following Theorem:\looseness-1 
\begin{theorem}
\label{thm:max_min_sum_combined_bound}
For any $\lambda \in  (0,1)$ \textsc{CombSfaSwp} solves Problem~\ref{prob:maxmin-genmix} with a factor $\max\{\frac{\beta \alpha}{\bar{\lambda}\beta+\alpha},\lambda\beta \}$
 in the \emph{heterogeneous} case, and $\max\{\min\{\alpha,\frac{1}{m}\},\frac{\beta \alpha}{\bar{\lambda}\beta + \alpha}, \lambda \beta\}$ in the \emph{homogeneous} case.
 Similarly, \textsc{CombSlbSmp} solves Problem~\ref{prob:minmax-genmix}
 with a factor $\min\{\frac{m\alpha}{m\bar{\lambda}+\lambda},\beta(m\bar{\lambda}+\lambda) \}$
 in the \emph{heterogeneous} case, and  $\min\{m, \frac{m\alpha}{m\bar{\lambda}+\lambda}, \beta(m\bar{\lambda}+\lambda)\}$  in the \emph{homogeneous} case.\looseness-1 
\end{theorem}
\notarxiv{The drawback of \textsc{CombSfaSwp} and \textsc{CombSlbSmp} is that they do not explicitly exploit the trade-off between the average-case and worst-case objectives in terms of $\lambda$. 
%In fact these schemes almost ignore $\lambda$ in the procedure. 
To obtain more practically interesting algorithms, we also give \textsc{GeneralGreedSat} that generalizes \textsc{GreedSat} to solve Problem~\ref{prob:maxmin-genmix}. 
Similar to \textsc{GreedSat} we define an intermediate objective: $\bar{F}^c_{\lambda} (\pi) = \frac{1}{m}\sum_{i=1}^m \min\{\bar{\lambda} f_i(A^\pi_i) + \lambda \frac{1}{m}\sum_{j=1}^m f_j(A_j^{\pi}),c\}$ in \textsc{GeneralGreedSat}.  
Following the same algorithmic design as in \textsc{GreedSat}, 
\textsc{GeneralGreedSat} only differs from \textsc{GreedSat} in Line~\ref{line:swp_greedy_sat}, where the submodular welfare problem is defined on the new objective $\bar{F}^{c}_{\lambda}(\pi)$. 
%Let $\alpha$ be the guarantee for solving submodular welfare, \textsc{GeneralGreedSat} approximates Problem~\ref{prob:maxmin-genmix} with the following bounds:\looseness-1
%\begin{theorem}
%\label{thm:generalGreedSat_bounds}
%Given $\epsilon$, $\alpha$, and $0 \leq \lambda\leq 1$, \textsc{GeneralGreedSat} finds a partition $\hat{\pi}$ that satisfies the following:
%$\bar{\lambda}\min_{i} f_i(A_i^{\hat{\pi}}) + \lambda \frac{1}{m}\sum_{i=1}^m f_i(A_i^{\hat{\pi}}) \geq \lambda \alpha (OPT-\epsilon)$
%where $OPT= \max_{\pi\in \Pi} \bar{\lambda} \min_{i}f_i(A_i^{\pi}) + \lambda \frac{1}{m} \sum_{i=1}^m f_{i}(A_i^{\pi})$. 
%Moreover, let $F_{\lambda,i}(\pi) = \bar{\lambda}f_i(A_i^{\pi}) +\lambda \frac{1}{m}\sum_{j=1}^m f_j(A_j^{\pi})$. 
%Given any $0 < \delta < \alpha $, 
%there is a set $I\subseteq \{1,\dots,m\}$ such that 
%$|I| \geq \ceil{m( \alpha-\delta)}$ and 
%$F_{i,\lambda} (\hat{\pi}) \geq \max\{\frac{\delta}{1-\alpha + \delta}, \lambda \alpha\}(OPT-\epsilon),\forall i\in I$.
%\end{theorem}
In~\cite{datapartitionextend} we show that
\textsc{GeneralGreedSat} gives $\lambda/2$ approximation, while also yielding a bi-criterion guarantee that generalizes Theorem~\ref{thm:greedy_sat_bound}. 
In particular
\textsc{GeneralGreedSat} recovers the bicriterion guarantee as shown in Theorem~\ref{thm:greedy_sat_bound} when $\lambda=0$. In the case of $\lambda=1$, \textsc{GeneralGreedSat} recovers the $1/2$-approximation 
guarantee of the greedy algorithm for solving the submodular welfare problem, i.e., the average-case objective. Moreover an improved guarantee is achieved by
\textsc{GeneralGreedSat} as $\lambda$ increases. Details are given in~\cite{datapartitionextend}.\looseness-1
%Theorem~\ref{thm:generalGreedSat_bounds}  generalizes Theorem~\ref{thm:greedy_sat_bound} when $\lambda=0$, i.e., it recovers the bi-criterion guarantee in Theorem~\ref{thm:greedy_sat_bound} for the worst-case scenario ($\lambda=0$). Moreover Theorem~\ref{thm:generalGreedSat_bounds} implies that the factor of $\alpha$ for the average-case objective can almost be recovered by \textsc{GeneralGreedSat} if $\lambda=1$.
%It also gives an improved guarantee as $\lambda$ increases suggesting that Problem~\ref{prob:maxmin-genmix} becomes easier as the mixed objective weights more on the average-case objective. 
%We also point out that the optimality guarantee of \textsc{GeneralGreedSat}
%smoothly interpolates the two extreme cases in terms of $\lambda$. 

To solve Problem~\ref{prob:minmax-genmix} we generalize \textsc{\lovasz{}\!Round} leading to \textsc{General\lovasz{}\!Round}. 
Similar to \textsc{\lovasz{}\!Round} we relax each submodular objective as its convex relaxation using the \lovasz{} extension. 
Almost the same as \textsc{\lovasz{}\!Round}, 
\textsc{General\lovasz{}\!Round} only differs in Line~\ref{line:solve_relaxation_lovasz}, where Problem~\ref{prob:minmax-genmix} is relaxed as the following convex program:
%\begin{align}
$\min_{x_1,\dots,x_m \in [0,1]^n} \bar{\lambda}\max_{i} \tilde{f}_i(x_i) + \lambda \frac{1}{m}\sum_{j=1}^m \tilde{f}_j(x_j), \mbox{ s.t }  \sum_{i=1}^m x_i(j) \geq 1, \text{ for } j=1,\dots,n.$
%\end{align}
Following the same rounding procedure as \textsc{\lovasz{}\!Round}, \textsc{General\lovasz{}\!Round} is guaranteed to give an $m$-approximation for Problem~\ref{prob:minmax-genmix} with general $\lambda$. Details are given in~\cite{datapartitionextend}.}
%After solving for
%the fractional solution $\{x_i^*\}_{i=1}^m$ to the convex program, \textsc{General\lovasz{}\!Round} then rounds it to a partition using the same rounding technique as \textsc{\lovasz{}\Round}.
%The following Theorem holds:\looseness-1
%\begin{theorem}
%\label{thm:general_lovasz_round}
%\textsc{General\lovasz{}\!Round} is guaranteed to find a partition $\hat{\pi}\in \Pi$ such that
%$\max_i \bar{\lambda} f_i(A_i^{\hat{\pi}}) + \lambda \frac{1}{m} \sum_{j=1}^m f_j(A_j^{\hat{\pi}}) \leq m \min_{\pi\in \Pi} \max_i \bar{\lambda} f_i(A_i^{{\pi}}) + \lambda \frac{1}{m} \sum_{j=1}^m f_j(A_j^{{\pi}})$.
%\end{theorem}
%Theorem~\ref{thm:general_lovasz_round} generalizes Theorem~\ref{thm:lovaszRound_bound} when $\lambda=0$. Moreover we achieve a factor of $m$ by \textsc{General\lovasz Round} for any $\lambda$. Though the approximation guarantee is independent of $\lambda$ the algorithm naturally exploits the trade-off between the worst-case and average-case objectives in terms of $\lambda$. 

\arxiv{

\begin{algorithm}[htb]
\caption{\textsc{GeneralGreedSat}}
\label{alg:GeneralgreedSat}
{
\begin{algorithmic}[1]
\STATE Input: \arxiv{$\epsilon$, }$\{f_i\}_{i=1}^m$, $m$, $V$, $\lambda$, $\alpha$.\\
\STATE Let $\bar{F}^c_{\lambda}(\pi) = \frac{1}{m}\sum_{i=1}^m \min\{ \bar{\lambda}
f_i(A^\pi_i) + \lambda \frac{1}{m}\sum_{j=1}^m f_j(A_j^{\pi}),c\}$.\\
\label{line:def_F_c_function_generalGreedSat}
\STATE Let $c_{\min} = 0$,
$c_{\max} = \sum_{i=1}^m f_{i}(V)$\\
\WHILE {$ c_{\max} - c_{\min}   \geq \epsilon $}
\STATE $c = \frac{1}{2}(c_{\max} + c_{\min})$\\
\STATE $\hat{\pi}^c \in \argmax_{\pi \in \Pi} \bar{F}^c_{\lambda}(\pi)$ // solved by \textsc{GreedSwp} (Alg.~\ref{alg:greedSWP}) \\
\label{line:swp_general_greedy_sat}
\IF{$\bar{F}^c(\hat{\pi}^c) < \alpha c$}
\STATE $c_{\max} = c$ \\	
\ELSE 
\STATE $c_{\min} = c$;  $\hat{\pi} \leftarrow \hat{\pi}^c$\\
\ENDIF
\ENDWHILE    
\STATE Output: $\hat{\pi}$.
\end{algorithmic}}
\end{algorithm} 

\textbf{GeneralGreedSat: }The drawback of \textsc{CombSfaSwp} and \textsc{CombSlbSmp} is that they do not explicitly exploit the trade-off between the average-case and worst-case in terms of $\lambda$. 
%In fact these schemes almost ignore $\lambda$ in the procedure. 
To obtain more practically interesting algorithms, we first give \textsc{GeneralGreedSat} (See Alg.~\ref{alg:GeneralgreedSat}) that generalizes \textsc{GreedSat} to solve Problem~\ref{prob:maxmin-genmix} for general $\lambda$. 
The key idea of \textsc{GeneralGreedSat} is again to relax Problem~\ref{prob:maxmin-genmix} to a simpler submodular welfare problem (SWP). Similar to \textsc{GreedSat} we define an intermediate objective:
\begin{align}
\bar{F}^c_{\lambda} (\pi) = \frac{1}{m}\sum_{i=1}^m \min\{\bar{\lambda} f_i(A^\pi_i) + \lambda \frac{1}{m}\sum_{j=1}^m f_j(A_j^{\pi}),c\}.
\end{align} 
%It can be verified that the combinatorial optimization problem $\max_{\pi\in \Pi} \bar{F}^c_{\lambda}(\pi)$
%This intermediate objective can also be interpreted as: $$\bar{F}_{\lambda}^c(\pi) = \frac{1}{m} \sum_{i=1}^m 	[\min\{\bar{\lambda}f_i(A_i^\pi),c\} + \lambda f_i(A_i^{\pi})].$$ 
%Defining each term of $\bar{F}_{\lambda}^c(\pi)$ as $f_i^{\prime}(A) = \min\{\bar{\lambda}f_i(A),c\} + \lambda f_i(A)$, it is easy to see that $f_i(A)$ is a monotone submodular function, and we have $\bar{F}^c_\lambda (\pi) = \frac{1}{m}\sum_{i=1}^m f^\prime_i(A_i^{\pi})$ as the sum of submodular functions defined on each block of the partition. 
It is easy to verify that the combinatorial optimization problem $\max_{\pi\in \Pi} \bar{F}^c_{\lambda} (\pi)$ (Line~\ref{line:swp_general_greedy_sat}) can be formulated as the submodular welfare problem, for which we can solve efficiently with \textsc{GreedSwp} (see Alg.~\ref{alg:greedSWP}). 
Defining $\alpha$ as the optimality guarantee of the algorithm for solving Line~\ref{line:swp_general_greedy_sat} \textsc{GeneralGreedSat} solves Problem~\ref{prob:maxmin-genmix} with the following bounds:
\begin{theorem}
\label{thm:generalGreedSat_bounds}
Given $\epsilon>0$, $0\leq \alpha\leq 1$, and $0 \leq \lambda\leq 1$, \textsc{GeneralGreedSat} finds a partition $\hat{\pi}$ that satisfies the following:
\begin{align}
\bar{\lambda}\min_{i} f_i(A_i^{\hat{\pi}}) + \lambda \frac{1}{m}\sum_{i=1}^m f_i(A_i^{\hat{\pi}}) \geq \lambda \alpha (OPT-\epsilon)
\label{eqn:general_greed_sat_first_bound}
\end{align}
%\begin{itemize}
%\item $\bar{\lambda}\min_{i} f_i(A_i^{\hat{\pi}}) + \lambda \frac{1}{m}\sum_{i=1}^m f_i(A_i^{\hat{\pi}}) \geq \lambda \alpha OPT$,
%\item At least $\ceil{m( \alpha-\delta)}$ blocks receive utility at least $\max\{\frac{\delta}{1-\alpha + \delta}, \lambda \alpha \} ( OPT-\epsilon)$,
%\end{itemize}
where $OPT= \max_{\pi\in \Pi} \bar{\lambda} \min_{i}f_i(A_i^{\pi}) + \lambda \frac{1}{m} \sum_{i=1}^m f_{i}(A_i^{\pi})$. 

Moreover, let $F_{\lambda,i}(\pi) = \bar{\lambda}f_i(A_i^{\pi}) +\lambda \frac{1}{m}\sum_{j=1}^m f_j(A_j^{\pi})$. 
Given any $0 < \delta < \alpha $, 
there is a set $I\subseteq \{1,\dots,m\}$ such that 
$|I| \geq \ceil{m( \alpha-\delta)}$ and 
\begin{align}
F_{i,\lambda} (\hat{\pi}) \geq \max\{\frac{\delta}{1-\alpha + \delta}, \lambda \alpha\}(OPT-\epsilon),\forall i\in I.
\label{eqn:general_greed_sat_second_bound}
\end{align}
\end{theorem}
Eqn~\ref{eqn:general_greed_sat_second_bound} in  Theorem~\ref{thm:generalGreedSat_bounds}  reduces to Theorem~\ref{thm:greedy_sat_bound} when $\lambda=0$, i.e., it recovers the bi-criterion guarantee in Theorem~\ref{thm:greedy_sat_bound} for the worst-case scenario ($\lambda=0$). Eqn~\ref{eqn:general_greed_sat_first_bound} in Theorem~\ref{thm:generalGreedSat_bounds} implies that $\alpha$-approximation for the average-case objective can almost be recovered by \textsc{GeneralGreedSat} if $\lambda=1$.
Moreover Theorem~\ref{thm:generalGreedSat_bounds} shows that the guarantee of \textsc{GeneralGreedSat} improves as $\lambda$ increases suggesting that Problem~\ref{prob:maxmin-genmix} becomes easier as the mixed objective weights more on the average-case objective. 
We also point out that the approximation guarantee of \textsc{GeneralGreedSat}
smoothly interpolates the two extreme cases in terms of $\lambda$.

 \begin{algorithm}[htb]
{
\begin{algorithmic}[1]
\caption{\textsc{General\lovasz{}\!Round}}
\label{alg:general_lovasz_ext_alg}
\STATE Input: $\{f_i\}_{i=1}^m$, $\{\tilde{f}_i\}_{i=1}^m$, $\lambda$, $m$, $V$.\\
{
\STATE Solve 
\begin{align}
&\min_{x_1,\dots,x_m \in [0,1]^n} \max_{i} \bar{\lambda} \tilde{f}_i(x_i) + \lambda \frac{1}{m}\sum_{j=1}^m \tilde{f}_j(x_j), \mbox{ s.t }  \sum_{i=1}^m x_i(j) \geq 1, \text{ for } j=1,\dots,n
\label{eqn:convex_relax_general_lovasz_round}
\end{align}
for $\{x^*_i\}_{i=1}^m$ via convex relaxation. \\
}
\STATE Rounding: Let $A_1=,\dots,=A_m=\emptyset$.\\
\label{line:start_rounding_general}
\FOR{$j=1,\dots, n$}
\STATE $\hat{i}\in \argmax_{i} x^*_i(j)$; $A_{\hat{i}} = A_{\hat{i}} \cup j$\\
\ENDFOR
\label{line:end_rounding_general}
\STATE Output $\hat{\pi} = \{A_i\}_{i=1}^m$.
\end{algorithmic}}
\end{algorithm}

\textbf{General\lovasz{}\!Round: }Next we focus on designing practically more interesting algorithms for Problem~\ref{prob:minmax-genmix} with general $\lambda$. In particular we generalize \textsc{\lovasz{}\!Round} leading to \textsc{General\lovasz{}\!Round} as shown in Alg.~\ref{alg:general_lovasz_ext_alg}.
%) that achieves an approximation factor of $m$. 
%For the homogeneous setting of Problem~\ref{prob:minmax-genmix} we also introduce an intuitive and efficient greedy heuristic -- \textsc{GeneralGreedMin}.  
Sharing the same idea with \textsc{\lovasz{}\!Round}, \textsc{General\lovasz{}\!Round} first relaxes Problem~\ref{prob:minmax-genmix} as a convex program (defined in Eqn~\ref{eqn:convex_relax_general_lovasz_round}) using the \lovasz extension of each submodular objective. 
Given the fractional solution to the convex program $\{x_i^*\}_{i=1}^m$, the algorithm then rounds it to a partition using the $\theta$-rounding technique (Line~\ref{line:start_rounding_general}-~\ref{line:end_rounding_general}). Note the rounding technique used for \textsc{General\lovasz{}\!Round} is the same as in \textsc{\lovasz{}\!Round}. 
The following Theorem holds:\looseness-1
\begin{theorem}
\label{thm:general_lovasz_round}
\textsc{General\lovasz{}\!Round} is guaranteed to find a partition $\hat{\pi}\in \Pi$ such that
\begin{align}
\max_i \bar{\lambda} f_i(A_i^{\hat{\pi}}) + \lambda \frac{1}{m} \sum_{j=1}^m f_j(A_j^{\hat{\pi}}) \leq m \min_{\pi\in \Pi} \max_i \bar{\lambda} f_i(A_i^{{\pi}}) + \lambda \frac{1}{m} \sum_{j=1}^m f_j(A_j^{{\pi}}).  
\end{align}
\end{theorem}
Theorem~\ref{thm:general_lovasz_round} generalizes Theorem~\ref{thm:lovaszRound_bound} when $\lambda=0$. Moreover we achieve a factor of $m$ by \textsc{General\lovasz Round} for any $\lambda$. Though the approximation guarantee is independent of $\lambda$ the algorithm naturally exploits the trade-off between the worst-case and average-case objectives in terms of $\lambda$. The drawback of
\textsc{General\lovasz{}\!Round} is that it requires high order polynomial queries of the \lovasz{} extension of the submodular objectives, hence is not computationally feasible for even medium sized tasks. Moreover, if we restrict ourselves to the homogeneous setting ($f_i$'s are identical), it is easy to verify that arbitrary partitioning already achieves a guarantee of $m$ while Problem~\ref{prob:minmax-genmix}, in general, cannot be approximated better than $m$ as shown in Theorem~\ref{thm:hardness_SLB}. 

\textbf{GeneralGreedMin: }
In this case, we should resort to intuitive heuristics that are scalable to large-scale applications to solve Problem~\ref{prob:minmax-genmix} with general $\lambda$. To this end we design a greedy heuristic called \textsc{GeneralGreedMin} (see Alg.~\ref{alg:greed_min_heuristic_combined}). 
\begin{algorithm}[htb]
{
\begin{algorithmic}[1]
\caption{\textsc{GeneralGreedMin}}
\label{alg:greed_min_heuristic_combined}
\STATE Input: $f$, $m$, $V$, $0 \leq \lambda \leq 1$; \\ 
\STATE Solve $S_{\text{seed}} \in \argmax_{S\subseteq V; |S| = m} f(S)$ for $m$ seeds with $S_{\text{seed}} = \{s_1,\dots, s_m\}$. \\
\STATE Initialize each block $i$ by the seeds as $A_i \leftarrow \{s_i\}, \forall i$.\\
\STATE Initialize a counter as $k=m$ and $R=V\setminus S_{\text{seed}}$.\\
\WHILE{$R\neq \emptyset$}
\IF{$k\leq (1-\lambda) |V|$}
\STATE $j^*\in \argmin_{j} f(A_j)$ \\ 
\STATE $a^* \in \min_{a\in R} f(a|A_{j^*})$\\
\label{line:greedy_step1_heuristic}
\STATE $A_{j^*}\leftarrow A_{j^*} \cup {a^*}; R\leftarrow R\setminus a^*$
\ELSE
\FOR{$i=1,\dots, m$}
\STATE $a_i^* \in \argmin_{a\in R} f(a|A_{i})$
\label{line:greedy_step2_heuristic}
\ENDFOR
\STATE $j^*\in \argmin_{i=1,\dots, m} f(a_i^* | A_{i})$;
\STATE $A_{j^*} \leftarrow A_{j^*} \cup a^*$; $R\leftarrow R\setminus a_{j^*}^*$\\
\ENDIF
\STATE $k = k+1$;\\
\ENDWHILE
\STATE Output $\{A_i\}_{i=1}^m$.
\end{algorithmic}}
\end{algorithm}

%As shown in Theorem~\ref{thm:hardness_SLB},  Problem~\ref{prob:minmax-genmix} generally cannot be approximated better than a factor of $m$, while arbitrary partitioning already achieves a guarantee of $m$. The proposed approaches or the existing algorithms for either extremal case of Problem~\ref{prob:minmax-genmix} are not scalable for this task. For instance, \textsc{\lovasz{}\!Round} solves the continuous relaxation of the discrete objective and requires high order polynomial queries of the \lovasz extension valuation. \textsc{MMin}, though much more scalable, deals with the modular version of the objective, for which the LP-based algorithm may still be computationally unfeasible~\cite{lenstra1990approximation}. Similarly, existing approaches for the average-case objective (Problem~\ref{prob:minmax-genmix} with $\lambda=1$) either requires solving instances of Queyranne's algorithm~\cite{narasimhan2005q} or a convex relaxation of \lovasz extensions~\cite{chekuri2011approximation,zhao2004generalized}. In both cases,  they are not scalable to this task. Given the computation limitation, we resort to a simple and scalable heuristic to solve Problem~\ref{prob:minmax-genmix}. The heuristic is described in Alg~\ref{alg:greed_min_heuristic_combined}. 
The algorithm first solves a constrained submodular maximization on $f$ to obtain a set $S_{\text{seed}}$ of $m$ seeds. Since $f$ is submodular, maximizing $f$ always leads to a set of diverse seeds, where the diversity is measured by the objective $f$.  
We initialize each block $A_i$ with one seed from $S_{\text{seed}}$. 
%Denote $R=V\setminus S_{\text{seed}}$ as the set of unassigned items.  
Defining $k$ as the number of items that have already been assigned. 
The main algorithm consists of two phases. In the first phase ($k \leq (1-\lambda) |V|$), we, for each iteration, assign the item that has the smallest marginal gain to the block whose valuation is the smallest. 
Since the functions are all monotone, any additions to a block can
(if anything) only increase its value. Such procedure inherently minimizes the worst-case objective, since it chooses the minimum valuation block
to add to in order to keep the maximum valuation block from
growing further.
In the second phase ($k > \lambda |V|$), we 
assign an item such that its marginal gain is the smallest among all remaining items and all blocks. The greedy procedure in this phase, on the hand, is suitable for minimizing the average-case objective, since it, in each iteration, assigns an item so that the valuation of the average-case objective increases the least. The trade-off between the worse-case and the average-case objectives is controlled by $\lambda$, which is used as the input argument to the algorithm. In particular, $\lambda$ controls the fraction of the iterations in the algorithm to optimize the average-case objective.  
When $\lambda=1$, the algorithm solely focuses on the average-case objective, while only the worst-case objective is minimized if $\lambda=0$. 

In general \textsc{GeneralGreedMin} requires $O(m|V|^2)$ function valuations, which may still be computationally difficult for large-scale applications. In practice, one can relax the condition in Line~\ref{line:greedy_step1_heuristic} and~\ref{line:greedy_step2_heuristic}. Instead of searching among all items in $R$, one can, in each round, randomly select a subset $\hat{R}\subseteq R$  and choose an item with the smallest marginal gain from only the subset $\hat{R}$. The resultant computational complexity is reduced to $O(m |\hat{R}| |V|)$ function valuations. Empirically we observe that \textsc{GeneralGreedMin} can be sped up more than 100 times by this trick without much performance loss.

}
{

\section{Experiments}
\label{sec:experiments}

\arxivalt{In this section we empirically evaluate the algorithms proposed for \theproblems{}. We first compare the performance of the various algorithms discussed in this paper on a synthetic data set. We then evaluate some of the scalable algorithms proposed for \theproblems{} on large-scale real-world data partitioning applications including distributed ADMM, distributed neural network training, and lastly unsupervised image segmentation tasks.
}
{We conclude in this section by empirically evaluating the algorithms
proposed for \theproblems{} on real-world data partitioning
applications including distributed ADMM, distributed deep neural network training, and lastly unsupervised image
segmentation tasks. \looseness-1}

\arxiv{
\subsection{Experiments on Synthetic Data}
In this section we evaluate separately on four different cases: Problem~\ref{prob:maxmin-genmix} with $\lambda=0$ (SFA), Problem~\ref{prob:minmax-genmix} with $\lambda=0$ (SLB), Problem~\ref{prob:maxmin-genmix} with $0<\lambda<1$, and Problem~\ref{prob:minmax-genmix} with $0<\lambda<1$. 
Since some of the algorithms, such as the Ellipsoidal Approximations~\cite{goemans2009approximating} and \lovasz relaxation algorithms, are computationally intensive, we restrict ourselves to only $40$ data instances, i.e., $|V|=40$. For simplicity we only evaluate on the homogeneous setting ($f_i$'s are identical). For each case we test with two types of submodular functions: facility location function, and the set cover function. 
The facility location function is defined as follows:
\begin{align}
f_{\text{fac}}(A) = \sum_{v\in V} \max_{a\in A} s_{v,a},
\end{align}
where $s_{v,a}$ is the similarity between item $v$ and $a$ and is symmetric, i.e., $s_{v,a}=s_{a,v}$ for any pair of $v$ and $a$. We define $f_{\text{fac}}$ on a complete similarity graph with each edge weight $s_{v,a}$ sampled uniformly and independently from $[0,1]$. The set cover function $f_{\text{sc}}$
is defined by a 
bipartite graph between $V$ and $U$ ($|U|=40$), where we define an edge between an item $v\in V$ and a key $u\in U$ independently with probability $p=0.2$. 

\begin{figure}[htb]
\centering
\begin{subfigure}[]{0.46\linewidth}
\caption{Problem~\ref{prob:maxmin-genmix} on $f_{\text{fac}}$ with $\lambda=0$}
\label{fig:prob1_lambda0_fac}
\vspace{-0.25cm}
\includegraphics[width=0.98\textwidth]{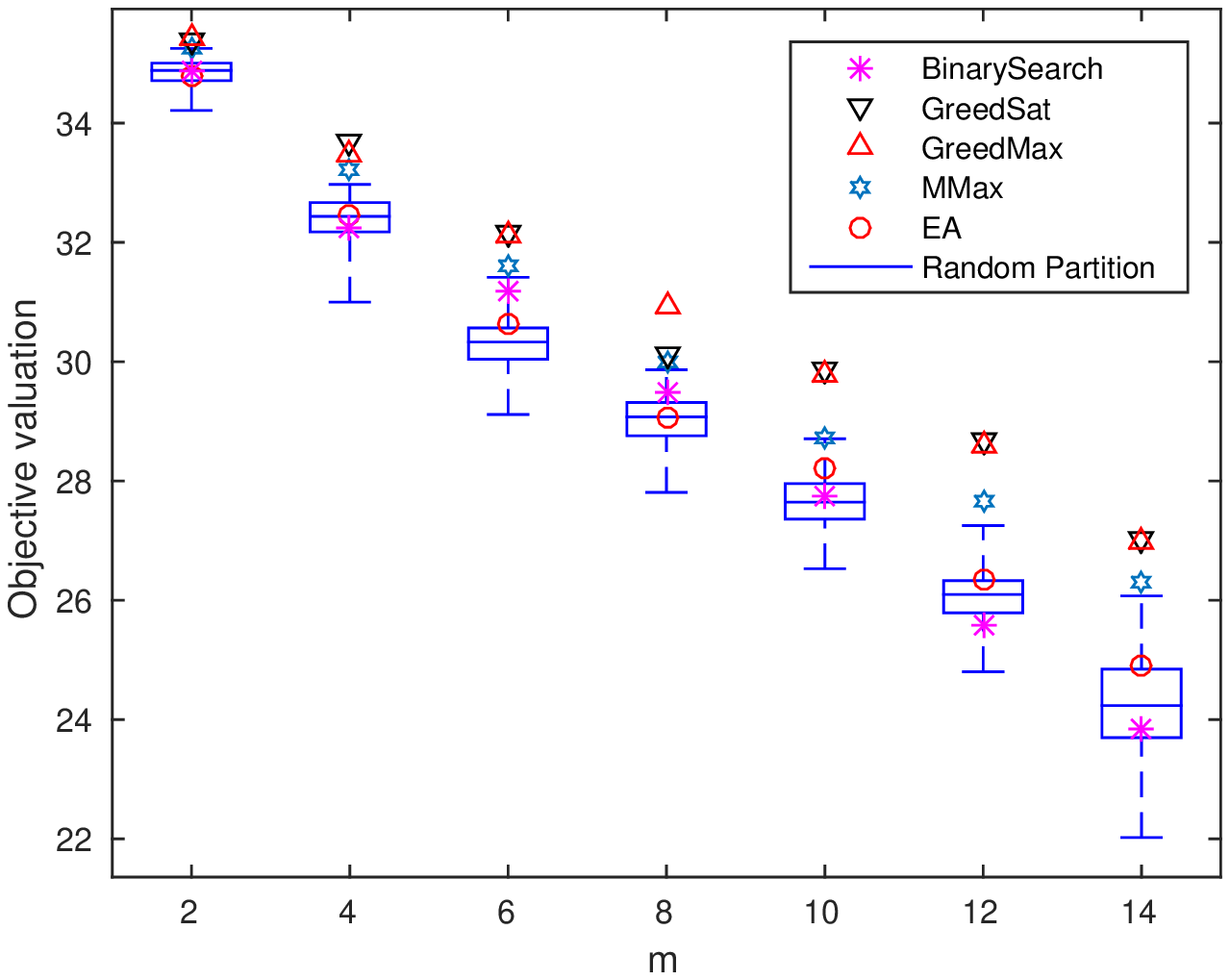}
\end{subfigure}
\begin{subfigure}[]{0.46\linewidth}
\caption{Problem~\ref{prob:maxmin-genmix} on $f_{\text{sc}}$ with $\lambda=0$}
\label{fig:prob1_lambda0_sc}
\vspace{-0.25cm}
\includegraphics[width=0.98\textwidth]{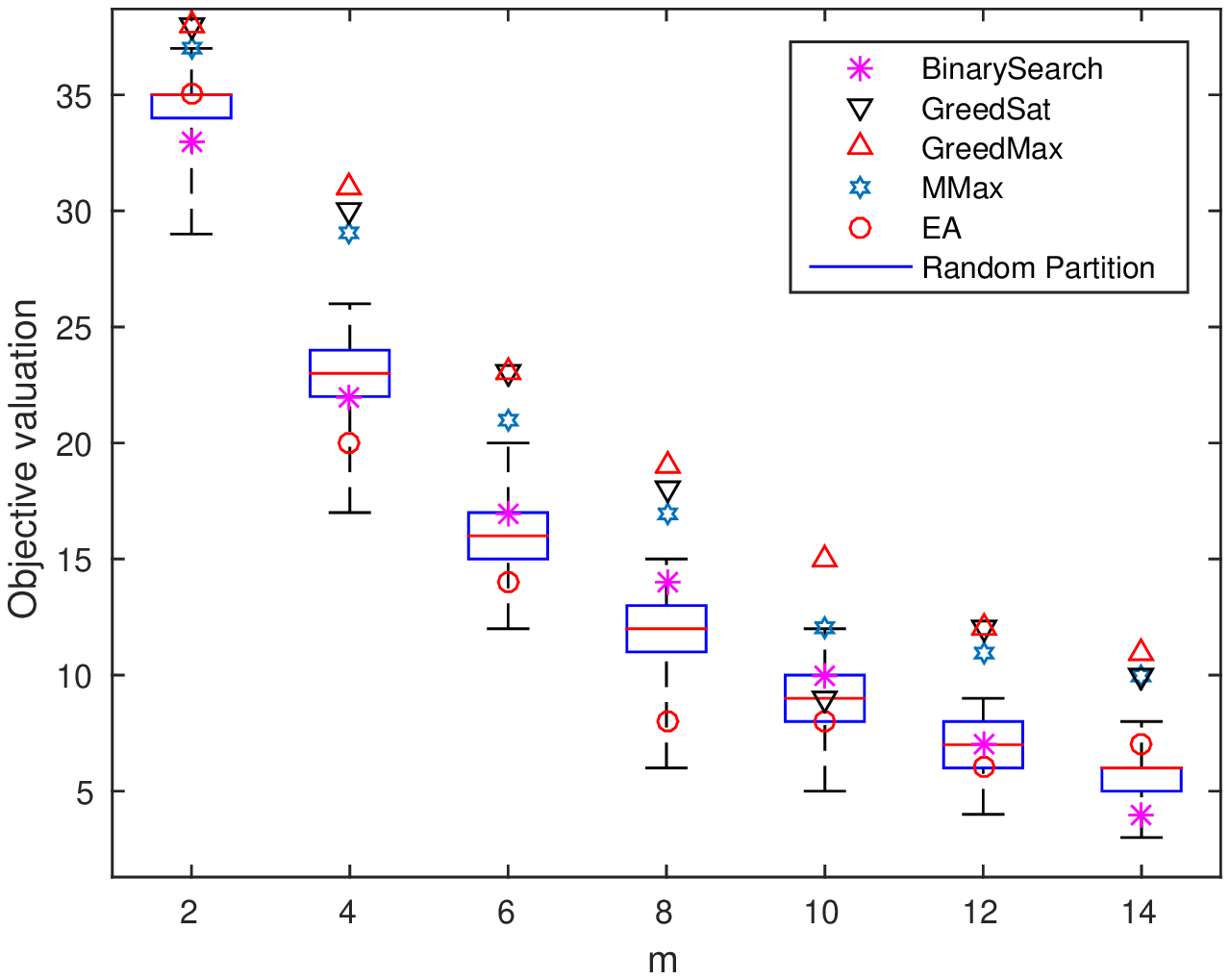}
\end{subfigure}
\begin{subfigure}[]{0.46\linewidth}
\caption{Problem~\ref{prob:maxmin-genmix} on $f_{\text{fac}}$ with varying $\lambda$}
\label{fig:prob1_general_lambda_fac}
\vspace{-0.25cm}
\includegraphics[width=0.98\textwidth]{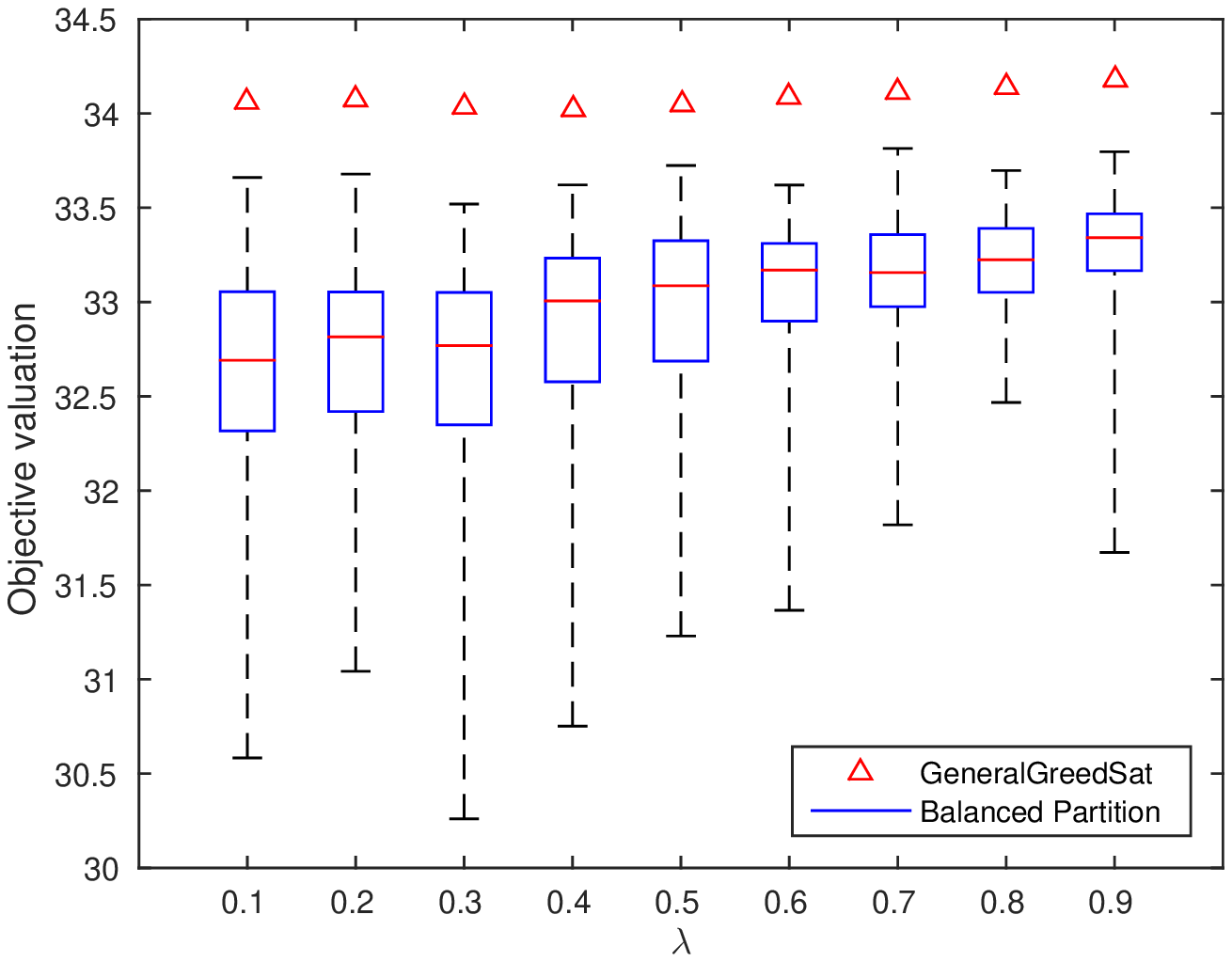}
\end{subfigure}
\begin{subfigure}[]{0.46\linewidth}
\caption{Problem~\ref{prob:maxmin-genmix} on $f_{\text{sc}}$ with varying $\lambda$}
\label{fig:prob1_general_lambda_sc}
\vspace{-0.25cm}
\includegraphics[width=0.98\textwidth]{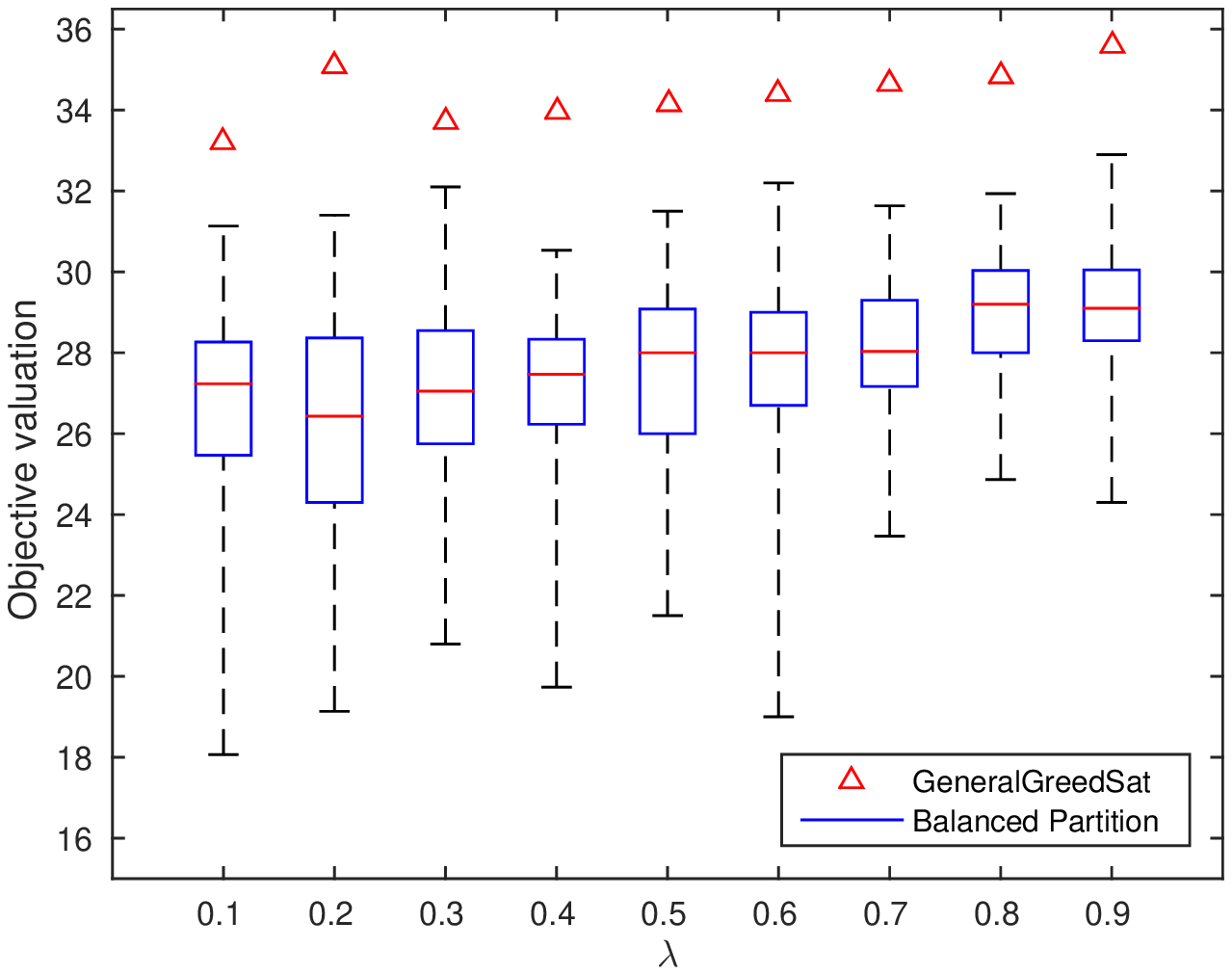}
\end{subfigure}
\caption{Synthetic experiments on Problem~\ref{prob:maxmin-genmix} with $\lambda=0$ on facility location function (a) and set cover function (b). Problem~\ref{prob:maxmin-genmix} with general $0<\lambda<1$ on facility location function (c) and set cover function (d).}
\end{figure}
\paragraph{Problem~\ref{prob:maxmin-genmix}}

For $\lambda=0$, i.e., SFA, we compare among 6 algorithms: \textsc{GreedMax}, \textsc{GreedSat}, \textsc{MMax}, Balanced Partition (\textsc{BP}), Ellipsoid Approximation (\textsc{EA})~\cite{goemans2009approximating}, and Binary Search algorithm (\textsc{BS})~\cite{khot2007approximation}. Balanced Partition method simply partitions the ground set $V$ into $m$ blocks such that the size of each block is balanced and is either $\ceil{\frac{|V|}{m}}$ or $\lfloor \frac{|V|}{m} \rfloor$. We run 100 randomly generated instances of the balanced partition method. 
\textsc{GreedSat} is implemented with the choice of the hyperparameter $\alpha=1$. We compare the performance of these algorithms in Figure~\ref{fig:prob1_lambda0_fac} and~\ref{fig:prob1_lambda0_sc}, where we vary the number of blocks $m$ from 2 to 14. 
The three proposed algorithms (\textsc{GreedMax}, \textsc{GreedSat}, and \textsc{MMax}) significantly and consistently outperform all baseline methods for both $f_{\text{fac}}$ and $f_{\text{sc}}$. Among the proposed algorithms we observe that \textsc{GreedMax}, in general, yields the superior performance. Given the empirical success, computational efficiency, and tight theoretical guarantee, we suggest \textsc{GreedMax} as the first choice of algorithm to solve SFA under the homogeneous setting.

Next we evaluate Problem~\ref{prob:maxmin-genmix} with general $0<\lambda<1$.  Baseline algorithms for SFA
such as Ellipsoidal Approximations, Binary Search do not apply to the mixed scenario. Similarly the proposed algorithms such as \textsc{GreedMax}, \textsc{MMax} do not simply generalize to this scenario. We therefore only compare \textsc{GeneralGreedSat} with the Balanced Partition as a baseline. The results are summarized in Figure~\ref{fig:prob1_general_lambda_fac} and~\ref{fig:prob1_general_lambda_sc}. We observe that \textsc{GeneralGreedSat} consistently and significantly outperform even the best of 100 instances of the baseline method for all cases of $\lambda$.

\begin{figure}[htb]
\centering
\begin{subfigure}[]{0.46\linewidth}
\caption{Problem~\ref{prob:minmax-genmix} on $f_{\text{fac}}$ with $\lambda=0$}
\label{fig:prob2_lambda0_fac}
\vspace{-0.25cm}
\includegraphics[width=0.98\textwidth]{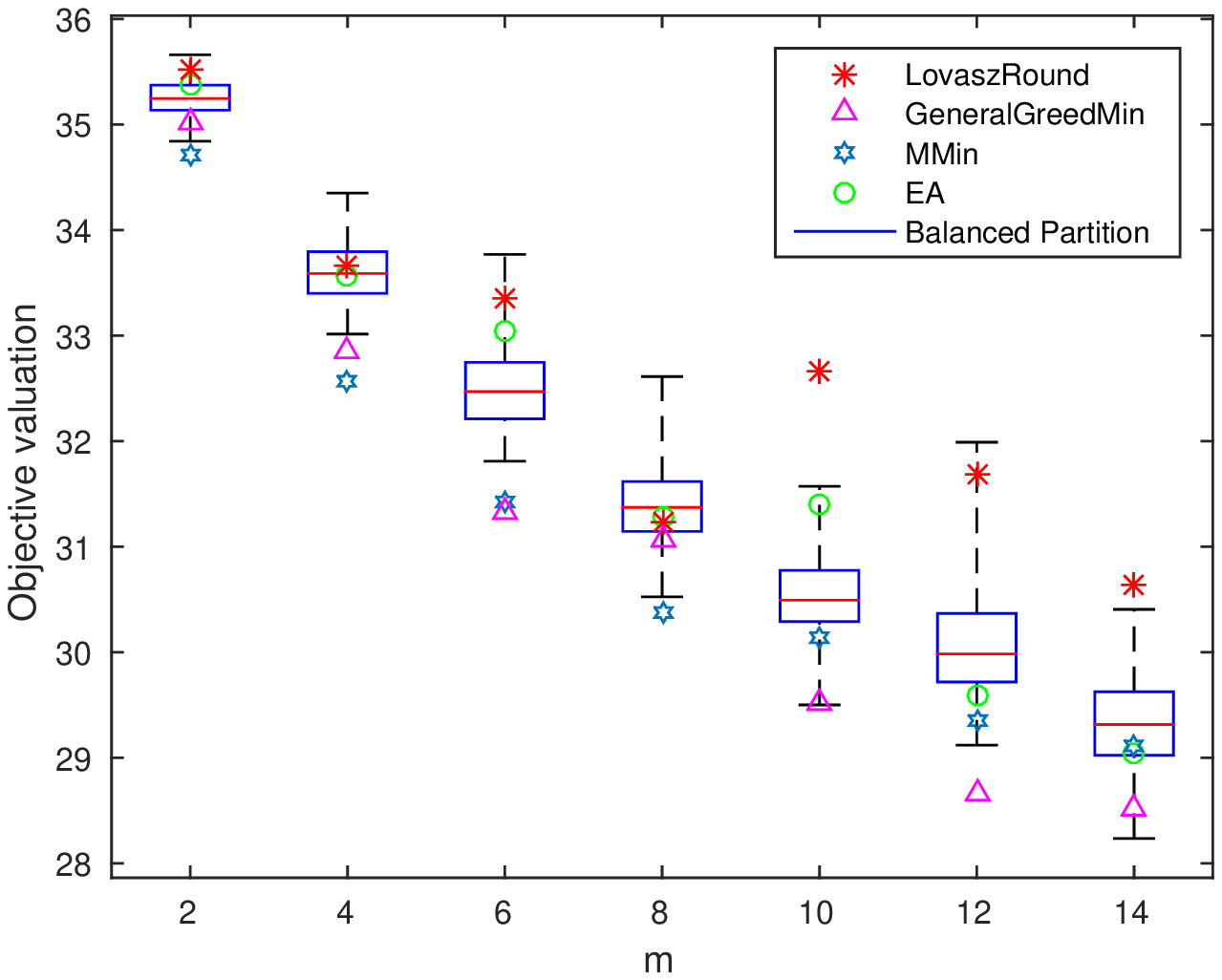}
\end{subfigure}
\begin{subfigure}[]{0.46\linewidth}
\caption{Problem~\ref{prob:minmax-genmix} on $f_{\text{sc}}$ with $\lambda=0$}
\label{fig:prob2_lambda0_sc}
\vspace{-0.25cm}
\includegraphics[width=0.98\textwidth]{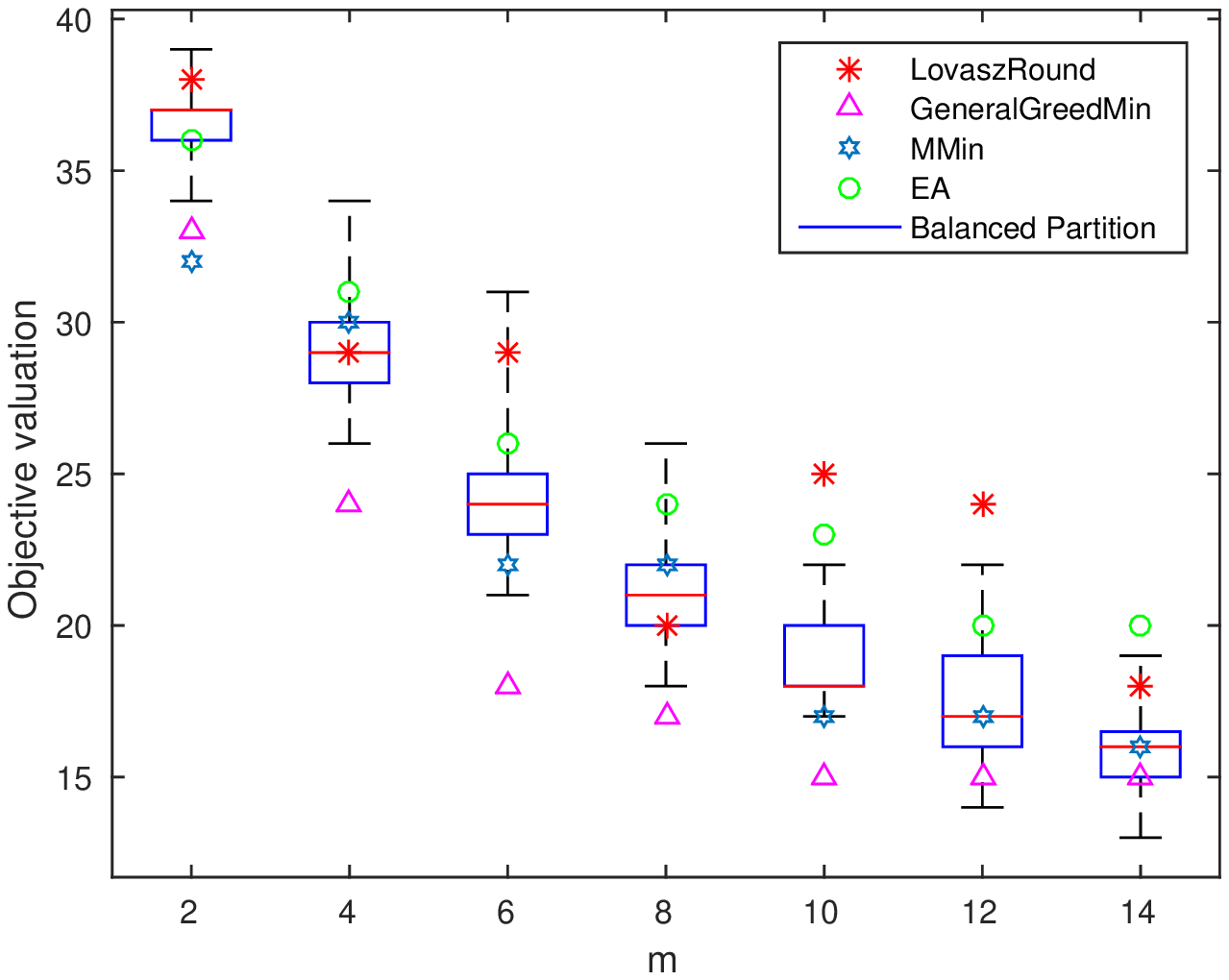}
\end{subfigure}
\begin{subfigure}[]{0.46\linewidth}
\caption{Problem~\ref{prob:minmax-genmix} on $f_{\text{fac}}$ with varying $\lambda$}
\label{fig:prob2_general_lambda_fac}
\vspace{-0.25cm}
\includegraphics[width=0.98\textwidth]{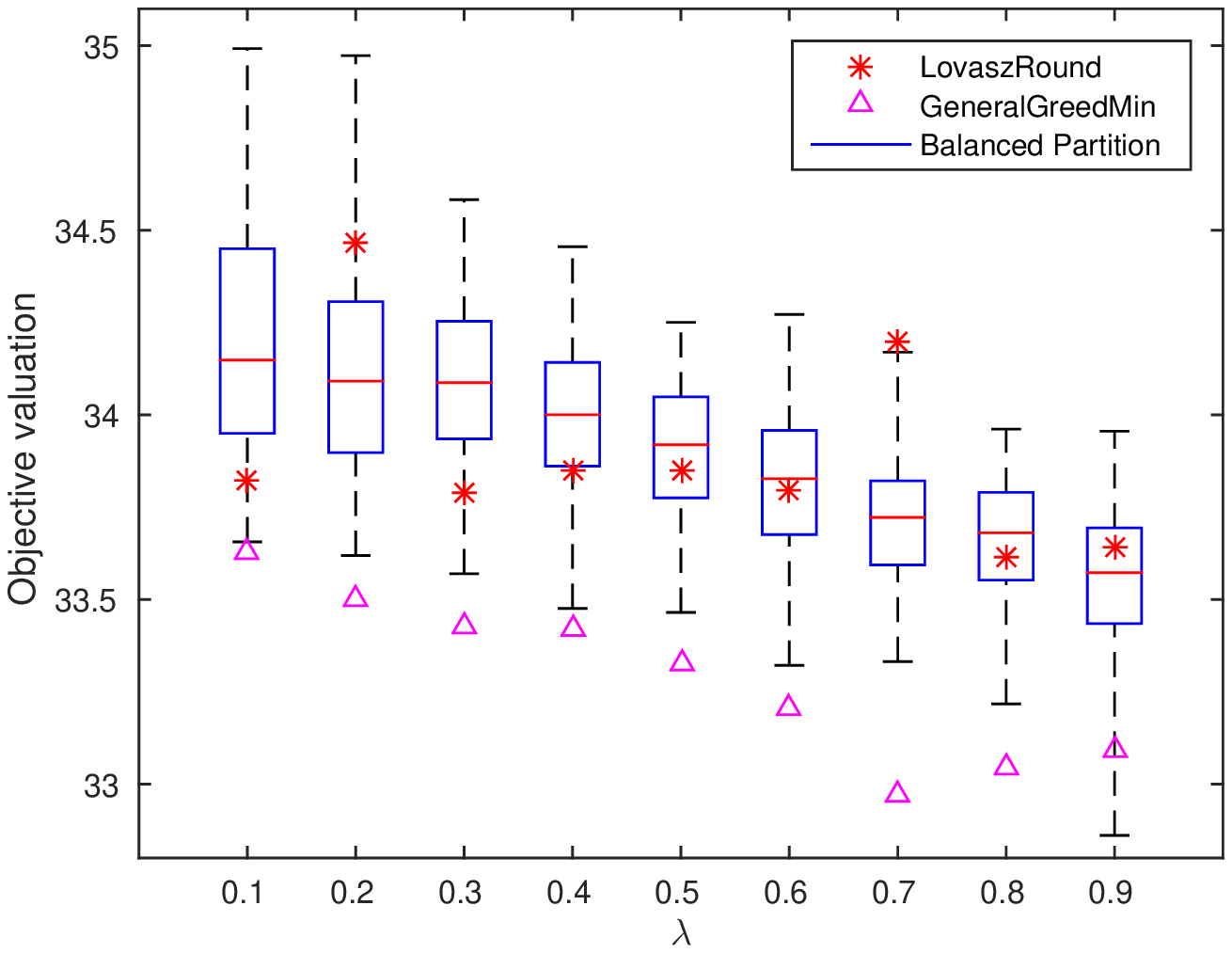}
\end{subfigure}
\begin{subfigure}[]{0.46\linewidth}
\caption{Problem~\ref{prob:minmax-genmix} on $f_{\text{sc}}$ with varying $\lambda$}
\label{fig:prob2_general_lambda_sc}
\vspace{-0.25cm}
\includegraphics[width=0.98\textwidth]{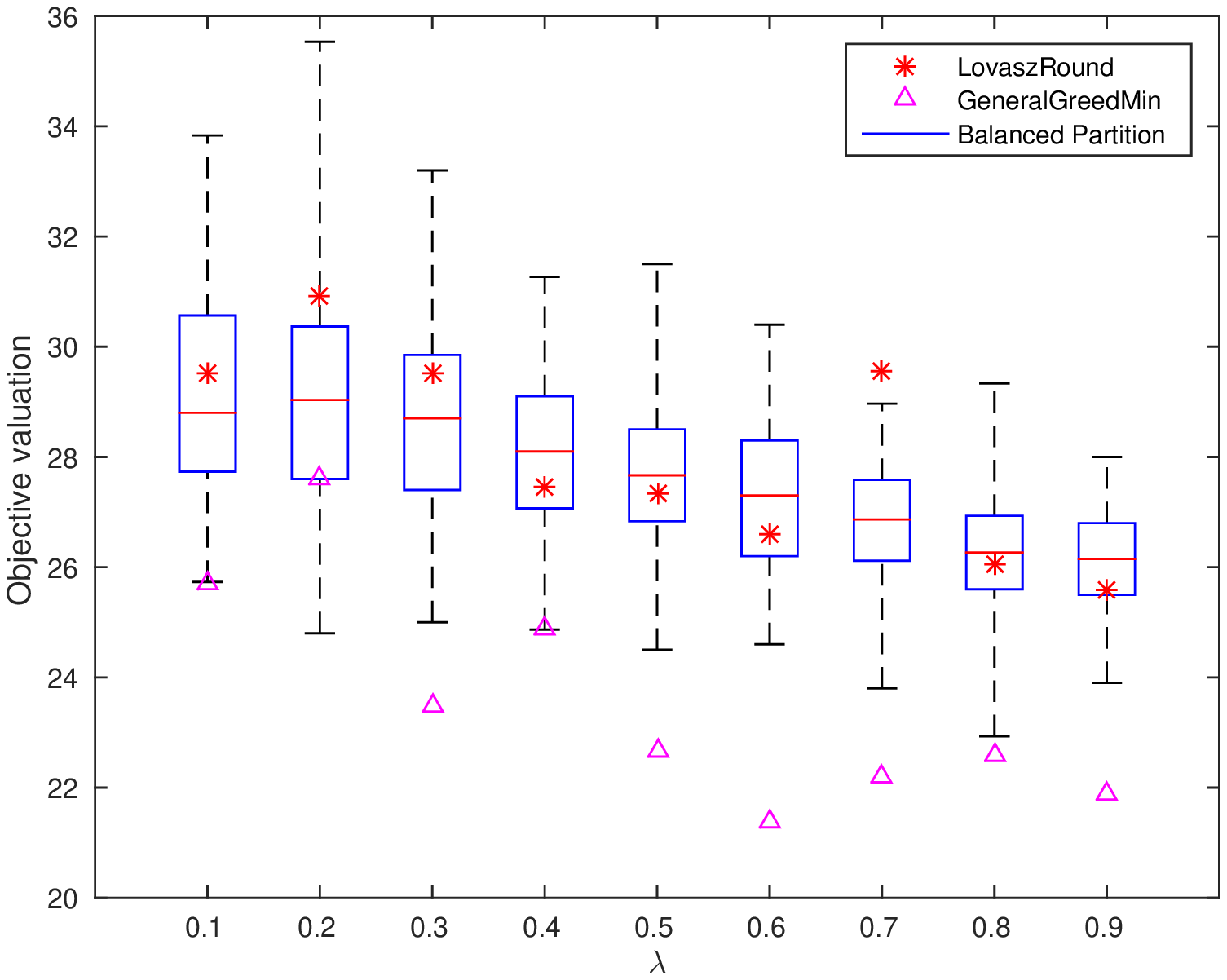}
\end{subfigure}
\caption{Synthetic experiments on Problem~\ref{prob:minmax-genmix} with $\lambda=0$ on facility location function (a) and set cover function (b). Problem~\ref{prob:minmax-genmix} with general $0<\lambda<1$ on facility location function (c) and set cover function (d).}
\end{figure}

\paragraph{Problem~\ref{prob:minmax-genmix}} 
For $\lambda=0$, i.e., SLB, we compare among 5 algorithms: \textsc{\lovasz{}\!Round}, \textsc{MMin}, \textsc{GeneralGreedMin}, Ellipsoid Approximation (\textsc{EA})~\cite{goemans2009approximating}, and Balanced Partition~\cite{svitkina2011submodular}. We implement \textsc{GeneralGreedMin} with the input argument $\lambda=0$. 
%The idea of \textsc{EA} is to approximate each submodular function with its ellipsoidal approximation and reduces the problem to its modular version. 
%Balanced partition method, on the other hand, arbitrarily partitions the ground set $V$ into $m$ blocks such that the size of each block is balanced and is at most $\ceil{\frac{|V|}{m}}$.
We also run $100$ randomly generated instances of the Balanced Partition method as a baseline. 
We show the results in Figure~\ref{fig:prob2_lambda0_fac} and~\ref{fig:prob2_lambda0_sc}. Among all five algorithms \textsc{MMin} and \textsc{GeneralGreedMin}, in general, perform the best. Between \textsc{MMin} and \textsc{GeneralGreedMin} we observe that \textsc{GeneralGreedMin} performs marginally better, especially on $f_{\text{sc}}$.
%For several cases of $m$, \textsc{MMin} and \textsc{GeneralGreedMin} outperform the best of $100$ instances of balanced partition, suggesting the efficacy of these two methods. 
The computationally intensive algorithms, such as Ellipsoid Approximation and \textsc{\lovasz{}\!Round}, do not perform well, though they carry better worst-case approximation factors for the heterogeneous setting. 

Lastly we evaluate Problem~\ref{prob:minmax-genmix} with general $0<\lambda<1$. Since \textsc{MMin} and Ellipsoid Approximation do not apply for the mixed scenario, we test only on \textsc{General\lovasz{}\!Round}, \textsc{GeneralGreedMin}, and Balanced Partition. Again we test on 100 instances of randomly generated balanced partitions. 
We vary $\lambda$ in this experiment.
The results are shown in Figure~\ref{fig:prob2_general_lambda_fac} and~\ref{fig:prob2_general_lambda_sc}. The best performance is consistently achieved by \textsc{GeneralGreedMin}.

}

%\kai{Kai}{Need to redo the plots for the Journal version.}
%\kai{Kai}{Add another set of experiments on the synthetic and small sized data set. In this set of experiments, we only show the results in terms of the objective valuations.}
%\kai{Kai}{We also need to restructure the experimental section, and add a separate section for conclusions.}
\arxiv{
\subsection{Problem~\ref{prob:maxmin-genmix} for Distributed Training}}
\notarxiv{\vspace{-.3\baselineskip}}
\arxivalt{
In this section we focus on applications of Problem~\ref{prob:maxmin-genmix} to real-wold machine learning problems. In particular we examine how a partition obtained by solving Problem~\ref{prob:maxmin-genmix} with certain instances of submodular functions perform for distributed training of various statistical models.

\paragraph{Distributed Convex Optimization: }
We first consider data partitioning for distributed convex optimization. 
We evaluate the distributed convex optimization on a text categorization task. We use 20 Newsgroup data set\arxiv{~\footnote{Data set is obtained at http://qwone.com/$\sim$jason/20Newsgroups/}}, which consists of 18,774 articles divided almost evenly across 20 classes. The text categorization task is to classify an article into one newsgroup (of twenty) to which it was posted. 
We randomly split 2/3 and 1/3 of the whole data as the training and test data. The task is solved as a multi-class classification problem, which we formulate as an $\ell_2$ regularized logistic regression. We solve this convex optimization problem in a distributive fashion, where the data samples are partitioned and distributed across multiple machines. In particular we implement an ADMM 
algorithm as described in~\cite{boyd2011distributed} to solve the distributed convex optimization problem. Given a partition $\pi$ of the training data into $m$ separate clients, in each iteration, ADMM first assigns each client $i$ to solve an $\ell_2$ regularized logistic regression on its block of data $A_i^{\pi}$ using L-BFGS, aggregate the solutions from all $m$ clients according to the ADMM update rules, and then sends the aggregated solution back to each client. This iterative procedure is carried out so that solutions on all clients converge to a consensus, which is the global solution of the overall large-scale convex optimization problem.
%\shengjie{Kai}{Add more details on how ADMM is implemented.}

We formulate the data partitioning problem as an instance of SFA (Problem~\ref{prob:maxmin-genmix} with $\lambda=0$) under the homogeneous setting. In the experiment, we solve the data partitioning using \textsc{GreedMax}, since it is efficient and attains the tightest guarantee among all algorithms proposed for this setting.
Note, however, \textsc{GreedSat} and \textsc{MMax} may also be used in the experiment. 
We model the utility of a data subset using the feature-based submodular function~\cite{wei2014-supervised-large-icassp,wei2015-submodular-data-active,tschiatschek2014learning}, which has the form:
\begin{align}
f_{\text{fea}}(A) = \sum_{u\in \mathcal{U}}m_u(V)\log m_u(A),
\end{align}
where $\mathcal{U}$ is the set of ``features'', $m_u(A)=\sum_{a\in A}m_u(a)$ with $m_u(a)$ measuring the degree that the article $a$ possesses the feature $u\in \mathcal{U}$. 
In the experiments, we define $\mathcal{U}$ as the set of all words occurred in the entire data set and $m_u(a)$ as the number of occurrences of the word $u\in \mathcal{U}$ in the article $a$. $f_{\text{fea}}$ is in the form of a sum of concave over modular functions, hence is monotone submodular~\cite{stobbe10efficient}. 
The class of feature-based submodular function has been widely applied to model the utility of a data subset on a number of tasks, including speech data subset selection~\cite{wei2014-supervised-large-icassp, wei2014-unsupervised-icassp}, and image summarization~\cite{tschiatschek2014learning}. 
Moreover $f_{\text{fea}}$ has been shown in~\cite{wei2015-submodular-data-active} to model the log-likelihood of a data subset for a \Naive Bayes classifier.
%~\cite{wei2015-submodular-data-active} also empirically observes that $f_{\text{fea}}$ models well for training a logistic regression classifier. 

We compare the submodular partitioning with the random partitioning for $m=5$ and $m=10$. 
We test with $10$ instances of random partitioning. 
We first examine how balanced the sizes of the blocks yielded by submodular partitioning are. This is important since if the block sizes vary a lot in a partition, the 
computational loads across the blocks are imbalanced, and the actual efficiency of the parallel system is significantly reduced. Fortunately we observe that submodular partitioning yields very balanced partition. 
%Let $p_i(\pi) = \frac{|A_i^{\pi}|}{|V|}$ be the fraction of items assigned to block $i$ of the partition. We measure how balanced a partition $\pi$ is as the entropy of the distribution $p(\pi)$. \shengjie{Kai}{We may also need to add the entropy of the block sizes in here, and compare it with random. Hopefully, they are very comparable.} 
The sizes of all blocks in the resultant partitioning for $m=5$ range between $2,225$ and $2,281$. In the case of $m=10$ the maximum block is of size $1,140$, while the smallest block has $1,109$ items.

%$\min_i |A_i| = 2225$, $\max_i |A_i| = 2281$, and $|V| / m = 2253.8$; for $m=10$, $\min_i |A_i| = 1109$, $\max_i |A_i| = 1140$, and $|V| / m = 1126.9$
\begin{figure}[htb]
\minipage{0.48\textwidth}
\includegraphics[width=\linewidth, trim=4.1cm 8.5cm 4.1cm 8.7cm, clip=true]{20news_p5.pdf}
\endminipage
\minipage{0.48\textwidth}
  \includegraphics[width=\linewidth, trim=4.1cm 8.5cm 4.1cm 8.7cm, clip=true]{20news_p10_contrived.pdf}
%  \captionsetup{labelformat=empty}
   \endminipage
 \caption{Distributed convex optimization tested on 20Newsgroups.}
 \label{fig:ADMM_20News_res}
\end{figure}

The comparison between submodular partitioning and random partitioning  in terms of the accuracy of the model attained at any iteration is shown in Fig~\ref{fig:ADMM_20News_res}. 
%\shengjie{Kai}{We may also need to plot the accuracy of the model obtained from each block.}
For $m=10$ we also run an instance on an adversarial partitioning, where each block is formed by grouping every two of the 20 classes in the training data. 
We observe submodular partitioning converges faster than the random partitioning, both of which perform significantly better than the adversarial partition. 
In particular significant and consistent improvement over the best of 10 random instances is achieved by the submodular partition across all iterations when $m=5$.

%\everyone{Jeff}{9/11: how balanced are the blocks of the partitions? I.e., do some blocks have many more or less? Should also plot the entropy of the (normalized) block size to measure how balanced they are. The reason this is important is that if the partitions are balanced, in terms of sizse, we'll still get good parallel speedup. If they are very imbalanced (even if they are ``informationally'' balanced via $f$), then the actuall running time will be reduced.}
%\jeff{Kai}{The block sizes by the submodular partitioning seem to be well balanced. We will add the entropy of the block sizes into the paper. }

}
{\textbf{ADMM:}
We first consider data partitioning for distributed convex optimization. 
The evaluation task is text categorization on the 20 Newsgroup data set\arxiv{~\footnote{Data set is obtained at http://qwone.com/$\sim$jason/20Newsgroups/}}, which consists of 18,774 articles divided almost evenly across 20 classes. \arxiv{The goal is to classify an article into one newsgroup (of twenty) to which it was posted. We randomly split 2/3 and 1/3 of the whole data as the training and test data.} We formulate the multi-class classification as an $\ell_2$ regularized logistic regression, which is solved by ADMM implemented as~\cite{boyd2011distributed}. \shengjie{Kai}{Add more details on how ADMM is implemented.} 
We run 10 instances of random partitioning on the training data as a baseline. \notarxiv{In this case, we use the feature based function (same as the one used in~\cite{wei2015-submodular-data-active}), in the homogeneous setting of Problem~\ref{prob:maxmin-genmix} (with $\lambda = 0$). We use \textsc{GreedMax} as the partitioning algorithm. In Figure~\ref{fig:ADMM_20News_res}, we observe that the resulting partitioning performs much better than a random partitioning (and significantly better than an adversarial partitioning, formed by grouping similar items together).
More details are given in~\cite{datapartitionextend}.
}
\looseness-1
}

\notarxiv{\vspace{-.3\baselineskip}}
\arxivalt{
\textbf{Distributed Deep Neural Network Training: }
Next we evaluate our framework on distributed neural network training.
We test on two tasks: 1) handwritten digit recognition on the MNIST database\arxiv{~\footnote{Data set is obtained at yann.lecun.com/exdb/mnist}}; 2) phone classification on the TIMIT data.  

The data for the handwritten digit recognition task consists of 60,000 training and 10,000 test samples. Each data sample is an image of handwritten digit. The training and test data are almost evenly divided into 10 different classes. 
For the phone classification task, the data consists of 1,124,823 training and 112,487 test samples. Each sample is a frame of speech. The training data is divided into 50 classes, each of which corresponds to a phoneme. The goal of this task to classify each speech sample into one of the 50 phone classes.

A $4$-layer DNN model is applied to the MNIST experiments, and we train a $5$-layered DNN for the TIMIT experiments.
We apply the same distributed training procedure for both tasks. 
Given a partitioning of the training data, we distributively solve $m$ instances of sub-problems in each iteration. We define each sub-problem on a separate block of the data. We employ the stochastic gradient descent as the solver on each instance of the sub-problem. In the first iteration we use a randomly generated model as the initial model shared among the $m$ sub-problems. Each sub-problem is solved with $10$ epochs of the stochastic gradient decent training. We then average the weights in the $m$ resultant models to obtain a consensus model, which is used as the initial model for each sub-problem in the successive iteration. Note that this distributed training scheme is similar to the ones presented in \cite{povey2014parallel}. \shengjie{Kai}{Please double check if the procedure described here is the same as what you implemented. Also please include more details about the implementation if needed.}

The submodular partitioning for both tasks
is obtained by solving the homogeneous case of Problem~\ref{prob:maxmin-genmix} ($\lambda=0$) using \textsc{GreedMax} on a form of clustered facility location, as proposed and used in~\cite{wei2015-submodular-data-active}. The function is defined as follows:
\begin{align}
f_{\text{c-fac}}(A) = \sum_{y\in \mathcal{Y}}\sum_{v\in V^y}\max_{a\in A\cap V^y} s_{v,a}
\end{align}
where $s_{v,a}$ is the similarity measure between sample $v$ and $a$, $\mathcal{Y}$ is the set of class labels, and $V^y$ is the set of samples in $V$ with label $y\in \mathcal{Y}$. Note $\{V^y\}_{y\in \mathcal{Y}}$ forms a disjoint partitioning of the ground set $V$. In both the MNIST and TIMIT experiments we compute the similarity $s_{v,a}$ as the RBF kernel between the feature representation of $v$ and $a$. ~\cite{wei2015-submodular-data-active} show that $f_{\text{c-fac}}$ models the log-likelihood of a data subset for a Nearest Neighbor classifier. They also empirically demonstrate the efficacy of $f_{\text{c-fac}}$ in the case of neural network based classifiers. 

\begin{figure}[htb]
\centering
%\caption*{\textbf{Distributed neural network training tested on MNIST and TIMIT}}
\minipage{0.46\textwidth}
  \includegraphics[width=\linewidth, trim=3.8cm 8.5cm 4.1cm 8.7cm, clip=true]{mnist_p5.pdf}
%\shengjie{Jeff}{fix this file/missing figure}
    \includegraphics[width=\linewidth, trim=3.8cm 8.5cm 4.1cm 8.7cm, clip=true]{mnist_p10.pdf}
  \caption{MNIST}
  \label{fig:DNN_MNIST_res}
\endminipage
\minipage{0.46\textwidth}
  \includegraphics[width=\linewidth, trim=3.8cm 8.5cm 4.1cm 8.7cm, clip=true]{timit_p30.pdf}
    \includegraphics[width=\linewidth, trim=3.8cm 8.5cm 4.1cm 8.7cm, clip=true]{timit_p40.pdf}
  \caption{TIMIT}
  \label{fig:DNN_timit_res}
\endminipage
\end{figure}

\shengjie{Kai}{Again, we need to add more plots for these set of experiments: (1) the individual performance instead of using the distributed procedure. Just to show each block tends to perform well. (2) show the block sizes in the resultant partitioning. }
Similar to the ADMM experiment we also observe that submodular partitioning yields very balanced partitions in all cases of this experiment. In the cases of $m=5$ and $m=10$ for the MNIST data set the sizes of the blocks in the resultant submodular partitioning are within the range $[11991, 12012]$ and $[5981, 6019]$, respectively. For the TIMIT data set, the block sizes range between $37,483$ and $37510$ in the case of $m=30$, and the range of $[28121, 28122]$ is observed for $m=40$.

%For TIMIT and $m=30$, we have $\min_i |A_i| = 37483$, $\max_i |A_i| = 37510$, and $|V| / m = 37498.1$; for $m=40$, we have $\min_i |A_i| = 28121$, $\max_i |A_i| = 28122$, and $|V| / m = 28121.6$. For MNIST and $m=5$, we have $\min_i |A_i| = 11991$, $\max_i |A_i| = 12012$, and $|V| / m = 12000$; for $m=10$, we have $\min_i |A_i| = 5981$, $\max_i |A_i| = 6019$, and $|V| / m = 6000$.
We also run 10 instances of random partitioning as a baseline.
The comparison between submodular partitioning and random partitioning  in terms of the accuracy of the model attained at any iteration is shown in Fig~\ref{fig:DNN_MNIST_res} and~\ref{fig:DNN_timit_res}. 
% As shown in Figure~\ref{fig:DNN_MNIST_res} and~\ref{fig:DNN_timit_res}, the submodular partitioning outperforms the random baseline. 
 The adversarial partitioning, which is formed by grouping items with the same class, cannot even be trained in both cases. Consistent and significant improvement is again achieved by submodular partitioning for all cases. 
}
{\textbf{Distributed Deep Neural Network (DNN) Training: }
Next we evaluate our framework on distributed deep neural network (DNN) training.
We test on two tasks: 1) handwritten digit recognition on the MNIST database\arxiv{~\footnote{Data set is obtained at yann.lecun.com/exdb/mnist}}, which consists of 60,000 training and 10,000 test samples; 2) phone classification on the TIMIT data, which has 1,124,823 training and 112,487 test samples.  
A $4$-layer DNN model is applied to the MNIST experiment, and we train a $5$-layer DNN for TIMIT.
%We use, for MNIST experiment, a DNN model, which consists of two convolution layers followed by two fully connected layers. 
\arxivalt{Given a partition of the training data, in each iteration, the distributed learning scheme first assigns each client to perform stochastic gradient descent on its block of data, averages the resultant models, and then sends it back. The stochastic gradient descent is performed with $10$ epochs of the training data in each iteration. Note that this distributed training scheme is similar to the ones presented in \cite{povey2014parallel}. For both experiments the submodular partitioning 
is obtained by solving the homogeneous case of Problem~\ref{prob:maxmin-genmix} ($\lambda=0$) using \textsc{GreedMax} on a form of clustered facility location (as proposed and used in~\cite{wei2015-submodular-data-active})}{For both experiments the submodular partitioning 
is obtained by solving the homogeneous case of Problem~\ref{prob:maxmin-genmix} ($\lambda=0$) using \textsc{GreedMax} on a form of clustered facility location (as proposed and used in~\cite{wei2015-submodular-data-active}).
We perform distributed training using an averaging stochastic gradient descent scheme, similar to the one in~\cite{povey2014parallel}}.  \arxivalt{Shown to model the utility of a data subset for training Nearest Neighbor classifiers~\cite{wei2015-submodular-data-active}, the clustered facility location function is submodular and has the form: $f_{\text{c-fac}}(A) = \sum_{y\in \mathcal{Y}}\sum_{v\in V^y}\max_{a\in A\cap V^y} s_{v,a}$, where $s_{v,a}$ is the similarity measure between sample $v$ and $a$, $\mathcal{Y}$ is the set of class labels, and $V^y$ is the set of samples in $V$ with label $y\in \mathcal{Y}$.}{}
%~\cite{wei2015-submodular-data-active} theoretically show that $f_{\text{c-fac}}$ models the utility of data subset for training Nearest Neighbor classifiers, and also empirically demonstrate the efficacy of $f_{\text{c-fac}}$ in the case of DNN based classifiers. 
We also run 10 instances of random partitioning as a baseline. As shown in Figure~\ref{fig:DNN_MNIST_res} and~\ref{fig:DNN_timit_res}, the submodular partitioning outperforms the random baseline. An adversarial partitioning, which is formed by grouping items with the same class, in either case, cannot even be trained.
\looseness-1
} 

\arxivalt{
}
{
}

\arxivalt{
}
{
\begin{wrapfigure}{r}{0.70\textwidth} 
\vspace{-2.5ex}
    \centerline{\includegraphics[width=0.7\textwidth,]{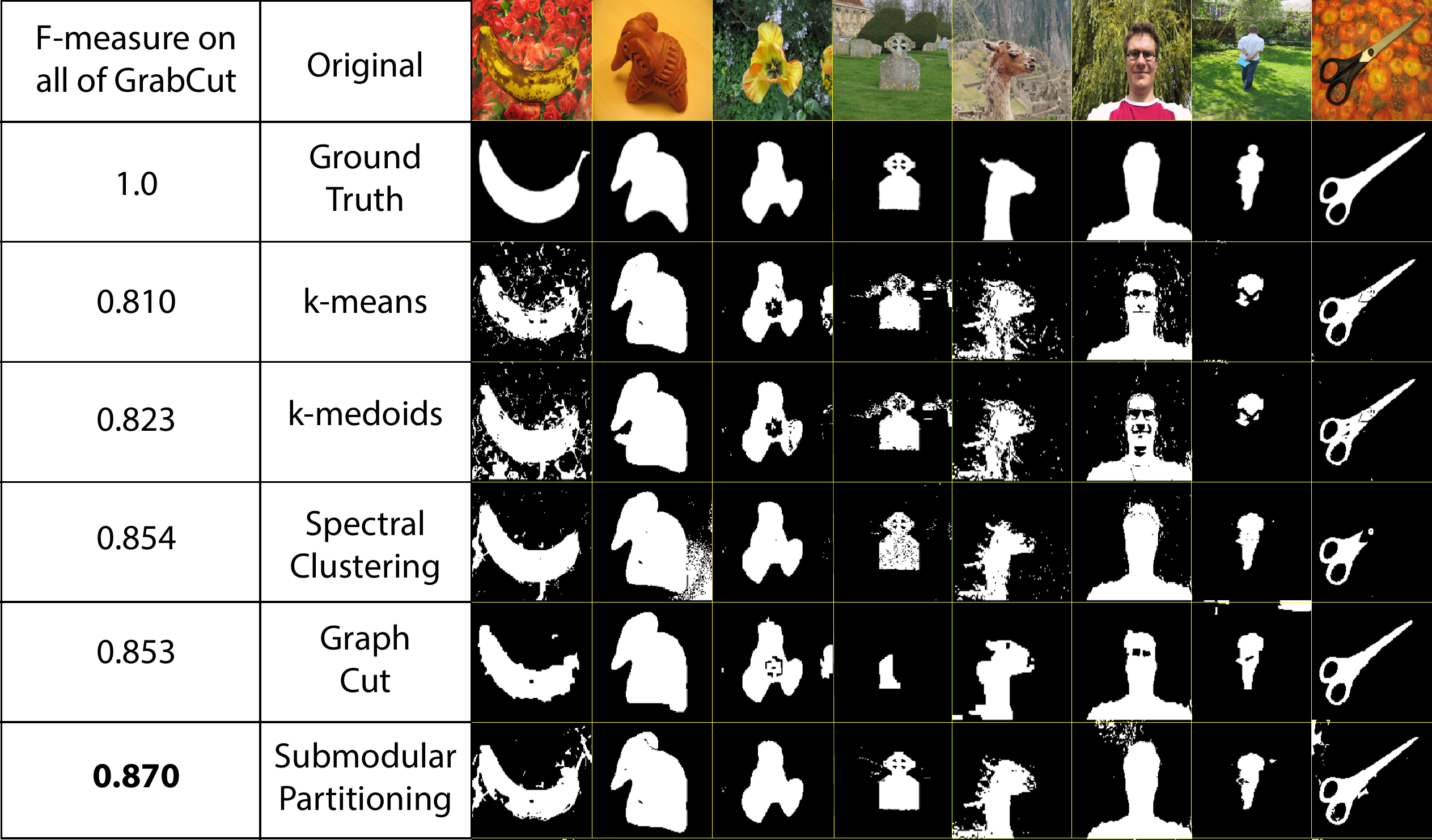}}
    \caption{Unsupervised image segmentation (right: some examples).}
    \label{fig:comparison_segmentation_example_images}
  \vspace{-2.6ex}
\end{wrapfigure} }

\arxivalt{
\subsection{Problem~\ref{prob:minmax-genmix} for Unsupervised Image Segmentation}}
{\notarxiv{\vspace{-.3\baselineskip}}}
\arxivalt{
Lastly we test the efficacy of \minmaxprob{} on the task of unsupervised image
segmentation. We evaluate on the Grab-Cut data set, which consists of 30 color images. Each image has  
ground truth foreground/background labels.
 By ``unsupervised'', we
mean that no labeled data at any time in supervised or semi-supervised
training, nor any kind of interactive segmentation, was used in
forming or optimizing the objective. 
In our experiments, the image segmentation task is solved as unsupervised clustering of the pixels, where the goal is to obtain a partitioning of the pixels such that the majority of the pixels in each block share either the same foreground or the background labels.

Let $V$ be the ground set of pixels of an image,  $\pi$ be an $m$-partition of the image, and $\{y_v\}_{v\in V}$ as the pixel-wise ground truth label ($y_v=\{0,1\}$ with 0 being background and 1 being foreground). We measure the performance of the partition $\pi$ in the following two steps: (1) for each block $i$, predict $\hat{y}_v$ for all the pixels $v\in A_i^{\pi}$ in the block as either $0$ or $1$ having larger intersection with the ground truth labels, 
i.e., predict $\hat{y}_v = 1,\forall v\in A_i^{\pi}$, if $\sum_{v\in A_{i}^{\pi}} \mathbbm{1}\{y_v = 1\} \geq \sum_{v\in A_{i}^{\pi}} \mathbbm{1}\{y_v = 0\} $, and predict $\hat{y}_v = 0,\forall v\in A_i^{\pi}$ otherwise. 
(2) report the performance of the partition $\pi$ as the F-measure of the predicted labels $\{\hat{y}_v\}_{v\in V}$ relative to the ground truth label $\{y_v\}_{v\in V}$.

In the experiment we first preprocess the data by downsampling each image by a factor $0.25$ for testing efficiency.  
We represent each pixel $v$ as 5-dimensional features $x_v\in \mathbb{R}^5$, including the RGB values and pixel positions. We normalize each feature within $[0,1]$.
To obtain a segmentation of each image we solve an instance of Problem~\ref{prob:minmax-genmix} ($0<\lambda<1$) under the homogeneous setting using \textsc{GeneralGreedMin} (Alg.~\ref{alg:greed_min_heuristic_combined}). We use the facility location function $f_{\text{fac}}$ as the objective for Problem~\ref{prob:minmax-genmix}. The similarity $s_{v,a}$ between the pixels $v$ and $a$ is computed as $s_{v,a} = C - \|x_v - x_a\|_2$ with $C=\max_{v,v^\prime \in V} \|x_v-x_v^\prime\|_2$ being the maximum pairwise Euclidean distance. 
%Since $f_{\text{fac}}$ is known as the objective for $k$-medoids clustering~\cite{krause2010discriminative}, it naturally fits the unsupervised image segmentation task, which is essentially a clustering problem. 
Since the facility location function $f_{\text{fac}}$ is defined on a pairwise similarity graph, which requires $O(|V|^2)$ memory complexity. It becomes computationally infeasible for medium sized images. Fortunately a facility location function that is defined on a sparse $k$-nearest neighbor similarity graph performs just as well with $k$ being very sparse~\cite{wei2014fast}. In the experiment, we instantiate $f_{\text{fac}}$ by a sparse $10$-nearest neighbor sparse graph, where each item $v$ is connected only to its $10$ closest neighbors.

A number of unsupervised methods are tested as baselines in the experiment, including 
$k$-means, $k$-medoids, graph cuts~\cite{Boykov2004} and spectral clustering~\cite{von2007tutorial}. We use the RBF kernel sparse similarity matrix as the input for spectral clustering.
The sparsity of the similarity matrix is $k$ and the width parameter of the RBF kernel $\sigma$. We test with various choices of $\sigma$ and $k$ and find that the setting of $\sigma=1$ and $k=20$ performs the best, with which we report the results. 
For graph cuts, we use the MATLAB implementation~\cite{Bagon2006}, which has a smoothness parameter $\alpha$. We tune $\alpha=0.3$ to achieve the best performance and report the result of graph cuts using this choice. 

\begin{figure}[htb]
\minipage{0.46\textwidth}%
\includegraphics[width=\linewidth]{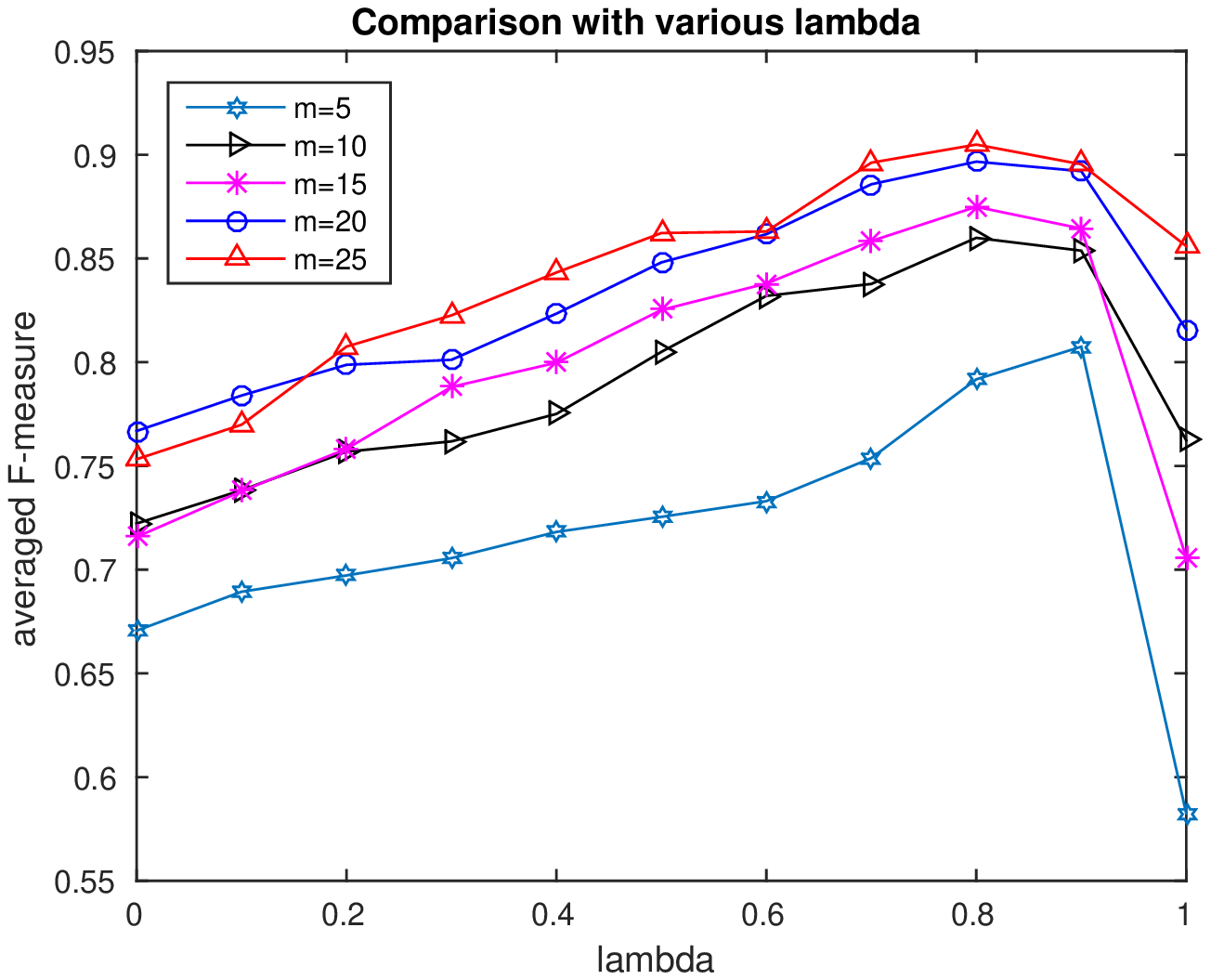}
\caption{•}
\label{fig:image_seg_lambda}
 \endminipage
\minipage{0.46\textwidth}%
\includegraphics[width=\linewidth]{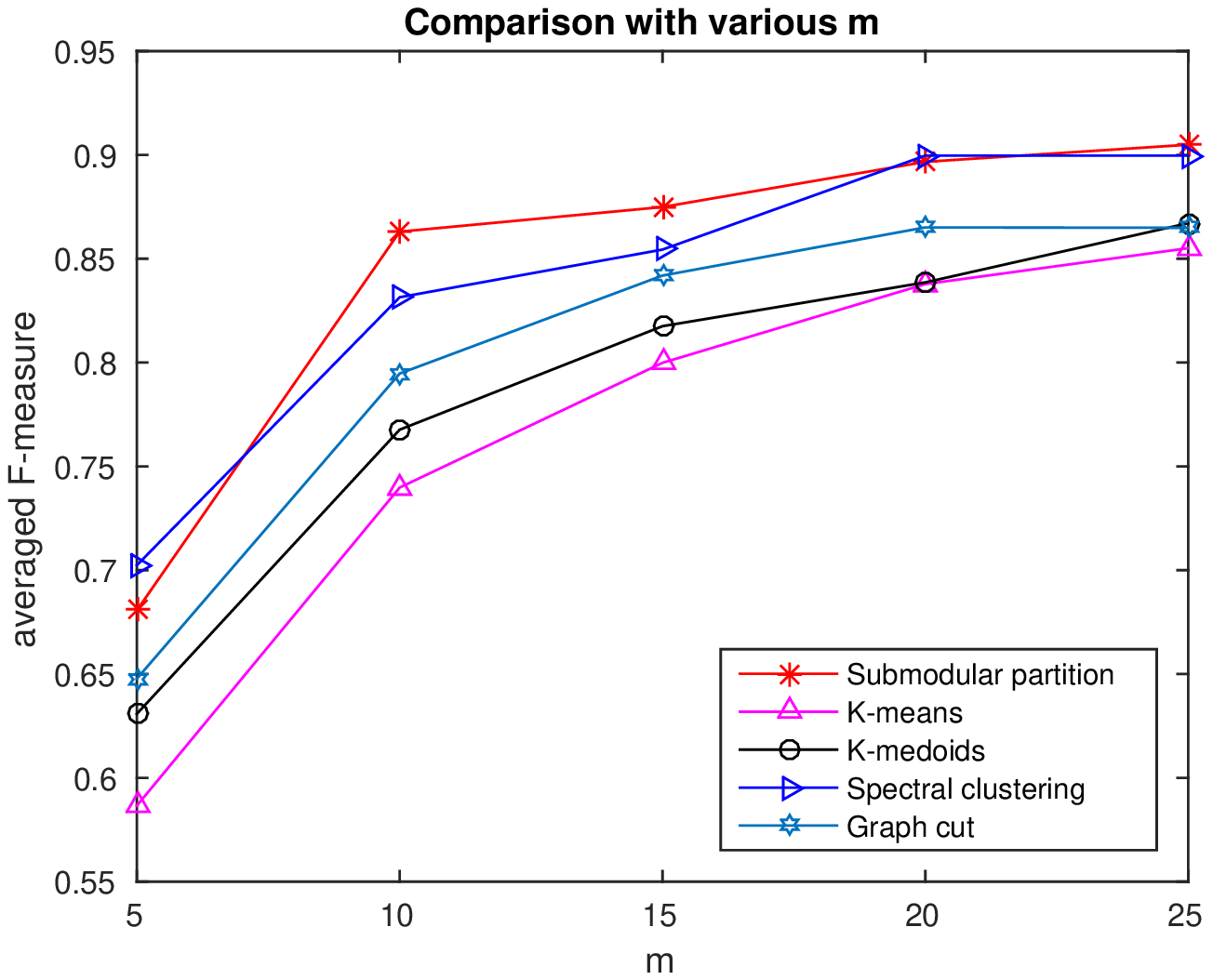}
\caption{}
\label{fig:image_seg_m}
 \endminipage
%\caption{•}
\end{figure}

The proposed image segmentation method involves a hyperparameter $\lambda$, which controls the trade-off between the worst-case objective and the average-case objective. 
First we examine how the performance of our method varies with different choices of $\lambda$ in Figure~\ref{fig:image_seg_lambda}. The performance is measured as the averaged $F$-measure of a partitioning method over all images in the data set. 
%Fixing $m=20$ we plot the average F-measure 
Interestingly we observe that the performance smoothly varies as $\lambda$ increases from 0 to 1. In particular the best performance is achieved when $\lambda$ is within the range  $[0.7,0.9]$. It suggests that using only the worst-case or the average-case objective does  not suffice for the unsupervised image segmentation / clustering task, and an improved result is achieved by mixing these two extreme cases. 
In the subsequent experiments we show only the result of our method with $\lambda=0.2$. Next we compare the proposed approach with baseline methods on various $m$ in Figure~\ref{fig:image_seg_m}. In general, each method improves as $m$ increases. Submodular partitioning method performs the best on almost all cases of $m$. 
%\kai{Kai}{Add the plot showing the performance on various $m$.}
Lastly we show in Figure~\ref{fig:comparison_segmentation_example_images} example segmentation results on several example images as well as averaged F-measure in the case of $m=15$. We observe that submodular partitioning, in general, leads to less noisy and more coherent segmentation in comparison to the baselines. 
\begin{figure}[htb]
    \centerline{\includegraphics[width=0.7\textwidth,]{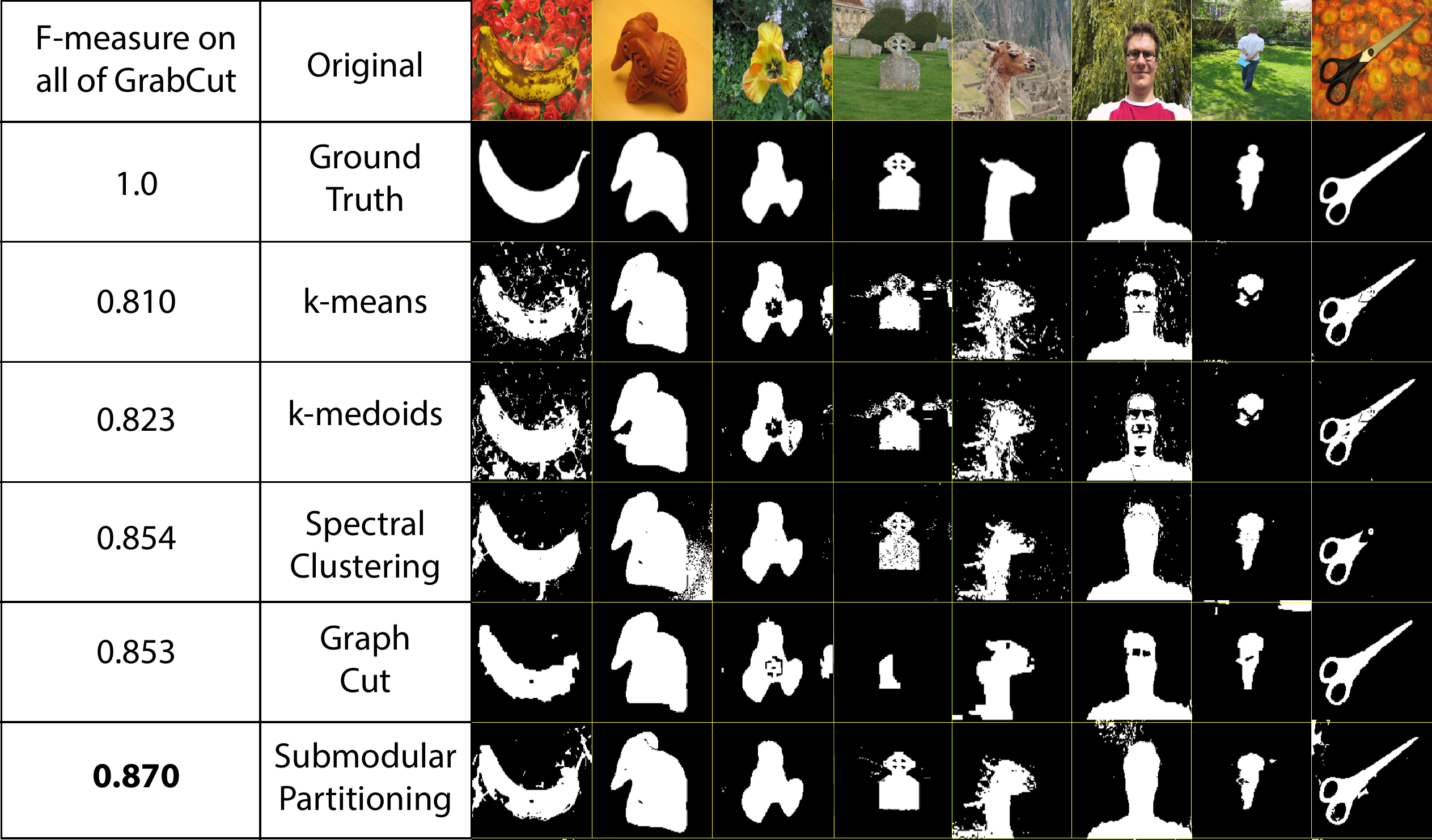}}
    \caption{Unsupervised image segmentation (right: some examples).
\jeff{Kai}{We need to redo the table with new set of F-measure results, since we use a different way of mapping from partition to binary segmentation.}    }
    \label{fig:comparison_segmentation_example_images}
\end{figure}
}
{\textbf{Unsupervised Image Seg- mentation:} We test the efficacy of \minmaxprob{} on unsupervised image
segmentation over the GrabCut data set (30 color images and their
ground truth foreground/background labels). By ``unsupervised'', we
mean that no labeled data at any time in supervised or semi-supervised
training, nor any kind of interactive segmentation, was used in
forming or optimizing the objective.  
The submodular partitioning for each image is obtained by solving the homogeneous case of Problem~\ref{prob:minmax-genmix} ($\lambda=0.8$) using a modified variant of \textsc{GreedMin} on the facility location function. 
We compare our method against the other
unsupervised methods $k$-means, $k$-medoids, spectral clustering, and graph cuts. 
Given an $m$-partition of an image and its ground truth labels, we assign each 
of the $m$ blocks either to the foreground or
background label having the larger intersection. In Fig.~\ref{fig:comparison_segmentation_example_images} we show 
example segmentation results after this mapping on several example images as well
as averaged F-measure (relative to ground truth) over the whole data set. 
\notarxiv{More details are given in \cite{datapartitionextend}.}\looseness-1}

%\notarxiv{As can be seen, in all the above cases submodular partitioning is scalable and yields improved results.}
%We show these mappings 

\notarxiv{
\textbf{Acknowledgments:} 
This material is based upon work supported by the National Science Foundation under
Grant No.\ IIS-1162606, 
% Rishabh, always use this IIS grant
the National Institutes of Health under award
R01GM103544, 
% Kai, always use this R01 grant for citations
and by a Google, a Microsoft, and an Intel research award. R.\ Iyer
acknowledges support from a Microsoft Research Ph.D Fellowship.
This work was supported in part by TerraSwarm, one of six centers of STARnet, a Semiconductor Research Corporation program sponsored by MARCO and DARPA.
}

\section{Conclusions}
\label{sec:conclusions}
In this paper, we considered two novel mixed robust/average-case submodular partitioning problems, which generalize four well-known problems: submodular fair allocation (SFA), submodular load balancing (SLB), submodular welfare problem (SWP), and submodular multiway partition (SMP). 
While the average case problems, i.e., SWP and SMP, admit efficient and tight algorithms, existing approaches for the worst case problems, i.e., SFA and SLB, are, in general, not scalable. We bridge this gap by providing several new algorithms that not only scale to large data sets but that also achieve comparable theoretical guarantees. Moreover we provide a number of efficient frameworks for solving the general mixed robust/average-case submodular partitioning problems. 
%These frameworks generalize the proposed algorithms for solving the worst-case partitioning problems or use them as the building blocks. 
We also demonstrate that submodular partitioning is applicable in a number of machine learning problems involving distributed optimization, computational load balancing, and unsupervised image segmentation. Lastly we empirically show the effectiveness of the proposed algorithms on these machine learning tasks.\looseness-1

\acks{This material is based upon work supported by the National Science Foundation under
Grant No.\ IIS-1162606, 
% Rishabh, always use this IIS grant
the National Institutes of Health under award
R01GM103544, 
% Kai, always use this R01 grant for citations
and by a Google, a Microsoft, and an Intel research award. R.\ Iyer
acknowledges support from a Microsoft Research Ph.D Fellowship.
This work was supported in part by TerraSwarm, one of six centers of STARnet, a Semiconductor Research Corporation program sponsored by MARCO and DARPA.}


\newpage

\appendix
\section*{Appendix}
\subsection*{Proof for Theorem~\ref{thm:submod_two_case}}
\noindent{\bf Theorem }{\it
If $f_1$ and $f_2$ are monotone submodular, 
$\min\{f_1(A), f_2(V\setminus A)\}$ is also submodular.
}
\begin{proof}
To prove the Theorem, we show a more general result: Let $f$ and $h$ be submodular, and $f-h$ be either monotone increasing or decreasing, then $g(S) = \min\{f(S), h(S)\}$ is also submodular. The Theorem follows by this result, since $f(S) = f_1(S)$ and $h(S) = f_2(V\setminus S)$ are both submodular and $f(S) - h(S) = f_1(S) - f_2(V\setminus S)$ is monotone increasing.

In order to show $g$ is submodular, we prove that $g$ satisfies the following:
\begin{align}
g(S) + g(T) \geq g(S\cap T) + g(S\cup T), \forall S,T \subseteq V
\label{eqn:proof_g_submodular}
\end{align}
If $g$ agrees with either $f$ or $h$ on both $S$ and $T$, and since 
\begin{align}
f(S) + f(T) \geq f(S\cup T) + f(S\cap T)\\
h(S) + h(T) \geq h(S\cup T) + h(S\cap T),\\
\end{align}
Eqn~\ref{eqn:proof_g_submodular} follows. 

Otherwise, w.l.o.g. we consider $g(S) = f(S)$ and $g(T) = h(T)$. For the case where $f-h$ is monotone non-decreasing, consider the following:
\begin{align}
g(S) +g(T) &= f(S) + h(T) 
\\ & \geq f(S\cup T) + f(S\cap T) -( f(T) - h(T)) \mbox{ // submodularity of $f$}
\\ &\geq f(S\cup T) + f(S\cap T) - (f(S\cup T)- h(S\cup T)) \mbox{ // monotonicity of $f-h$}
\\ & = f(S\cap T) + h(S\cup T)
\\ & \geq g(S\cap T) + g(S\cup T). 
\end{align}

Similarly, for the case where $f-h$ is monotone non-increasing, consider the following:
\begin{align}
g(S) +g(T) &= f(S) + h(T) 
\\ & \geq  h(S\cap T) + h(S\cup T) + (f(S) - h(S) ) \mbox{ // submodularity of $h$}
\\ &\geq h(S\cup T) + h(S\cap T) + (f(S\cup T)- h(S\cup T)) \mbox{ // monotonicity of $f-h$}
\\ & = h(S\cap T) + f(S\cup T)
\\ & \geq g(S\cap T) + g(S\cup T). 
\end{align}
\end{proof}

\subsection*{Proof for Theorem~\ref{thm:greedy_max_guarantee}}
{\bf Theorem }{\it
\arxivalt{
Under the homogeneous setting ($f_i=f$ for all $i$), \textsc{GreedMax} is guaranteed to find a partition $\hat{\pi}$ such that 
\arxivalt{
\begin{align}
\min_{i=1,\dots,m} f(A_i^{\hat{\pi}}) \geq \frac{1}{m} \max_{\pi \in \Pi }\min_{i=1,\dots,m}  f(A_i^\pi).
\end{align}
}
{$\min_{i=1,\dots,m} f(A_i^{\hat{\pi}}) \geq \frac{1}{m} \max_{\pi \in \Pi }\min_{i=1,\dots,m}  f(A_i^\pi)$.} 
}
{
\textsc{GreedMax} achieves a guarantee of $1/m$ under the homogeneous setting ($f_i=f$ for all $i$).\looseness-1
}
%Conversely, for any $m\geq 2$, there exists a submodualr function $f$, on which an instance of \textsc{GreedMax} achieves an approximation factor $1/2$.
}
\begin{proof}
We prove that the guarantee of $1/m$, in fact, even holds for a streaming version of the greedy algorithm (\textsc{StreamGreed}, see Alg.~\ref{alg:stream_greedy_SFA}). In particular, we show that \textsc{StreamGreed} provides a factor of $1/m$ for SFA under the homogeneous setting. Theorem~\ref{thm:greedy_max_guarantee} then follows since \textsc{GreedMax} can be seen as running \textsc{StreamGreed} with a specific order. 

\begin{algorithm}[htb]
\caption{\textsc{StreamGreed}}
\label{alg:stream_greedy_SFA}
Input: $V=\{v_1,v_2,\ldots,v_n\}$, $f$, $m$\\
Initialize: $A_1=,\dots,=A_m = \emptyset$, $k=1$ \\
\While {$k\leq n$}{
    $i^*\in \argmin_j f(A_j)$ \\	
	$A_{i^*}\leftarrow A_{i^*}\cup \{v_k\}$ \\	
	$k\leftarrow k+1$ \\
	}
\end{algorithm} 

To prove the guarantee for \textsc{StreamGreed}, we consider the resulting partitioning after an instance of \textsc{StreamGreed}: $\pi = (A^\pi_1\cup A^\pi_2,\dots, A^\pi_{m})$. For simplicity of notation, we write $A_i^\pi$ as $A_i$ for each $i$ in the remaining proof. We refer $OPT$ to the optimal solution, i.e., $OPT = \max_{\pi} \min_i f(A_i)$. 
W.l.o.g., we assume $f(A_1)=\min_i f(A_i)$.
Let $a_i$ be the last item to be chosen in block $A_i$ for $i=2,\dots,m$.

Claim 1: 
\begin{equation}
OPT\leq f(V\setminus \{a_2,\dots,a_m\})
\end{equation}

To show this claim, consider the following: 
If we enlarge the singleton value of $a_i, i=2,\ldots,m$, we obtain a new submodular function:  
\begin{equation}
f^\prime(A)=f(A)+\alpha \sum_{i=2}^m |A\cap a_i|,
\end{equation}
where $\alpha$ is sufficiently large. Then running \textsc{StreamGreed}  on $f^\prime$
with the same ordering of the incoming items leads to the same solution, since only the gain of the last added item for each block is changed.

Note that $f^\prime (A)\geq f(A),\forall A\subseteq V$, we then have $\max_{\pi} \min_{i} f^\prime (A_i^\pi) \geq OPT$. The optimal partitioning for $f^\prime$ can be easily obtained as $\pi^\prime=(V\setminus \{a_2,\dots,a_m\}, a_2,\dots,a_m)$. Therefore, we have that 
\begin{align}
OPT \leq \max_{\pi} \min_i f^\prime (A_i^\pi) = f^\prime (V\setminus \{a_2,\dots,a_m\}) = f(V\setminus \{a_2,\dots,a_m\}).
\end{align}

Lastly, we have that 
$f(A_1)\geq f(A_i\setminus a_i)$ for any $i=2,\dots,m$ due to the procedure of 
\textsc{StreamGreed}. Therefore we have the following: 
\begin{align}
f(A_1)& \geq 
\frac{1}{m}( f(A_1)+ \sum_{i=2}^m f(A_i\setminus a_i) ) \\
& \geq \frac{1}{m} f(V\setminus \{a_2,\dots,a_m\}) \text{ // submodularity of $f$}\\
& \geq \frac{1}{m} OPT \text{ // Claim 1}
\end{align}
\end{proof}

\subsection*{Proof for Theorem~\ref{thm:greedy_sat_bound}}
{\bf Theorem }{\it
Given $\epsilon$, $\alpha$ and any $0<\delta<\alpha$, \textsc{GreedSat} finds a partition such that at least $\ceil{m( \alpha-\delta)}$ blocks receive utility at least $\frac{\delta}{1-\alpha + \delta}( \max_{\pi}\min_{i}f_i(A_i^{\pi})-\epsilon)$. 
}

\begin{proof}
When \textsc{GreedSat} terminates, it identifies a $c_{\min}$ such that the returned solution $\hat{\pi}^{c_{\min}}$ satisfies $\bar{F}^{c_{\min}} (\hat{\pi}^{c_{\min}}) \geq \alpha c_{\min}$. Also it identifies a $c_{\max}$ such that the returned solution 
$\hat{\pi}^{c_{\max}}$ satisfies $\bar{F}^{c_{\max}} (\hat{\pi}^{c_{\max}}) < \alpha c_{\max}$. The gap between $c_{\max}$ and $c_{\min}$ is bounded by $\epsilon$, i.e., $c_{\max} - c_{\min} \leq \epsilon$.

Next, we prove that there does not exist any partitioning $\pi$ that satisfies $\min_{i}f_i(A_i^{\pi}) \geq c_{\max}$, i.e., $c_{\max} \geq \max_{\pi \in \Pi}\min_i f_i(A_i^{\pi})$. 

Suppose otherwise, i.e., $\exists \pi^*: \min_{i}f_i(A_i^{\pi^*}) = c_{\max}+\gamma$ with $\gamma \geq 0$. Let $c = c_{\max} + \gamma$, consider the intermediate objective $\bar{F}^{c} (\pi)= \frac{1}{m}\sum_{i=1}^m \min\{f_i(A^\pi_i),c\}$, we have that $\bar{F}^{c}(\pi^*) = c$. An instance of the algorithm for SWP on $\bar{F}^{c}$ is guaranteed to lead to a solution $\hat{\pi}^c$ such that $\bar{F}^{c}(\hat{\pi}^c) \geq \alpha c$. 
Since $c \geq c_{\max}$, it should follow that the returned solution $\hat{\pi}^{c_{\max}}$ for the value $c_{\max}$ also satisfies $\bar{F}^{c_{\max}}(\hat{\pi}) \geq \alpha c_{\max}$. However it contradicts with the termination criterion of \textsc{GreedSat}. Therefore, we prove that $c_{\max} \geq \max_{\pi \in \Pi}\min_i f_i(A_i^{\pi})$, which indicates that $c_{\min} \geq c_{\max} - \epsilon \geq \max_{\pi\in \Pi} \min_i f_i(A_i^\pi)-\epsilon$.

Let $c^*= \frac{c_{\max} + c_{\min}}{2}$ and the partitioning returned by running for $c^*$ be $\hat{\pi}$ (the final output partitioning from \textsc{GreedSat}). We have that
$\bar{F}^{c^*}(\hat{\pi}) \geq \alpha c^*$, we are going to show that for any $0<\delta<\alpha$, at least a $\ceil{m(\alpha-\delta)}$ blocks given by $\hat{\pi}$ receive utility larger or equal to $\frac{\delta}{1-\alpha+\delta} c^*$. 

Just to restate the problem: we say that the $i^{\text{th}}$ block is $(\alpha,\delta)$-good if $f_i(A_i^{\hat{\pi}}) \geq \frac{\delta}{1-\alpha + \delta} c^*$. Then the statement becomes: Given $ 0 < \delta < \alpha $,  there is at least $m\ceil{\alpha-\delta}$ blocks that are $(\alpha,\delta)$-good. 

To prove this statement,  
we assume, by contradiction, that there is strictly less than $\ceil{m(\alpha - \delta)}$ $(\alpha,\delta)$-good blocks. Denote the number of $(\alpha,\delta)$-good blocks as $m_{\text{good}}$. Then we have that $m_{\text{good}}\leq \ceil{m(\alpha -\delta)}-1 < m(\alpha - \delta)$.
Let $\theta=\frac{m_{\text{good}}}{m}$ be the fraction of $(\alpha, \delta)$ good blocks, then we have that $0 \leq \theta < (\alpha - \delta) < 1$. The remaining fraction $(1-\theta)$ of blocks are not good, i.e., they have valuation strictly less than $\frac{\delta}{1 - \alpha + \delta}c^*$. Then, consider the following:
\begin{align}
\bar{F}^{c^*}(\hat{\pi}) & = \frac{1}{m} \sum_{i=1}^m \min\{f_i(A_i^{\hat{\pi}}), c^*\} \\
& \stackrel{(a)}{<} \theta c^* + (1-\theta)  \frac{\delta}{1-\alpha + \delta} c^*
\\ & = \frac{\delta}{1-\alpha + \delta} c^* + \frac{1-\alpha}{1-\alpha + \delta} \theta c^*
\\ & \stackrel{(b)}{<} \frac{\delta}{1-\alpha + \delta} c^* + \frac{1-\alpha}{1-\alpha + \delta} (\alpha - \delta) c^*
\\ & = \alpha c^*
\end{align}
Inequality (a) follows since good blocks are upper bounded by $c^*$, and not good blocks have values upper bounded by $\frac{\delta}{1-\alpha + \delta}c^*$. Inequality (b) follows by the assumption on $\theta$. This therefore contradicts the assumption that $\bar{F}^{c^*}(\hat{\pi}) \geq \alpha c^*$, 
hence the statement is true. 

This statement can also be proved using a different strategy. Let $f_i^{c^*} = \min\{f_i(A_i^{\hat{\pi}}), c^*\}$ and $f_i = f_i(A_i^{\hat{\pi}})$ for all $i$. For any $0\leq \beta\leq 1$ the following holds:
\begin{align}
\alpha c^* \leq \bar{F}^{c^*}(\hat{\pi}) =  \frac{1}{m}\sum_i f^{c^*}_i  \leq \frac{1}{m} f_i
= \frac{1}{m}\sum_{i : f_i < \beta c^*} f_i
   + \frac{1}{m}\sum_{i: f_i \geq \beta c^*} f_i
 < \frac{1}{m}m_\text{bad}\beta c^*+ \frac{1}{m}m_\text{good}c^*
\end{align}
where $m = m_\text{bad}+m_\text{good}$ and $m_\text{good}$ are the number that are $\beta$-good (i.e., $i$ is $\beta$-good if $f_i \geq \beta c^*$). The goal is to place a lower bound on $m_{\text{good}}$. From the above
\begin{align}
\alpha  < (1 - \frac{m_{\text{good}}}{m}) \beta + \frac{m_{\text{good}}}{m} 
\end{align}
which means 
\begin{align}
m_{\text{good}} \geq \ceil{\frac{ \alpha  - \beta}{ 1 - \beta} m} 
\end{align}
Let $\beta = \frac{\delta}{1-\alpha + \delta}$, the statement immediately follows.

Note $c^* = \frac{c_{\min} + c_{\max}}{2} \geq c_{\max} - \epsilon \geq \max_{\pi \in \Pi} \min_i f_i(A_i^{\pi}) - \epsilon$. Combining pieces together, we have shown that at least $\ceil{m(\alpha - \delta)}$ blocks given by $\hat{\pi}$ receive utility larger or equal to $\frac{\delta}{1-\alpha + \delta} (\max_{\pi\in\Pi}\min_i f_i(A_i^{\pi}) - \epsilon)$.
\end{proof}

\subsection*{Proof for Theorem~\ref{thm:MMax_bound}}
{\bf Theorem }{\it
\textsc{MMax} achieves a worst-case guarantee of $O( \min_i \frac{1+(|{A}^{\hat{\pi}}_i|-1)(1-\kappa_{f_i}({A}^{\hat{\pi}}_i))}{|{A}^{\hat{\pi}}_i| 
\sqrt{m} \log^3 m})$,
 where $\hat{\pi}=({A^{\hat{\pi}}_1},\cdots, 
A^{\hat{\pi}}_{m})$ is the partition obtained by the algorithm, and $\kappa_f(A) = 1-\min_{v\in V} \frac{f(v|A\setminus v)}{f(v)}\in [0,1]$  is the curvature of a submodular function $f$ at $A\subseteq V$.
}
\begin{proof}
We assume the approximation factor of the algorithm for solving the modular version of Problem 1 is $\alpha = O(\frac{1}{\sqrt{m}\log^3 m})$~\cite{asadpour2010approximation}.
%It suffices to consider only the first iteration, where we apply the same simple modular proxy $h_i$ for each $f_i$. 
For notation simplicity, we write $\hat{\pi} = (\hat{A}_1,\dots, \hat{A}_m)$ as the resulting partition after the first iteration of \textsc{MMax}, and $\pi^* = (A_1^*,\dots, A_m^*)$ as its optimal solution. Note that first iteration suffices to yield the performance guarantee, and the subsequent iterations are designed so as to improve the empirical performance. Since the proxy function for each function $f_i$ used for the first iteration are the simple modular upper bound with the form: $h_i(X) = \sum_{j\in X} f_i(j) $. 

Given the curvature of each submodular function $f_i$, we can tightly bound a submodular function $f_i$ in the following form~\cite{curvaturesubmodular}:
\begin{align}
f_i(X)\leq h_i(X) \leq \frac{|X|}{1+(|X|-1)(1-\kappa_{f_i}(X))} f_i(X), \forall X\subseteq V
\label{eqn:tight_bound_modular_upper}
\end{align}

Consider the following:
\begin{align}
\min_i f_i(\hat{A}_i) & \geq \min_i \frac{1}{\frac{|\hat{A}_i|}{1+(|\hat{A}_i| - 1) (1-\kappa_{f_i} (\hat{A}_i))}} h_i(\hat{A}_i)
\\ & \geq \min_i \frac{1}{\frac{|\hat{A}_i|}{1+(|\hat{A}_i| - 1) (1-\kappa_{f_i} (\hat{A}_i))}}  \min_i h_i(\hat{A}_i) \\ & 
\geq \alpha \min_i \frac{1+(|\hat{A}_i| - 1)(1-\kappa_{f_i}(\hat{A}_i))}{|\hat{A}_i|} \min_i h_i(A_i^*) \\ & 
\geq \alpha \min_i \frac{1+(|\hat{A}_i| - 1)(1-\kappa_{f_i}(\hat{A}_i))}{|\hat{A}_i|} \min_i f_i(A_i^*)
\\ & = O(\min_i \frac{1+(|\hat{A}_i|-1)(1-\kappa_{f_i}(\hat{A}_i))}{|\hat{A}_i|\sqrt{m}\log^3 m}) \min_i f_i(A^*_i).
\end{align}
\end{proof}

\subsection*{Proof for Theorem~\ref{thm:iterative_improvement_for_MMAX}}
{\bf Theorem }{\it
%\label{thm:iterative_improvement_for_MMAX}
Suppose there exists an algorithm for solving the modular version of SFA with an approximation factor $\alpha\leq 1$, we have that
\begin{align}
\min_{i} f_i(A_i^{\pi_t}) \geq \alpha \min_{i} f_i(A_i^{\pi_{t-1}}).
\end{align}
}

\begin{proof}
Consider the following:
\begin{align}
\min_{i} f_i(A^{\pi_{t-1}}_i) &= \min_{i} h_i(A^{\pi_{t-1}}_i) // \mbox{ tightness of modular lower bound.}
\\ & \leq \alpha \min_{i} h_i(A_{i}^{\pi_t})
 // \mbox{ approximation factor of the modular SFA. }
\\ & \leq \alpha h_j(A_{j}^{\pi_t}) // j\in \argmin_{i} f_i(A_i^{\pi_t})
\\ & \leq \alpha f_{j}(A_j^{\pi_t}) // h_j(A_{j}^{\pi_{t-1}}) \mbox{ upper bounds } f_j \mbox{ everywhere.}
\\ & = \alpha \min_i f_i(A_i^{\pi_t})
\end{align}
\end{proof}

\subsection*{Proof for Theorem~\ref{thm:hardness_SLB}}
{\bf Theorem }{\it
For any $\epsilon>0$, 
SLB cannot be approximated to a factor of $(1-\epsilon)m$ for any $m=o(\sqrt{n/\log n})$ with polynomial number of queries even under the homogeneous setting. 
}
\begin{proof}
We use the same proof techniques as in~\cite{svitkina2008submodular}.
Consider two submodular functions:
\begin{align}
& f_1(S) = \min\{|S|, \alpha\}; \\
& f_2 (S) = \min\{ \sum_{i=1}^m \min\{\beta, |S\cap V_i|\}, \alpha \};
\end{align}
where $\{V_i\}_{i=1}^m$ is a uniformly random partitioning of $V$ into $m$ blocks, 
 $\alpha = \frac{n}{m}$ and $\beta = \frac{n}{m^2(1-\epsilon)} $. To be more precise about the uniformly random partitioning, we assign each item into any one of the $m$ blocks with probability $1/m$. 
It can be easily verified that $OPT_1 = \min_{\pi\in \Pi}\max_i  f_1(A_i^\pi) = n/m$ and $OPT_2 = \min_{\pi\in \Pi}\max_i f_2(A_i^\pi) = \frac{n}{m^2(1-\epsilon)}$. The gap is then $\frac{OPT_1}{OPT_2} = m(1-\epsilon)$. 
 
Next, we show that $f_1$ and $f_2$ cannot be distinguished with $n^{\omega(1)}$ number of queries.

Since $f_1(S) \geq f_2(S)$ holds for any $S$, this is equivalent as showing 
$P\{f_1(S) > f_2(S)\} < n^{-\omega(1)} $.
% is exponentially small for any one set $S$ over the random partition $\pi(V)$.  

As shown in~\cite{svitkina2008submodular}, $P\{f_1(S) > f_2(S)\}$ is maximized when $|S| = \alpha$. 
It suffices to consider only the case of $|S| = \alpha$ as follows:
 \begin{align}
 P\{f_1(S) > f_2(S) : |S| = \alpha\} = P \{\sum_{i=1}^m \min\{\beta, |S\cap V_i|\} < \alpha : |S| = \alpha\}
 \end{align}
 The necessary condition for $\sum_{i=1}^m \min\{\beta, |S\cap V_i|\} < \alpha$ is that $|S\cap V_i| > \beta$ is satisfied for some $i$. Using the Chernoff bound, we have that for any $i$, it holds $P\{|S\cap V_i| > \beta \} \leq e^{-\frac{\epsilon^2 n}{3m^2}} = n^{-\omega(1)}$ when $m=o(\sqrt{n/\log n})$. Using the union bound, it holds that the probability for any one block $V_i$ such that $|S\cap V_i|>\beta$ is also upper bounded by $n^{-\omega(1)}$. 
  Combining all pieces together, we have the following:
 \begin{align}
 P\{f_1(S) > f_2(S) \} \leq n^{-\omega (1)}. 
 \end{align}

Finally, we come to prove the Theorem. Suppose the goal is to solve an instance of SLB with $f_2$. Since $f_1$ and $f_2$ are hard to distinguish with polynomial number of function queries, any polynomial time algorithm for solving $f_2$ is equivalent to solving for
$f_1$.  However, the optimal solution for $f_1$ is $\alpha = \frac{n}{m}$, whereas the optimal solution for $f_2$ is $\beta = \frac{n}{m^2 (1-\epsilon)}$. Therefore, no polynomial time algorithm can find a solution with a factor $m(1-\epsilon)$ for SLB in this case. 
\end{proof}

\subsection*{Proof for Theorem~\ref{thm:lovaszRound_bound}}
{\bf Theorem }{\it
\arxivalt{
\textsc{\lovaszRound} is guaranteed to find a partition $\hat{\pi} \in \Pi$ such that $\max_{i}f_i(A_i^{\hat{\pi}}) \leq m \min_{\pi \in \Pi} \max_{i} f_i(A_i^{\pi})$.
}
{
\textsc{\lovaszRound} achieves a worst-case approximation factor $m$.
}
}
\begin{proof}
It suffices to bound the performance loss at the step of rounding the fractional solution $\{x^*_i\}_{i=1}^m$, or equivalently, the following:
\begin{align}
\max_i \tilde{f}_i(x_i^*) \geq  \frac{1}{m} \max_i f_i(A_i),
\label{eqn: m_bound}
\end{align}
where $\{A_i\}_{i=1}^m$ is the resulting partitioning after the rounding. 
To show Eqn~\ref{eqn: m_bound}, it suffices to show that $\tilde{f}_i(x_i^*) \geq \frac{1}{m} f_i(A_i)$ for all $i=1,\dots,m$. 
%It can be easily verified that if $\hat{f}_i(x_i^*) \geq \frac{1}{m} f_i(A_i)$ holds for all $i=1,\dots,m$, Eqn~\ref{eqn: m_bound} follows. 
Next, consider the following:
\begin{align}
f_i(A_i) &= \tilde{f}_i (1_{A_i})  = m \tilde{f}_i(\frac{1}{m} 1_{A_i}) // \mbox { positive homogeneity of \lovasz{} extension}
\end{align}
For any item $v_j\in A_i$, we have $x^*_i(j) \geq \frac{1}{m}$, since $\sum_{i=1}^m x^*_i(j) \geq 1$ and $x^*_i(j) = \max_{i^\prime} x_{i^\prime} (j)$. Therefore, we have $\frac{1}{m} 1_{A_i} \leq x_i^*$. Since $f_i$ is monotone, its extension $\tilde{f}_i$ is also monotone. As a result, $f_i(A_i) = m \tilde{f}_{i}(\frac{1}{m}1_{A_i}) \leq m \tilde{f}_i (x_i^*)$.

\end{proof}

\subsection*{Proof for Theorem~\ref{thm:MMin_bound}}

{\bf Theorem }{\it
\textsc{MMin} achieves a worst-case guarantee of $(2\max_i \frac{|A_i^{\pi^*}|}{1+(|A_i^{\pi^*}|-1)(1-\kappa_{f_i}(A_i^{\pi^*}))})$, where $\pi^* = (A_1^{\pi^*}, \cdots, A_m^{\pi^*})$ denotes the optimal partition.
}
\begin{proof}
Let $\alpha=2$ be the approximation factor of the algorithm for solving the modular version of Problem 2~\cite{lenstra1990approximation}. 
For notation simplicity, we write $\hat{\pi} = (\hat{A}_1,\dots, \hat{A}_m)$ as the resulting partition after the first iteration of \textsc{MMin}, and $\pi^* = (A_1^*,\dots, A_m^*)$ as its optimal solution. Again the first iteration suffices to yield the performance guarantee, and the subsequent iterations are designed so as to improve the empirical performance. Since the supergradients for each function $f_i$ used for the first iteration are the simple modular upper bound with the form: $h_i(X) = \sum_{j\in X} f_i(j) $, we can again tightly bound a submodular function $f_i$ in the following form:
\begin{align}
f_i(X)\leq h_i(X) \leq \frac{|X|}{1+(|X|-1)(1-\kappa_{f_i}(X))} f_i(X), \forall X\subseteq V
\end{align}
Consider the following:
\begin{align}
\max_{i} f_i(\hat{A}_i) &\leq \max_{i}h_i(\hat{A}_i)
\\ &  \leq \alpha \max_{i}h_i(A_i^*) 
\\& \leq \alpha \max_i \frac{|A_i^*|}{1+(|A_i^*|-1)(1-\kappa_{f_i}(A_i^*))} f_i(A_i^*) 
\\ & \leq \alpha \max_i \frac{|A_i^*|}{1+(|A_i^*|-1)(1-\kappa_{f_i}(A_i^*))}  \max_i f_i(A_i^*) 
\end{align}
\end{proof}

\subsection*{Proof for Theorem~\ref{thm:iterative_improvement_for_MMIN}}
{\bf Theorem }{\it
Suppose there exists an algorithm for solving the modular version of SLB with an approximation factor $\alpha\geq 1$, we have for each iteration $t$ that
\begin{align}
\max_{i} f_i(A_i^{\pi_t}) \leq \alpha \max_{i} f_i(A_i^{\pi_{t-1}}).
\end{align}
}
\begin{proof}
The proof is symmetric to the one for Theorem~\ref{thm:iterative_improvement_for_MMAX}.
\end{proof}

\subsection*{Proof for Theorem~\ref{thm:max_min_sum_combined_bound}}
We prove separately for Problem~\ref{prob:maxmin-genmix} and Problem~\ref{prob:minmax-genmix}. 

{\bf Theorem }{\it
Given an instance of Problem~\ref{prob:maxmin-genmix} with $0 < \lambda < 1$, \textsc{CombSfaSwp} provides an approximation guarantee of $\max\{\min\{\alpha,\frac{1}{m}\},\frac{\beta \alpha}{\bar{\lambda} \beta + \alpha}, \lambda \beta\}$ in the \emph{homogeneous} case, and a factor of $\max\{\frac{\beta \alpha}{\bar{\lambda} \beta + \alpha}, \lambda \beta\}$ in the \emph{heterogeneous} case, where $\alpha$ and $\beta$ are the approximation factors of \textsc{AlgWC} and \textsc{AlgAC} for SFA and SWP respectively. 
}
%\begin{theorem}
%\end{theorem}
\begin{proof}
We first prove the result for heterogeneous setting. For notation simplicity we write the worst-case objective as $F_1(\pi)=\min_{i=1,\ldots,m}f(A_i^{\pi})$ and the average-case objective as $F_2(\pi)=\frac{1}{m}\sum_{i=1,\ldots,m}f(A_i^{\pi})$. 

Suppose \textsc{AlgWC} outputs a partition $\hat{\pi}_1$ and \textsc{AlgAC} outputs a partition $\hat{\pi}_2$.  
Let $\pi^*\in \arg\max_{\pi\in\Pi}\bar{\lambda}F_1(\pi)+\lambda F_2(\pi)$ be the optimal partition.

We use the following facts: 

Fact1
\begin{align}
F_1(\hat{\pi}_1)\geq\alpha F_1(\pi)
\end{align}
Fact2
\begin{align}
F_2(\pi_2)\geq\beta F_2(\pi)
\end{align}
Fact3
\begin{align}
F_1(\pi)&\leq F_2(\pi)
\end{align}

Then we have that

\begin{align}
\bar{\lambda}F_1(\hat{\pi}_2)+\lambda F_2(\hat{\pi}_2) &\geq \lambda F_2(\hat{\pi}_2) \\
&\geq \lambda \beta F_2(\pi^*) \\
&\geq \lambda \beta \left[\bar{\lambda}F_1(\pi^*)+\lambda F_2(\pi^*)\right] \end{align}

and

\begin{align}
\bar{\lambda}F_1(\hat{\pi}_1)+\lambda F_2(\hat{\pi}_1)&\geq \mu\left[ \bar{\lambda}F_1(\hat{\pi}_1)+\lambda F_2(\hat{\pi}_1)\right]+(1-\mu)\left[ \bar{\lambda}F_1(\hat{\pi}_1)+\lambda F_2(\hat{\pi}_1)\right] \\
&\geq \mu\left[ \bar{\lambda}\alpha F_1(\pi^*)+\lambda \alpha F_1(\pi^*) \right]+(1-\mu)\left[0+\lambda \beta F_2(\pi^*)\right] \\
&\geq \frac{\mu\alpha}{\bar{\lambda}}\bar{\lambda}F_1(\pi^*)+(1-\mu) \beta \lambda F_2(\pi^*) \\
&\geq \min\{\frac{\mu\alpha}{\bar{\lambda}}, (1-\mu) \beta \}\left[\bar{\lambda}F_1(\pi^*)+\lambda F_2(\pi^*)\right]
\end{align}

$\min\{\frac{\mu\alpha}{\bar{\lambda}}, (1-\mu) \beta \}$ is a function over $0\leq \mu\leq 1$ and $\mu^*\in \arg\max_{\mu} \min\{\frac{\mu\alpha}{\bar{\lambda}}, (1-\mu) \beta \}$. It is easy to show

\begin{align}
\mu^*&=\frac{\bar{\lambda}\beta}{\bar{\lambda}\beta+\alpha} \\
\max_\mu\min\{\frac{\mu\alpha}{\bar{\lambda}}, (1-\mu) \beta \}&=\frac{\beta \alpha}{\bar{\lambda}\beta+\alpha}
\end{align}

\begin{align}
\bar{\lambda}F_1(\hat{\pi}_1)+\lambda F_2(\hat{\pi}_1)\geq \frac{\beta \alpha}{\bar{\lambda}\beta+\alpha}\left[\bar{\lambda}F_1(\pi^*)+\lambda F_2(\pi^*)\right]
\label{eqn:max_min_proof_general}
\end{align}

Taking the max over the two bounds leads to 
\begin{align}
\max\{\bar{\lambda}F_1(\hat{\pi}_1)+\lambda F_2(\hat{\pi}_1),\bar{\lambda}F_1(\hat{\pi}_2)+\lambda F_2(\hat{\pi}_2)\}\geq \max\{\frac{\beta \alpha}{\bar{\lambda}\beta+\alpha},\lambda\beta \} \max_{\pi\in\Pi}\bar{\lambda}F_1(\pi)+\lambda F_2(\pi)
\end{align}

Next we are going to show the result for the homogeneous setting. 
We have the following facts that hold for arbitrary partition $\pi$:
\begin{align}
F_1(\hat{\pi}_1)&\geq\alpha F_1(\pi),\ F_2(\hat{\pi}_2)\geq\beta F_2(\pi) \\
F_1(\pi)&\leq F_2(\pi),\ F_2(\pi_1)\geq \frac{1}{m} F_2(\pi)
\end{align}
Consider the following: 
\begin{align}
\bar{\lambda}F_1(\hat{\pi}_1)+\lambda F_2(\hat{\pi}_1)&\geq \alpha\bar{\lambda}F_1({\pi}^*)+\frac{\lambda}{m} F_2({\pi}^*) \\
&\geq\min\{\alpha, \frac{1}{m} \}\left[\bar{\lambda}F_1({\pi}^*)+\lambda F_2({\pi}^*) \right]
\end{align}
and
\begin{align}
\bar{\lambda}F_1(\hat{\pi}_2)+\lambda F_2(\hat{\pi}_2) &\geq \lambda F_2(\hat{\pi}_2) \\
&\geq \lambda \beta F_2(\pi^*) \\
&\geq \lambda \beta \left[\bar{\lambda} F_1(\pi^*)+\lambda F_2(\pi^*)\right] \end{align}
Taking the max over the two bounds and the result shown in Eqn~\ref{eqn:max_min_proof_general} gives the following:
\begin{align}
\max\{\bar{\lambda}F_1(\hat{\pi}_1)+\lambda F_2(\hat{\pi}_1),\bar{\lambda}F_1(\hat{\pi}_2)+\lambda F_2(\hat{\pi}_2)\}\geq \max\{\min\{\alpha,\frac{1}{m}\},\frac{\beta \alpha}{\bar{\lambda} \beta + \alpha},
 \lambda\beta \} \max_{\pi\in\Pi}\bar{\lambda}F_1(\pi)+\lambda F_2(\pi).
\end{align}

\end{proof}

\noindent{\bf Theorem }{\it
\textsc{CombSlbSmp} provides an approximation guarantee of $\min\{m, \frac{m\alpha}{m\bar{\lambda} + \lambda},\beta(m\bar{\lambda}+\lambda)\}$ in the \emph{homogeneous} case, and a factor of $\min\{\frac{m\alpha}{m\bar{\lambda}+\lambda},\beta(m\bar{\lambda}+\lambda) \}$ in the \emph{heterogeneous} case, for Problem~\ref{prob:minmax-genmix} with $0 \leq \lambda \leq 1$. 
}

\begin{proof}
Let $\hat{\pi}_1$ be the solution of \textsc{AlgWC} and $\hat{\pi}_2$ be the solution of \textsc{AlgAC}. 
Let $\pi^*\in \arg\min_{\pi\in\Pi}\bar{\lambda}F_1(\pi)+\lambda F_2(\pi)$ be the optimal partition. The following facts hold for all $\pi\in\Pi$:

Fact1
\begin{align}
F_1(\hat{\pi}_1)\leq\alpha F_1(\pi)
\end{align}
Fact2
\begin{align}
F_2(\hat{\pi}_2)\leq\beta F_2(\pi)
\end{align}
Fact3
\begin{align}
F_2(\pi)\leq F_1(\pi)\leq mF_2(\pi) 
\end{align}
Then we have the following: 
\begin{align}
\bar{\lambda}F_1(\hat{\pi}_1)+\lambda F_2(\hat{\pi}_1)&\leq F_1(\hat{\pi}_1) \\
&\leq \alpha F_1(\pi^*) \\
&\leq \frac{\alpha}{\bar{\lambda}+\frac{\lambda}{m}} \left[\bar{\lambda}F_1(\pi^*)+\frac{\lambda}{m} F_1(\pi^*)\right] \\
&\leq \frac{m\alpha}{m\bar{\lambda}+\lambda} \left[\bar{\lambda}F_1(\pi^*)+\lambda F_2(\pi^*)\right] 
\end{align}
and
\begin{align}
\bar{\lambda}F_1(\hat{\pi}_2)+\lambda F_2(\hat{\pi}_2) &\leq \beta m\bar{\lambda}F_2(\pi^*)+\beta\lambda F_2(\pi^*)\\
&\leq (m\bar{\lambda}+\lambda)\beta F_2(\pi^*) \\
&\leq (m\bar{\lambda}+\lambda)\beta\left[\bar{\lambda}F_1(\pi^*)+\lambda F_2(\pi^*)\right]\\
 \end{align}
Taking the minimum over the two leads to the following:
\begin{align}
\label{combine_min_max_and_min_sum_1}
\min\{\bar{\lambda}F_1(\hat{\pi}_1)+\lambda F_2(\hat{\pi}_1),\bar{\lambda}F_1(\hat{\pi}_2)+\lambda F_2(\hat{\pi}_2)\}\leq   \min\{\frac{m\alpha}{m\bar{\lambda}+\lambda},\beta(m\bar{\lambda}+\lambda) \} \min_{\pi\in\Pi}\bar{\lambda}F_1(\pi)+\lambda F_2(\pi)
\end{align}
Equation \ref{combine_min_max_and_min_sum_1} gives us a bound for both the homogeneous setting and the heterogeneous settings. 

Furthermore, in the homogeneous setting, for arbitrary partition $\pi$, we have 

\begin{align}
\label{combine_min_max_and_min_sum_2}
\bar{\lambda}F_1({\pi})+\lambda F_2({\pi}) \leq   m \min_{\pi\in\Pi}\bar{\lambda}F_1(\pi)+\lambda F_2(\pi)
\end{align} 
and we can tighten the bound for the homogeneous setting as follows: 
\begin{align}
\label{combine_min_max_and_min_sum_3}
\min\{\bar{\lambda}F_1(\hat{\pi}_1)+\lambda F_2(\hat{\pi}_1),\bar{\lambda}F_1(\hat{\pi}_2)+\lambda F_2(\hat{\pi}_2)\}&\leq   \min\{m,\frac{m\alpha}{m\bar{\lambda}+\lambda},\beta(m\bar{\lambda}+\lambda) \} \max_{\pi\in\Pi}\bar{\lambda}F_1(\pi)+\lambda F_2(\pi) 
\end{align}
\end{proof}

\subsection*{Proof for Theorem~\ref{thm:generalGreedSat_bounds}}
{\bf Theorem }{\it
Given $\epsilon$, $\alpha$, and, $0 \leq \lambda\leq 1$, \textsc{GeneralGreedSat} finds a partition $\hat{\pi}$ that satisfies the following:
\begin{align}
\bar{\lambda}\min_{i} f_i(A_i^{\hat{\pi}}) + \lambda \frac{1}{m}\sum_{i=1}^m f_i(A_i^{\hat{\pi}}) \geq \lambda \alpha (OPT-\epsilon),
\end{align}
%\begin{itemize}
%\item $\bar{\lambda}\min_{i} f_i(A_i^{\hat{\pi}}) + \lambda \frac{1}{m}\sum_{i=1}^m f_i(A_i^{\hat{\pi}}) \geq \lambda \alpha OPT$,
%\item At least $\ceil{m( \alpha-\delta)}$ blocks receive utility at least $\max\{\frac{\delta}{1-\alpha + \delta}, \lambda \alpha \} ( OPT-\epsilon)$,
%\end{itemize}
where $OPT= \max_{\pi\in \Pi} \bar{\lambda} \min_{i}f_i(A_i^{\pi}) + \lambda \frac{1}{m} \sum_{i=1}^m f_{i}(A_i^{\pi})$.

Moreover, let $F_{\lambda,i}(\pi) = \bar{\lambda}f_i(A_i^{\pi}) +\lambda \frac{1}{m}\sum_{j=1}^m f_j(A_j^{\pi})$. 
Given any $0 < \delta < \alpha $, 
there is a set $I\subseteq \{1,\dots,m\}$ such that 
$|I| \geq \ceil{m( \alpha-\delta)}$ and 
\begin{align}
F_{i,\lambda} (\hat{\pi}) \geq \max\{\frac{\delta}{1-\alpha + \delta}, \lambda \alpha\}(OPT-\epsilon),\forall i\in I.
\end{align}
}
\begin{proof}
Denote intermediate objective $\bar{F}^{c} (\pi) = \frac{1}{m}\sum_{i=1}^m \min\{ \bar{\lambda}f_i(A_i^{{\pi}}) + \lambda \frac{1}{m}\sum_{j=1}^m f_j(A_j^{{\pi}}), c\}$. Also we define the overall objective as $F(\pi) = \bar{\lambda} \min_{i}f_i(A_i^{\pi}) + \lambda \frac{1}{m} \sum_{i=1}^m f_{i}(A_i^{\pi})$. 
When the algorithm terminates, it identifies a $c_{\min}$ such that the returned solution $\hat{\pi}^{c_{\min}}$ satisfies $\bar{F}^{c_{\min}} (\hat{\pi}^{c_{\min}}) \geq \alpha c_{\min}$. Also it identifies a $c_{\max}$ such that the returned solution 
$\hat{\pi}^{c_{\max}}$ satisfies $\bar{F}^{c_{\max}} (\hat{\pi}^{c_{\max}}) < \alpha c_{\max}$. The gap between $c_{\max}$ and $c_{\min}$ is bounded by $\epsilon$, i.e., $c_{\max} - c_{\min} \leq \epsilon$.

Next, we prove that there does not exist any partitioning $\pi$ that satisfies $F(\pi) \geq c_{\max}$, i.e., $c_{\max} \geq OPT$. 

Suppose otherwise, i.e., $\exists \pi^*: F(\pi^*) = c_{\max}+\gamma$ with $\gamma \geq 0$. Let $c = c_{\max} + \gamma$, consider the intermediate objective $\bar{F}^{c} (\pi)$, we have that $\bar{F}^{c}(\pi^*) = c$. An instance of the algorithm for SWP on $\bar{F}^{c}$ is guaranteed to lead to a solution $\hat{\pi}^c$ such that $\bar{F}^{c}(\hat{\pi}^c) \geq \alpha c$. 
Since $c \geq c_{\max}$, it should follow that the returned solution $\hat{\pi}^{c_{\max}}$ for the value $c_{\max}$ also satisfies $\bar{F}^{c_{\max}}(\hat{\pi}) \geq \alpha c_{\max}$. However it contradicts with the termination criterion of \textsc{GreedSat}. Therefore, we prove that $c_{\max} \geq OPT$, which indicates that $c_{\min} \geq c_{\max} - \epsilon \geq OPT-\epsilon$.

Let $c^*= \frac{c_{\max} + c_{\min}}{2}$ and the partitioning returned by running for $c^*$ be $\hat{\pi}$ (the final output partitioning from the algorithm). We have that
$\bar{F}^{c^*}(\hat{\pi}) \geq \alpha c^*$.

Next we are ready to prove the Theorem: $F(\hat{\pi}) \geq \lambda \alpha$. For simplicity of notation, we rewrite $y_i = \bar{\lambda} f_i(A_i^{\hat{\pi}}) + \lambda \frac{1}{m} \sum_{j=1}^m f_j(A_j^{\hat{\pi}})$ and $x_i = \min\{\bar{\lambda} f_i(A_i^{\hat{\pi}}) + \lambda \frac{1}{m} \sum_{j=1}^m f_j(A_j^{\hat{\pi}}), c^*\} = \min\{y_i, c^*\}$ for each $i$. Furthermore, we denote the sample mean as $\bar{x} = \frac{1}{m}\sum_{i=1}^m x_i$ and $\bar{y} = \frac{1}{m} \sum_{i=1}^m y_i$.
Then, we have $F(\hat{\pi}) = \min_i y_i$ and $\bar{F}^{c^*}(\hat{\pi}) = \bar{x}$. We list the following facts to facilitate the analysis:
\begin{enumerate}
\item $0 \leq  x_i\leq c^*$ holds for all $i$;
\item $y_i \geq \lambda \bar{y}$ holds for all $i$;
\item $x_i \geq \lambda \bar{x}$ holds for all $i$;
\item $\bar{x} \geq \alpha c^*$;
\item $x_i = \min \{y_i, c^*\}, \forall i$.
\end{enumerate}
The second fact follows since 
\begin{align}
\bar{y} &= \frac{1}{m}\sum_{i=1}^m y_i 
\\ &= \frac{1}{m} \sum_{i=1}^m  \{\bar{\lambda} f_i(A_i^{\hat{\pi}}) + \lambda \frac{1}{m} \sum_{j=1}^m f_j(A_j^{\hat{\pi}})\}
\\ &= \frac{1}{m}\sum_{j=1}^m f_j(A_j^{\hat{\pi}}) \leq \frac{y_i}{\lambda}
\end{align}

Given the second fact, we can prove the third fact as follows. 
Let $i^* \in \argmin_{i} y_i$. 
By definition $x_i=\min\{y_i,c^*\}$, 
then $i^* \in \argmin_{i}x_i$. We consider the two cases: 

(1) $y_{i^*} \leq c^*$: In this case, we have that $x_{i^*} = y_{i^*}$. Since $x_i\leq y_i,\forall i$, it holds that $\bar{x}\leq \bar{y}$. The third fact follows as $x_{i^*} = y_{i^*} \geq \lambda \bar{y} \geq \lambda \bar{x}$. 

(2) $y_{i^*} > c^*$: In this case, $y_{i} \geq c^*$ holds for all $i$. As a result, we have $x_i = c^*, \forall i$. Therefore, $x_i = \bar{x} = c^* \geq \lambda c^*$.

Combining fact 3 and 4, it follows for each $i$:
\begin{align}
\bar{\lambda} f_i(A_i^{\hat{\pi}}) + \lambda \frac{1}{m} \sum_{j=1}^m f_j(A_j^{\hat{\pi}}) = y_i \geq x_i \geq \lambda \bar{x}  \geq \alpha \lambda c^* \geq \alpha \lambda (OPT - \epsilon).
\label{eqn:proof_individual_bound}
\end{align}
The first part of the Theorem is then proved.

The second part of the Theorem simply follows from the proof in Theorem~\ref{thm:greedy_sat_bound} and  Eqn~\ref{eqn:proof_individual_bound}.
\end{proof}

\subsection*{Proof for Theorem~\ref{thm:general_lovasz_round}}
{\bf Theorem }{\it
Define $F^\lambda (\pi) = \bar{\lambda} \max_{i}f_i(A_i^{{\pi}})  + \lambda \frac{1}{m} \sum_{i=1}^m f_i(A_i^{{\pi}})$ for any $0 \leq \lambda \leq 1$.
\textsc{General\lovasz Round} is guaranteed to find a partition $\hat{\pi} \in \Pi$ such that
\begin{align}
F^{\lambda}(\hat{\pi}) \leq m \min_{\pi\in \Pi} F^{\lambda}(\pi)
\end{align}
}
\begin{proof}
We essentially use the same proof technique in Theorem~\ref{thm:lovaszRound_bound} to show this result. 
After solving for the continuous solution $\{x^*_i\in \mathbb{R}^n\}_{i=1}^m$, 
the rounding step simply chooses for each $j=1,\dots, n$, assigns the item to the block $i^*\in \argmax_{i=1,\dots,m}x_i^*(j)$. 
We denote the resulting partitioning as $\hat{\pi} = \{A_i^{\hat{\pi}}\}_{i=1}^m$.

It suffices to bound the performance loss at the step of rounding the fractional solution $\{x^*_i\}_{i=1}^m$, or equivalently, the following:
\begin{align}
\tilde{f}_i(x_i^*) \geq  \frac{1}{m} f_i(A^{\hat{\pi}}_i),
\label{eqn:m_bound}
\end{align} 
Given Eqn~\ref{eqn:m_bound}, the Theorem follows since 
\begin{align}
F^{\lambda}(\pi^*) &\geq \bar{\lambda} \max_i \tilde{f}_i(x_i^*) + \lambda \frac{1}{m} \sum_{j=1}^m \tilde{f}_j(x_j^*)
\\ & \geq \frac{1}{m} [\bar{\lambda} \max_i f_i(A_i^{\hat{\pi}}) + \frac{\lambda}{m} \sum_{j=1}^m f_j(A_j^{\hat{\pi}})]
\\ & \geq \frac{1}{m} F^{\lambda}(\hat{\pi}).
\end{align}

To prove Eqn~\ref{eqn: m_bound}, consider the following:
\begin{align}
f_i(A^{\hat{\pi}}_i) &= \tilde{f}_i (1_{A^{\hat{\pi}}_i})  = m \tilde{f}_i(\frac{1}{m} 1_{A^{\hat{\pi}}_i}) // \mbox { positive homogeneity of \lovasz{} extension}
\end{align}

For any item $v_j\in A^{\hat{\pi}}_i$, we have $x^*_i(j) \geq \frac{1}{m}$, since $\sum_{i=1}^m x^*_i(j) \geq 1$ and $x^*_i(j) = \max_{i^\prime} x_{i^\prime} (j)$. Therefore, we have $\frac{1}{m} 1_{A_i} \leq x_i^*$. Since $f_i$ is monotone, its extension $\tilde{f}_i$ is also monotone. As a result, $f_i(A_i) = m \tilde{f}_{i}(\frac{1}{m}1_{A_i}) \leq m \tilde{f}_i (x_i^*)$.
\end{proof}

\bibliography{submod}

\end{document}